\documentclass[12pt]{article}
\usepackage{graphicx,amsmath}
\usepackage{units}
\usepackage{color}

\newlength{\dinwidth}
\newlength{\dinmargin}
\def\lapproxeq{\lower .7ex\hbox{$\;\stackrel{\textstyle                                                    
<}{\sim}\;$}}                                                    
\def\gapproxeq{\lower .7ex\hbox{$\;\stackrel{\textstyle                                                    
>}{\sim}\;$}}                                                    
\def\be{\begin{equation}}                                                    
\def\ee{\end{equation}}                                                    
\def\bea{\begin{eqnarray}}                      
\def\eea{\end{eqnarray}}

\def\GeV{\rm GeV}
\def\TeV{\rm TeV}

\def\sh{\hat s}
\def\sh2{{\hat s}^2}

\def\MS{\overline{\rm MS}}

\def\a{\alpha_S(M_Z^2)}

\begin{document}

\begin{flushright}                                                    
LCTS/2015-17  \\
IPPP/15/33  \\
DCPT/15/66 \\                                                    
\today \\                                                    
\end{flushright} 

\vspace*{0.5cm}

\begin{center}
{\Large \bf Uncertainties on $\alpha_S$ in the MMHT2014 global PDF }\\ 
\vspace*{0.5cm}{\Large \bf analysis and implications for SM predictions}\\

\vspace*{1cm}
L. A. Harland-Lang$^{a}$, A. D. Martin$^b$, P. Motylinski$^a$ and R.S. Thorne$^a$\\                                               
\vspace*{0.5cm}                                                    
                                                  
$^a$ Department of Physics and Astronomy, University College London, WC1E 6BT, UK \\           
$^b$ Institute for Particle Physics Phenomenology, Durham University, DH1 3LE, UK                                                    
                                                    
\vspace*{1cm}

\begin{abstract}
\noindent We investigate the uncertainty in the strong coupling $\alpha_S(M_Z^2)$
when allowing it to be a free parameter in the recent MMHT global
analyses of  deep-inelastic and related hard scattering data that was
undertaken to determine the parton distribution functions (PDFs) of the
proton. The analysis uses the standard framework of leading twist
fixed--order collinear factorisation in the ${\overline {\rm MS}}$
scheme. We study the constraints on $\alpha_S(M_Z^2)$ coming from 
individual data sets by repeating the NNLO and NLO fits
spanning the range $0.108$ to $0.128$ in units of $0.001$, 
making all PDFs sets available. The inclusion of the cross section for inclusive $t\bar{t}$
production allows us to explore the correlation
between the mass $m_t$ of the top quark and $\alpha_S(M_Z^2)$.
We find that the best fit values are $\alpha_S(M_Z^2)=0.1201\pm 0.0015$ and $0.1172\pm 0.0013$ 
at NLO and NNLO respectively, with the central values changing to $\a=0.1195$ and $0.1178$ when the world average 
of $\alpha_S(M_Z^2)$ is used as a data point. 
We investigate the interplay between the uncertainties on $\alpha_S(M_Z^2)$ and on the PDFs. In particular we calculate the cross sections
for key processes at the LHC and show how the uncertainties from
the PDFs and from $\alpha_S(M_Z^2)$ can be provided independently and be combined.

\end{abstract}
                                                   
\vspace*{0.5cm}                                                    
                                                    
\end{center}

\section{Introduction  \label{sec:1}} 
There has been a continual improvement in the precision and in the variety of 
the data for deep--inelastic and related hard--scattering processes.  Noteworthy 
additions in the years since the MSTW2008 analysis~\cite{MSTW} have been the 
HERA combined H1 and ZEUS data on the total~\cite{H1+ZEUS} and charm structure functions~\cite{H1+ZEUScharm}, and 
the variety of new data sets obtained at the LHC, as well as updated Tevatron 
data (for full references see~\cite{MMHT}). Moreover the procedures used in the global PDF analyses of these data have 
been refined, allowing the partonic structure of the proton to be determined 
with ever--increasing accuracy and reliability. These improvements are important 
as it is necessary to quantify the Standard Model background as accurately as 
possible in order to isolate possible experimental signals of New Physics.  One 
area that needs careful attention, at the present level of accuracy, is the 
treatment of the strong coupling, $\alpha_S$ itself, in the global analyses.  
Here we extend the recent MMHT2014 global PDF analysis~\cite{MMHT} to study the 
uncertainties 
on $\alpha_S$ and their implications for predictions for processes at the LHC.

\section{Treatment of $\a$ in the MMHT2014 analysis}

\begin{figure} [t]
\begin{center}
\vspace*{-1.0cm}
\includegraphics[height=9cm]{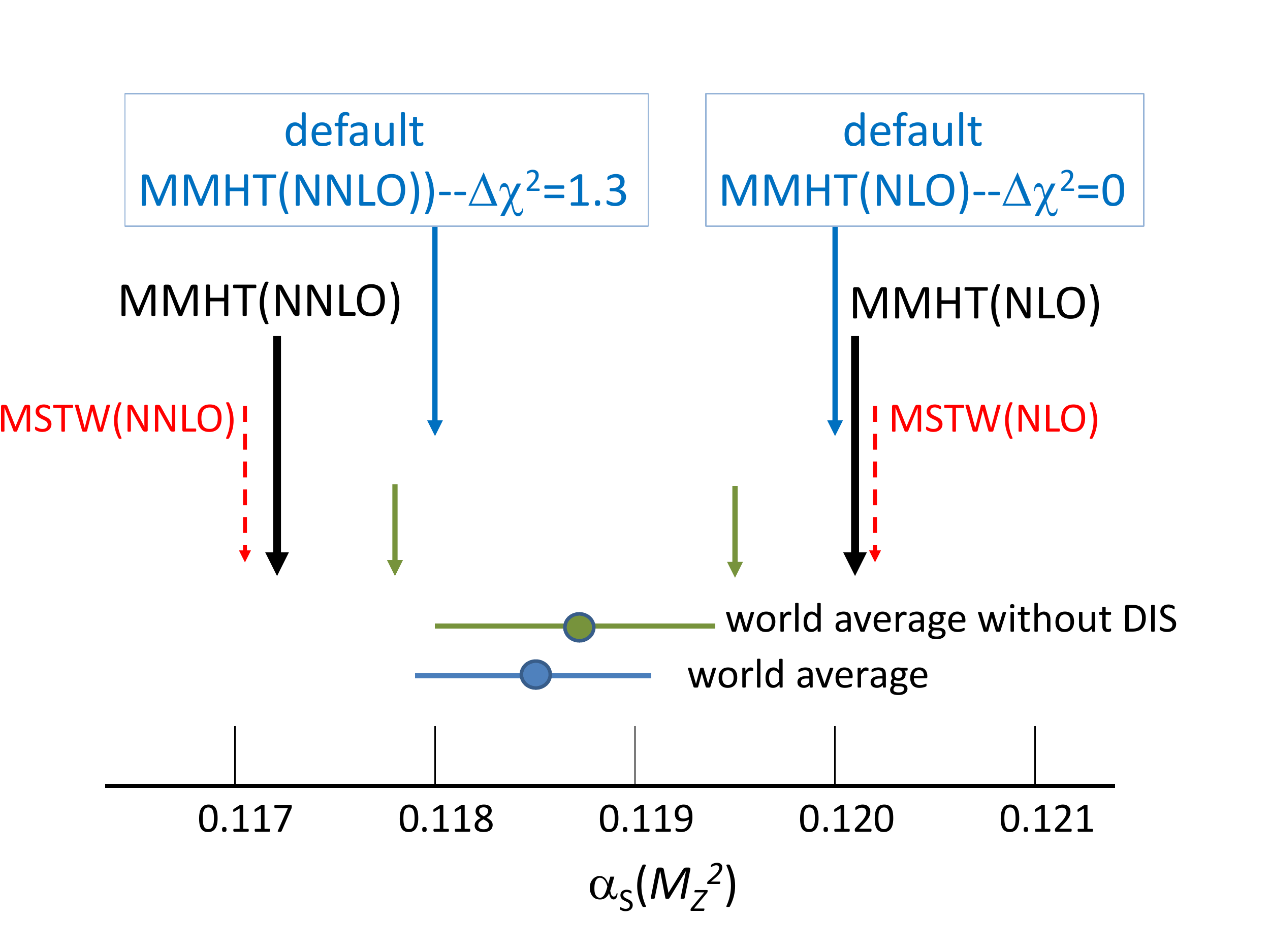} 
\caption{\sf The dark arrows indicate the optimal values of $\alpha_S(M_Z^2)$ 
found in NLO and NNLO fits of the MMHT2014 analysis \cite{MMHT}. The dashed 
arrows indicate the values found in the MSTW2008 analysis~\cite{MSTW}. We also show the world average value, which we note was obtained assuming, for 
simplicity, that the NLO and NNLO values are the same -- which, in principle, 
is not the case.  The short arrows are also of interest as they indicate the NLO and NNLO values which would have been obtained 
from the MMHT2014 global analyses if the world average value (obtained 
without including DIS data) were to be included in the fit. However, the default 
values $\alpha_{S,{\rm NLO}}=0.120$ and $\alpha_{S,{\rm NNLO}}=0.118$ were selected for the final MMHT2014 PDF sets `for ease of use'; indeed, the small values of $\Delta \chi^2$ are the minute changes in $\chi^2_{\rm global}$ in going from the optimal to these default fits. }
\label{fig:alpha}
\end{center}
\end{figure}

\begin{figure} [t]
\begin{center}
\vspace*{-1.0cm}
\includegraphics[height=6cm]{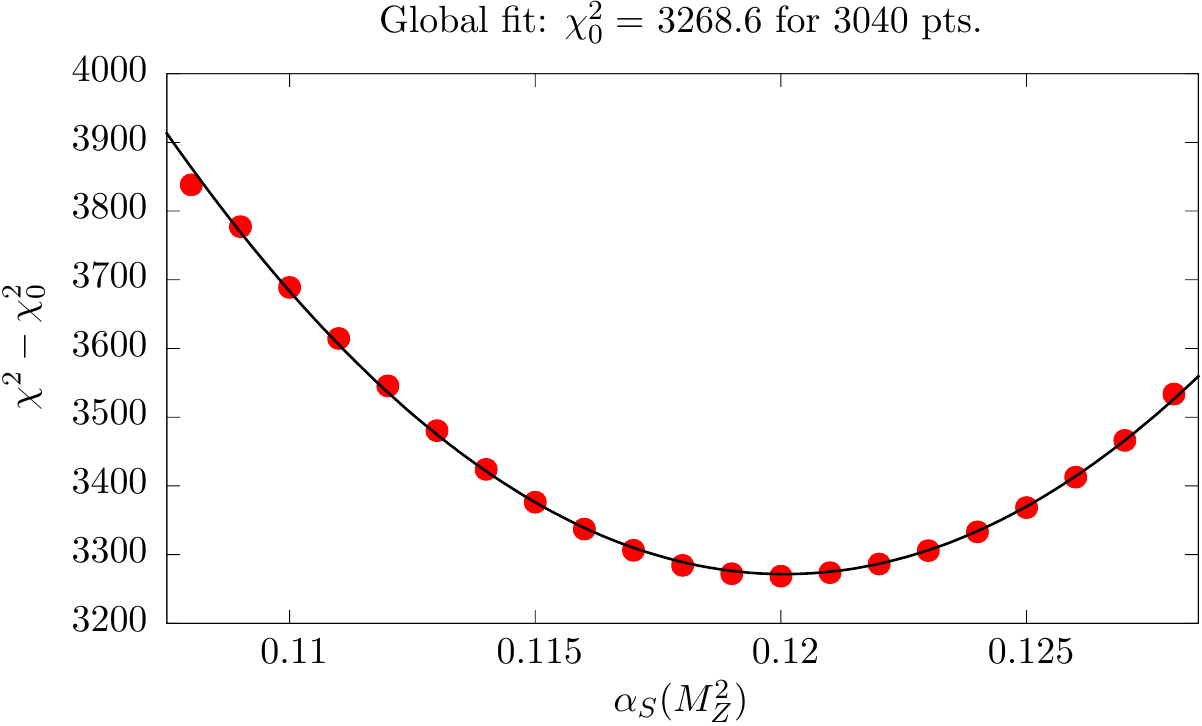}
\includegraphics[height=6cm]{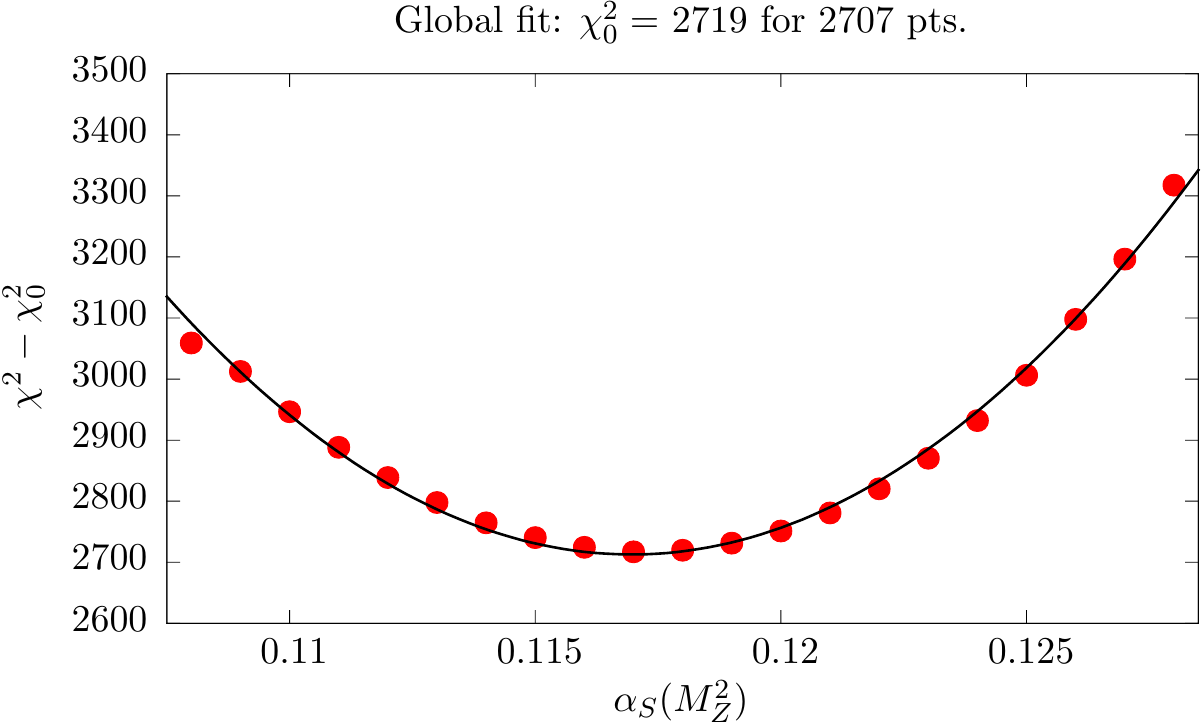}
\caption{\sf The upper and lower plots show total $\chi^2$ as a function of the value of the parameter $\a$ for the NLO and NNLO MMHT2014 fits, respectively.  }
\label{fig:total}
\end{center}
\end{figure}

We refer to Fig.~\ref{fig:alpha} for an overview of the treatment and of the 
values of $\alpha_S$ obtained in the MMHT2014 global PDF analysis \cite{MMHT}.
At both NLO and NNLO the value of $\a$ is allowed to vary just as another free parameter in the global fit. The best fit values are found to be
\bea
\alpha_{S,{\rm NLO}}(M_Z^2) & = & 0.1201 \label{eq:optNLO}\\
\alpha_{S,{\rm NNLO}}(M_Z^2) & = & 0.1172, \label{eq:optNNLO}
\eea
as indicated by the dark arrows in Fig.~\ref{fig:alpha}. The corresponding total 
$\chi^2$ profiles versus $\a$ are shown in Fig.~\ref{fig:total}. These plots 
clearly show the reduction in the optimum value of $\a$ as we go from the NLO to 
the NNLO analysis.  In the next section we show how the individual data sets 
contribute to make up this $\chi^2$ profile versus $\a$.


It is sometimes debated whether one should attempt to extract the value of 
$\alpha_S(M_Z^2)$ from the PDF global fits or to simply use a fixed value taken 
from elsewhere -- for example, to use the world average value~\cite{PDG2014}.
However, we believe that very useful information on the coupling can be obtained 
from PDF fits, and hence have performed fits where this is left as a free 
parameter. As the extracted values of $\alpha_S(M_Z^2)$ in the NLO and NNLO 
MMHT2014 analyses \cite{MMHT} reasonably bridge the world average of 
$\alpha_S(M_Z^2)=0.1185\pm 0.0006$~\cite{PDG2014}, we regard these as our best fits. We note it
is a common result in PDF analyses, and in other extractions of the strong 
coupling, for the best fit value to fall slightly as the order of the 
theoretical calculations increases.
However, in order to explore further, as well as leaving $\alpha_S(M^2_Z)$ as a 
completely independent parameter, the MMHT2014 analyses were repeated
including the world average value (without the inclusion of DIS data to avoid 
double counting) of $\alpha_S(M_Z^2)=0.1187\pm 0.0007$ as a data point in the fit. 
This changed the preferred values to 
\be
\alpha_{S,{\rm NLO}}(M^2_Z)=0.1195 \quad {\rm and} \quad \alpha_{S,{\rm NNLO}}(M^2_Z)=0.1178,
\ee 
as indicated by the short arrows in Fig~\ref{fig:alpha}. Each of these is about 
one standard deviation away from the world average, 
so our PDF fit is entirely consistent with the independent determinations of 
the coupling. Moreover, the quality of the fit to the data (other than the 
single `data'
point on $\alpha_S(M_Z^2)$) increases by about 1.5 units in $\chi^2$ at NLO and just 
over one unit at NNLO when the coupling was included as a data point.

However, ultimately for the use of PDF sets by external users 
it is preferable to present the sets at common (and hence `rounded') 
values of $\alpha_S(M_Z^2)$ in order to compare and combine with PDF 
sets from other groups, for example as in \cite{PDF4LHC1, PDF4LHC2, bench1, bench2}. At NLO in the MMHT2014 analysis \cite{MMHT} we hence chose $\alpha_S(M_Z^2)=0.120$ as the default 
value for which the PDF sets with full error eigenvectors are made available. 
This is 
essentially identical to the value for the best PDF fit when the coupling is 
free, and still very similar when the world average was included as a 
constraint. At NNLO, $\alpha_S(M_Z^2)=0.118$ was chosen as a rounded value, 
very near to both the best fit value and the world average, and the fit quality 
is still 
only 1.3 units in $\chi^2$ higher than that when the coupling was free. This 
is extremely close to the 
value determined when the world average is included as a data point. Hence, in 
MMHT2014 \cite{MMHT}, we chose
to use $\alpha_S(M_Z^2)=0.118$ as the default for NNLO PDFs, a value which is 
very consistent with the world average.  At NLO we also made a set available with 
$\alpha_S(M_Z^2)=0.118$, but in this case the $\chi^2$ increases by 17.5 units 
from the best fit value. In \cite{MMHT} we also made available PDF sets 
corresponding to the best fit for $\alpha_S(M_Z^2)$ values $\pm 0.001$ relative 
to the default values in order for users to determine the 
$\alpha_S(M_Z^2)$--uncertainty in predictions if so desired. We will return to 
the issue of PDF+$\alpha_S(M_Z^2)$ uncertainty later.       

Before we continue we should specify how the running of $\alpha_S(Q^2)$ is 
treated. There is more than one definition of the coupling commonly used in 
QCD phenomenology. Although the various prescriptions are all formally 
equivalent since they differ only at higher orders, numerical differences of 
the order of up to 1$\%$ can  occur. We use the definition based on the full 
solution of the renormalisation group equation, in $\MS$ scheme, at the 
appropriate order, with boundary condition defined by the value of 
$\alpha_S(M_Z^2)$. This is identical to the definition in public codes such as 
\textsc{pegasus}~\cite{Vogt:2004ns} and \textsc{hoppet}~\cite{Salam:2008qg}, 
and is now effectively the standard in PDF analyses.\footnote{In $\MS$ scheme
this involves discontinuities at flavour transition points at NNLO. 
For a suggestion for a smooth transition
in a physical scheme see~\cite{deOliveira:2013tya}.} It differs, for 
example, from solutions to the renormalisation group equations truncated at a 
particular order.    


\section{Description of data sets as a function of $\alpha_S$} \label{sec:desc}

The NNLO MMHT2014 global analysis \cite{MMHT} was based on a fit to 40 
different sets of data on deep--inelastic and related hard scattering 
processes. There were 10 different data sets of structure functions from the 
fixed--target charged lepton--nucleon experiments of the SLAC, BCDMS, NMC and 
E665 collaborations, six different neutrino data sets on $F_2,~xF_3$ and dimuon 
production from the NuTeV, CHORUS and CCFR collaborations, two Drell--Yan data 
sets from E886/NuSea, six different data sets from HERA involving the combined 
H1 and ZEUS structure function data, seven data sets from the Tevatron giving 
the measurements of inclusive jet, $W$ and $Z$ production by the CDF and D0 
collaborations and, finally, nine data sets from the ATLAS, CMS and LHCb 
collaborations at the LHC. In addition, the NLO fit also used jet data from the
ATLAS, CMS and H1 and ZEUS collaborations, which were not used at NNLO because it was judged
that at present there is not sufficient knowledge of the full jet  NNLO cross 
section; jet production at the Tevatron, on the other hand, is much nearer to threshold than at the LHC, so the threshold
approximation to the full NNLO calculation is much more likely to provide a reliable estimate in this case.
The goodness--of--fit quantity, $\chi^2_n$, for each of the data sets, 
$n=1,...40$, is given for the NLO and NNLO global fits in the 
Table 5 of~\cite{MMHT}, and the $\chi^2$ definition is explained in Section 
2.5 of the same article.  The references to all the data that are fitted are also given in~\cite{MMHT}.


In the NNLO global fit of~\cite{MMHT}, let us denote the contribution to the 
total $\chi^2$ from data set $n$ by $\chi^2_{n,0}$.
Here we explore the $\chi^2_n$ profiles as a function of $\alpha_S(M_Z^2)$ by 
repeating the global fit for different fixed values of $\alpha_S(M_Z^2)$ in the 
neighbourhood of the optimum value given in (\ref{eq:optNNLO}). The results 
are shown in Figs. 3--7, where
we plot the $\chi^2_n$ profiles for each data set $n$ as the difference from 
the value at the \emph{global} minimum, $\chi^2_{n,0}$, when varying 
$\alpha_S(M_Z^2)$.  The points ($\bullet$) in Figs. 3--7 are generated for fixed 
values of $\alpha_S(M_Z^2)$ between 0.108 and 0.128 in steps of 0.001.  These 
points are then fitted to a quadratic function of $\alpha_S(M_Z^2)$ shown by 
the continuous curves.  By definition we expect the profiles to satisfy 
$(\chi^2_n-\chi^2_{n,0})=0$ at $\alpha_S(M_Z^2)=0.1172$, corresponding to the 
value of $\alpha_S(M_Z^2)$ at the NNLO global minimum.  Ideally, a data set 
should show a quadratic minimum about this point.  Of course, in practice, the 
various data sets may pull, in varying degrees, to smaller or larger values 
of $\alpha_S(M_Z^2)$. There is a small amount of point--to--point fluctuation
for the values of $(\chi^2_n-\chi^2_{n,0})$, even near the minimum, but near
the minimum this is generally only at the level of fractions of a unit 
in $\chi^2$ for a given data set. The fluctuations become larger as we go to 
values of $\a$ far from the minimum, particularly for lower $\a$, mainly 
because changes in $\chi^2$ with small changes in $\a$ are becoming much greater. However, some of the ``jumps'' for individual sets
near $\a=0.108$ imply that the global minimum in $\chi^2$ for this choice
of $\a$ is rather flat in certain parameter directions, with some relatively
easy trade--off between the data sets which are poorly fit, and a transition
to a different, approximately degenerate global minimum occurring with a 
small change in $\a$. Indeed, we have verified that at $\a=0.108$ there is a 
local minimum where ``jumps'' are eliminated, but with slightly higher 
global $\chi^2$ than the result where there are ``jumps''. 
This highlights the fact that the PDF uncertainty
is difficult to define properly for a value of $\a$ which is far from optimal
and leads to many data sets being badly fit.     

We repeat this exercise at NLO. Then the profiles will satisfy  
$(\chi^2_n-\chi^2_{n,0})=0$ at $\alpha_S(M_Z^2)=0.1201$. We include in the plots the NLO points (as triangles) and show the 
corresponding quadratic fit by a dashed curve. In Fig. 8 we show the 
$\chi^2_n$ profiles for the LHC and HERA jet data that were included in the 
NLO fit. Here the bullet points and profile curve correspond to the 
NLO fit.  These data were not included in the NNLO fit.

The fixed--target structure function data in the first 14 plots in Figs. 3 and 
4 have been available for several years. These data play an important role in 
constraining the value of $\alpha_S(M_Z^2)$.
There is some tension between these data sets. The BCDMS (and also the 
E665) data prefer values of $\alpha_S(M_Z^2)$ around 0.110. On the other hand 
the NMC data prefer values around 0.122; and the SLAC  $F_2^{p,d}$ data prefer 
$\alpha_S(M_Z^2)$ values around 0.115 and 0.122 respectively. The neutrino 
$F_2$ and $xF_3$ data prefer $\alpha_S(M_Z^2) \sim 0.120$; while neutrino 
dimuon production has little dependence on $\alpha_S(M_Z^2)$, since the 
extra $B(D\to \mu)$ branching ratio parameter (see eq.(19) of \cite{MMHT}) 
can partially compensate for the changes in $\alpha_S(M_Z^2)$.

The NNLO corrections to the structure functions are positive and speed up the 
evolution, leading to smaller optimum values of  $\alpha_S(M_Z^2)$ than those 
at NLO, such that the spread of optimum values of $\alpha_S(M_Z^2)$ for the 
different data sets is somewhat reduced. Thus the overall fit to this subset 
of the data is marginally better at NNLO. The difference 
$\alpha_{S,{\rm NNLO}}<\alpha_{S,{\rm NLO}}$ is clearly evident in the majority of the 
corresponding plots.

The recent combined H1 and ZEUS structure function data from HERA prefer a 
value of $\alpha_S(M_Z^2)$ of about 0.120 at NNLO. Perhaps the only surprising 
result is the  $\alpha_S(M_Z^2)$ behaviour of the combined data for 
$F_2^{\rm charm}$ which prefers a very low value of  $\alpha_S(M_Z^2)$ at NNLO, 
whereas the uncombined data had a perfect quadratic behaviour about 0.118, see 
Fig.5 of \cite{MSTWalpha}. Note, however, that the combined data contains some 
points at the lowest $Q^2$ which were not available as an individual data set.
These data, particularly at low $Q^2$, are sensitive to the value of the charm 
mass $m_c$, and there is a correlation between its value and
$\alpha_s(M_Z^2)$ ~\cite{MSTWhf}. This will be studied again with 
the up--to--date data in a future article.

\begin{figure}
  \centering
  \includegraphics[width=0.45\textwidth]{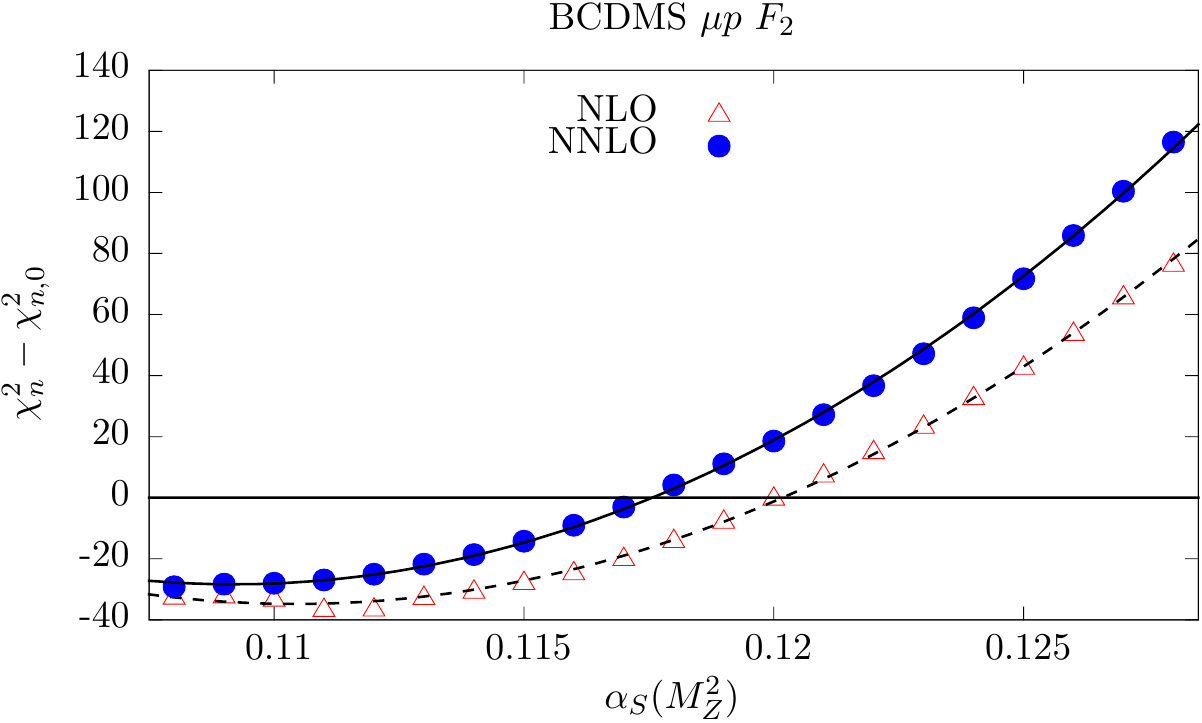}\qquad
    \includegraphics[width=0.45\textwidth]{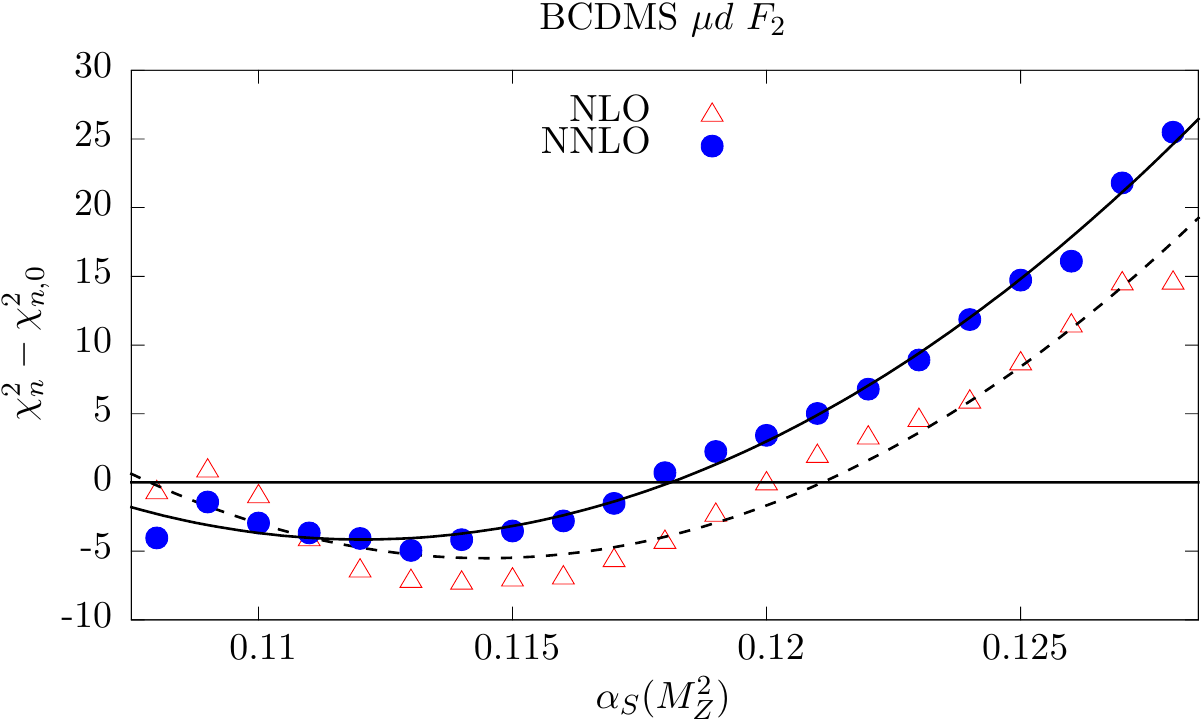}
      \includegraphics[width=0.45\textwidth]{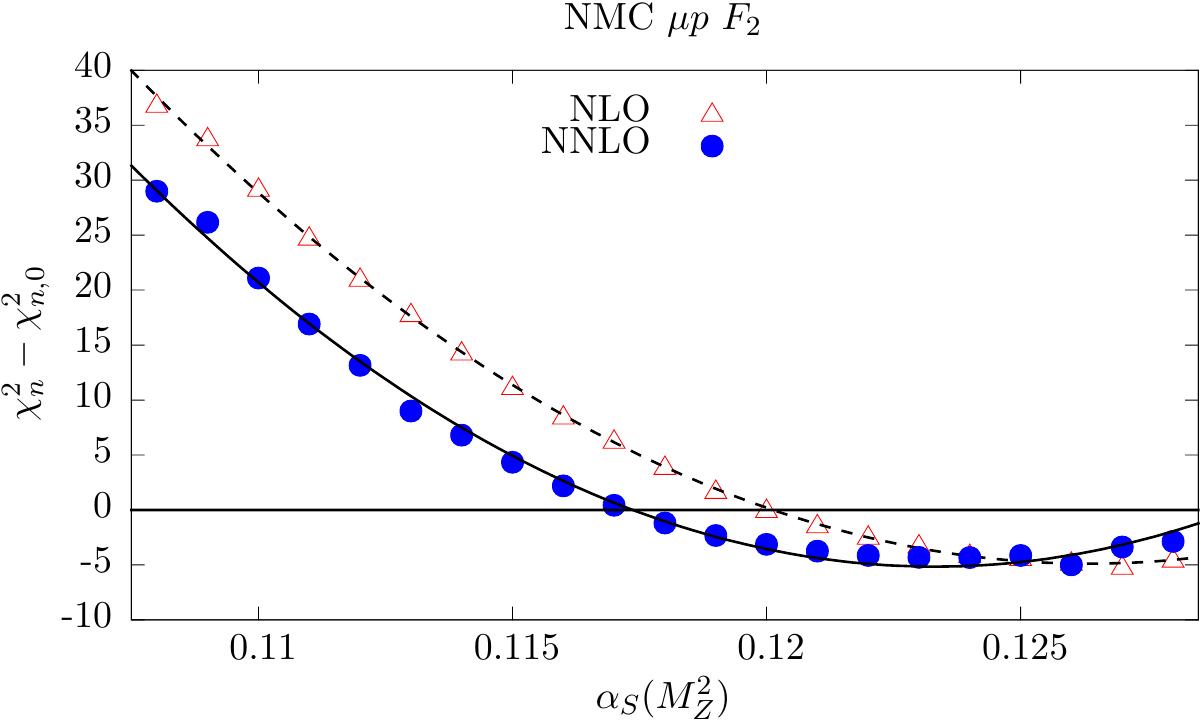}\qquad
    \includegraphics[width=0.45\textwidth]{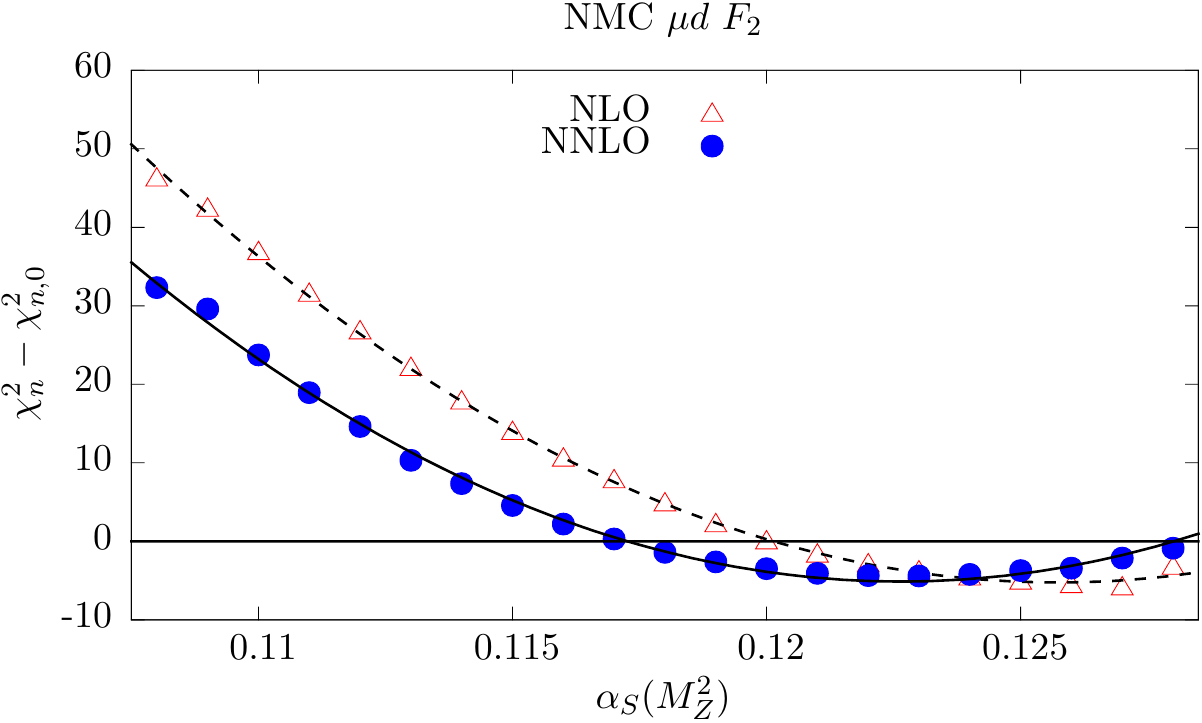}
          \includegraphics[width=0.45\textwidth]{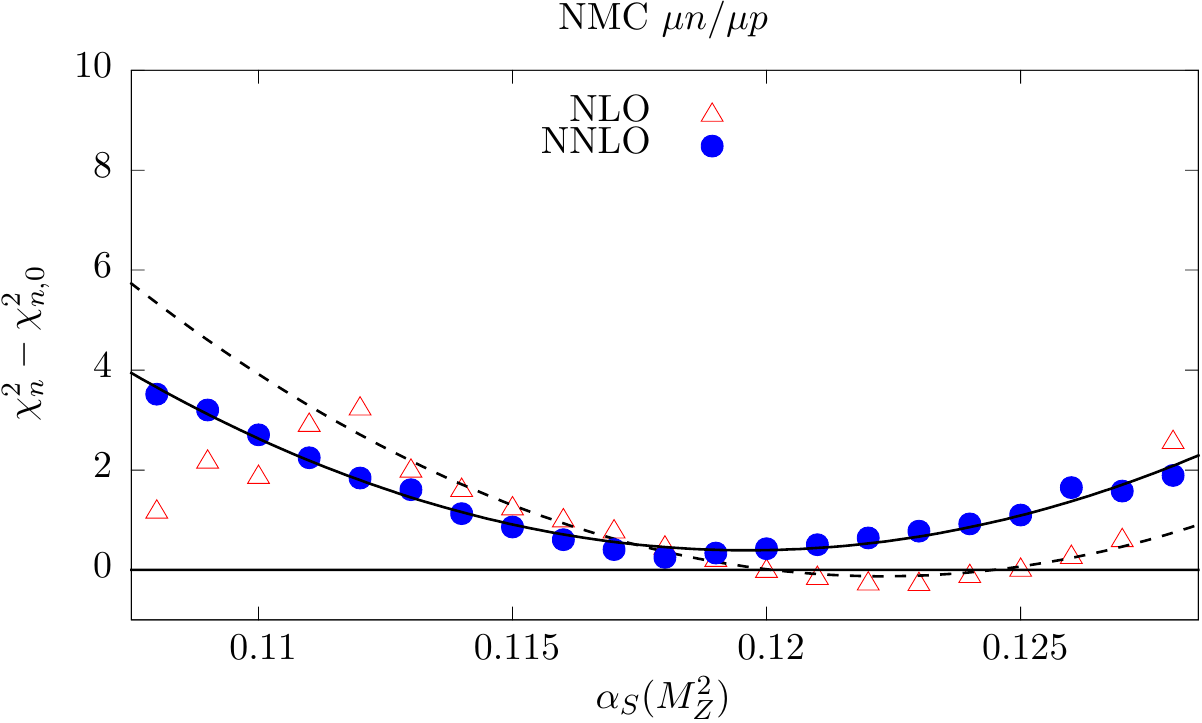}\qquad
    \includegraphics[width=0.45\textwidth]{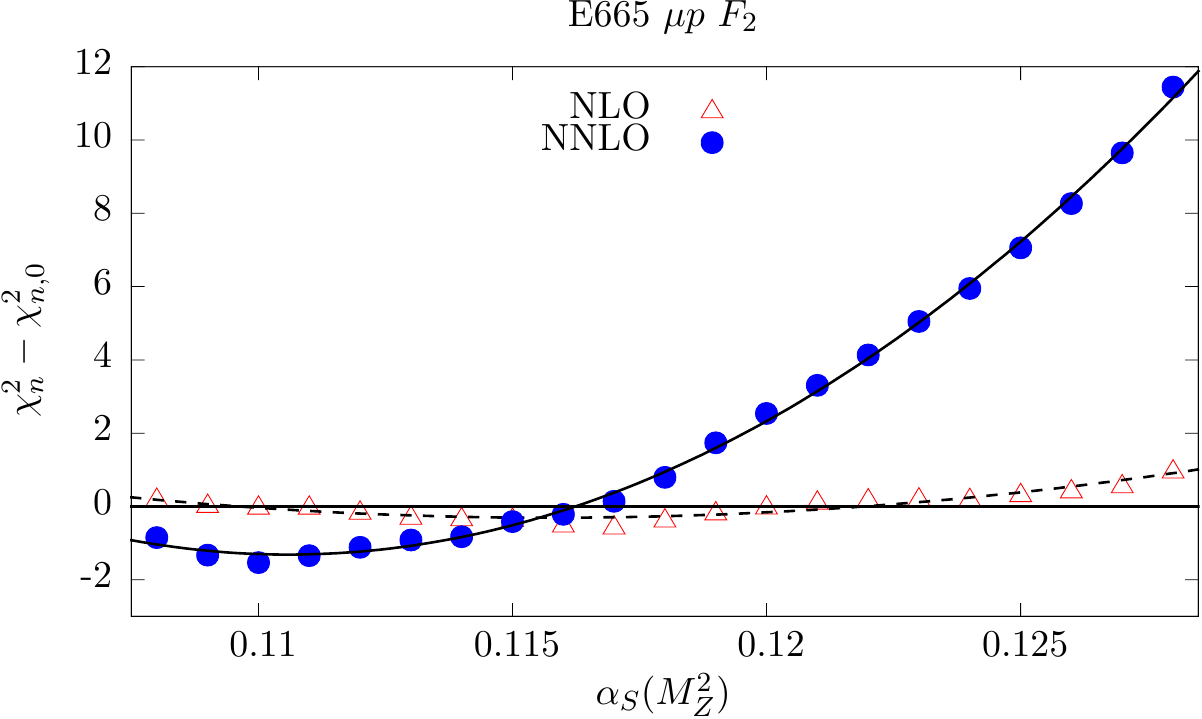}
              \includegraphics[width=0.45\textwidth]{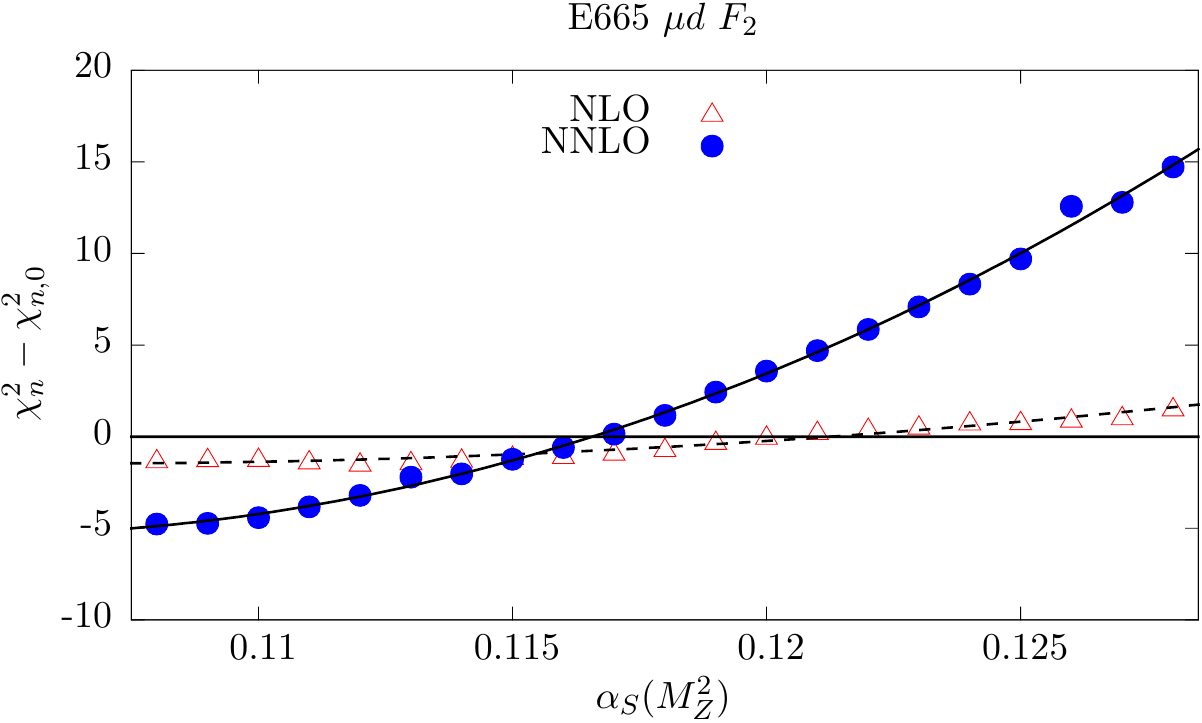}\qquad
    \includegraphics[width=0.45\textwidth]{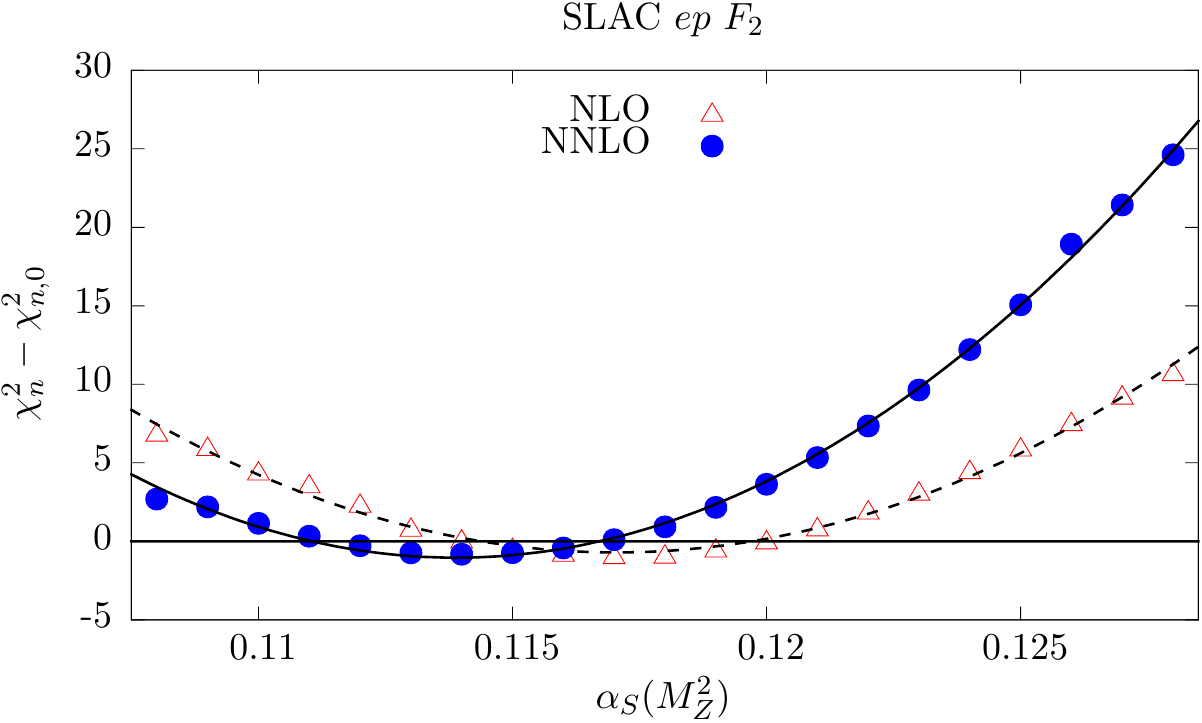}
  \caption{$\chi_n^2$ profiles obtained when varying $\alpha_S(M_Z^2)$ for the subset of data from deep--inelastic fixed--target experiments. The results from the NNLO global fits are shown by bullet points (and a continuous curve), while those from the NLO global fits are shown by triangles (and a dashed curve). The plots are continued in next figure.}
  \label{fig:nnlofixed1}
\end{figure}

\begin{figure}
  \centering
              \includegraphics[width=0.45\textwidth]{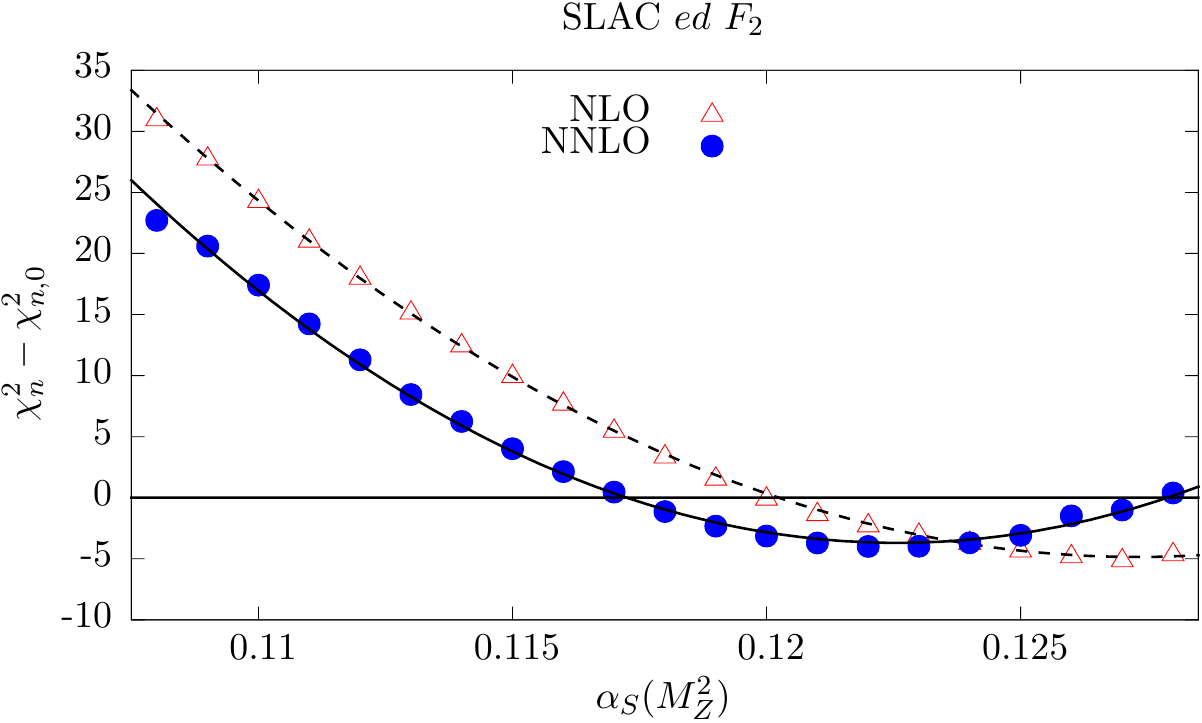}\qquad
    \includegraphics[width=0.45\textwidth]{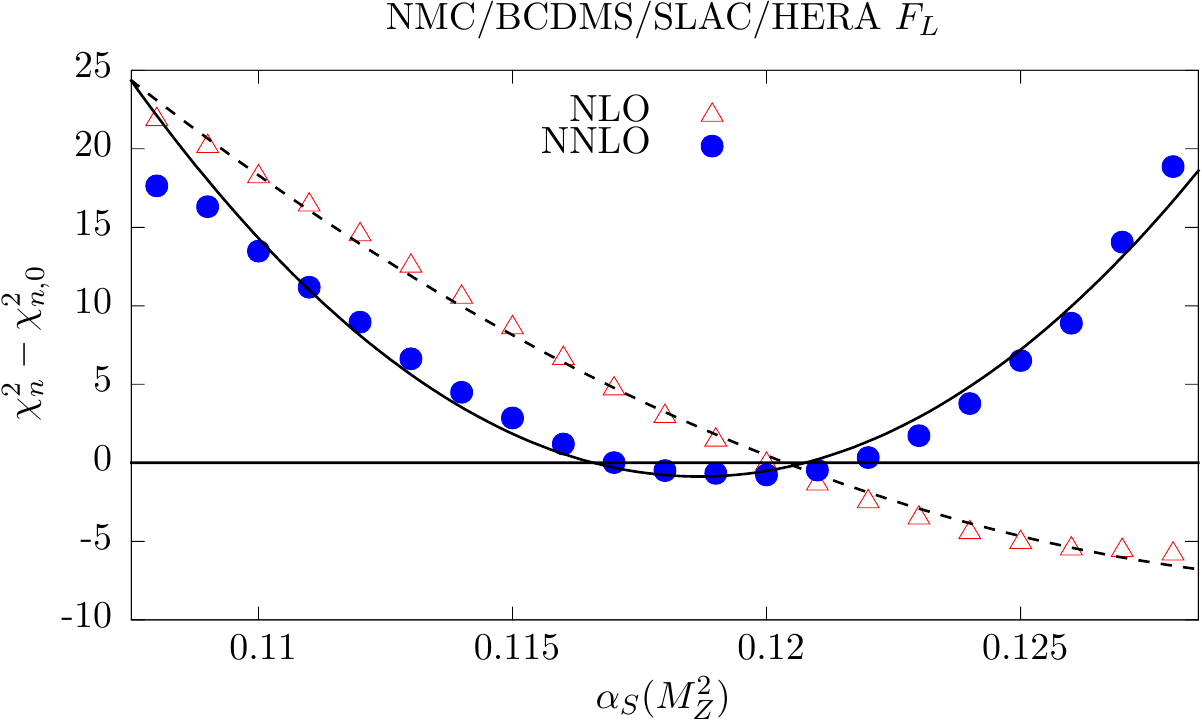}
                 \includegraphics[width=0.45\textwidth]{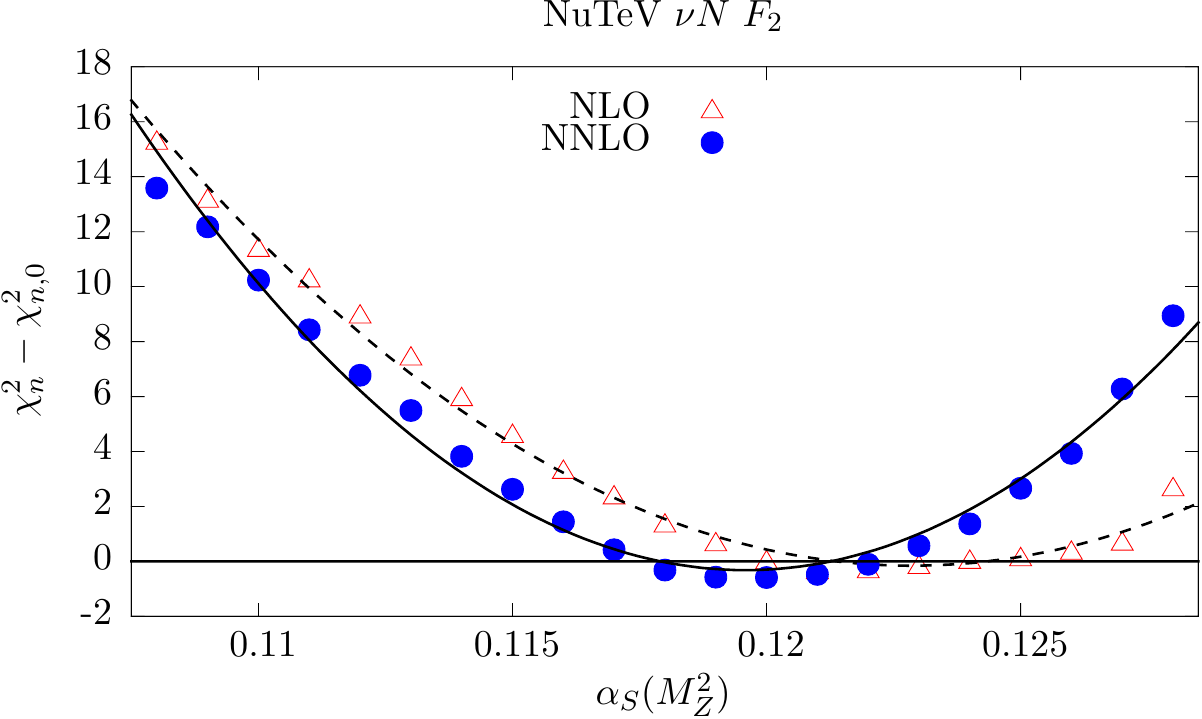}\qquad
    \includegraphics[width=0.45\textwidth]{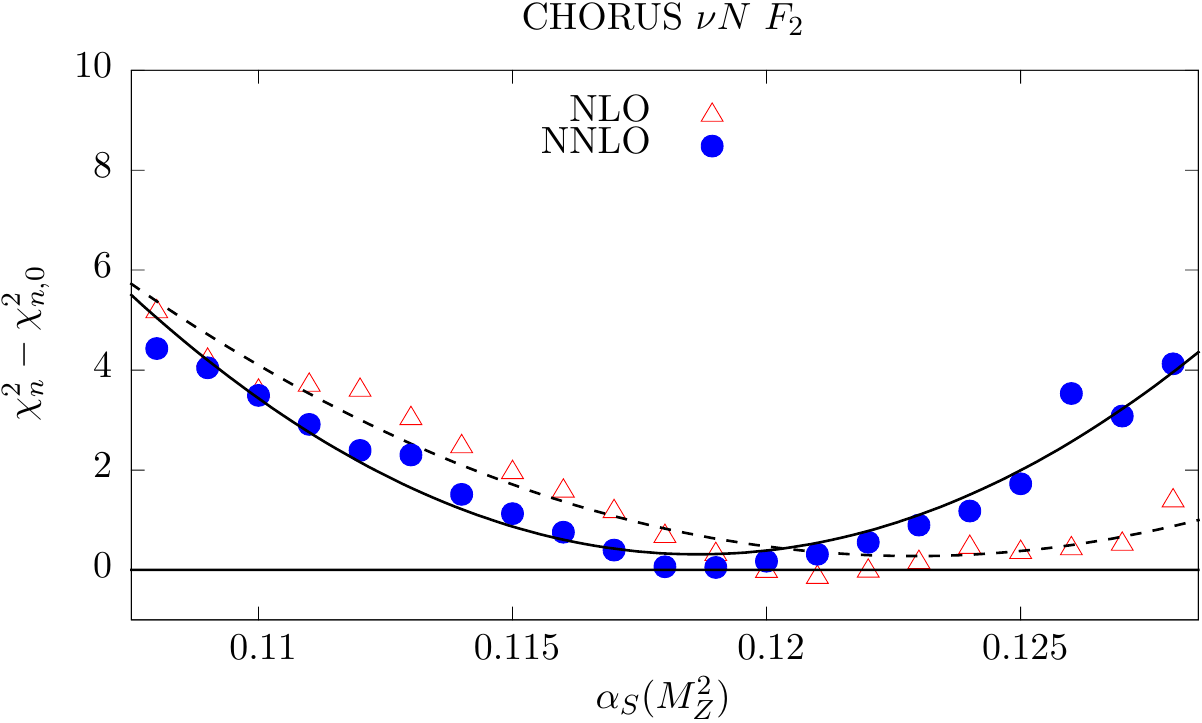}
                     \includegraphics[width=0.45\textwidth]{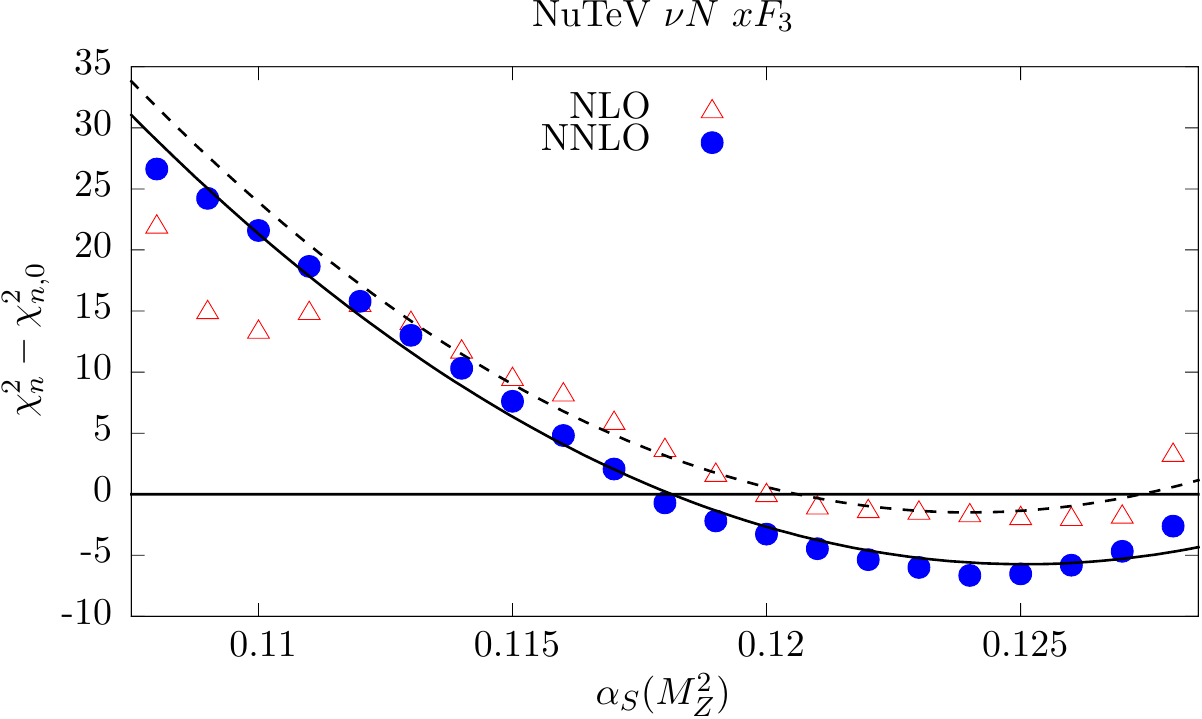}\qquad
    \includegraphics[width=0.45\textwidth]{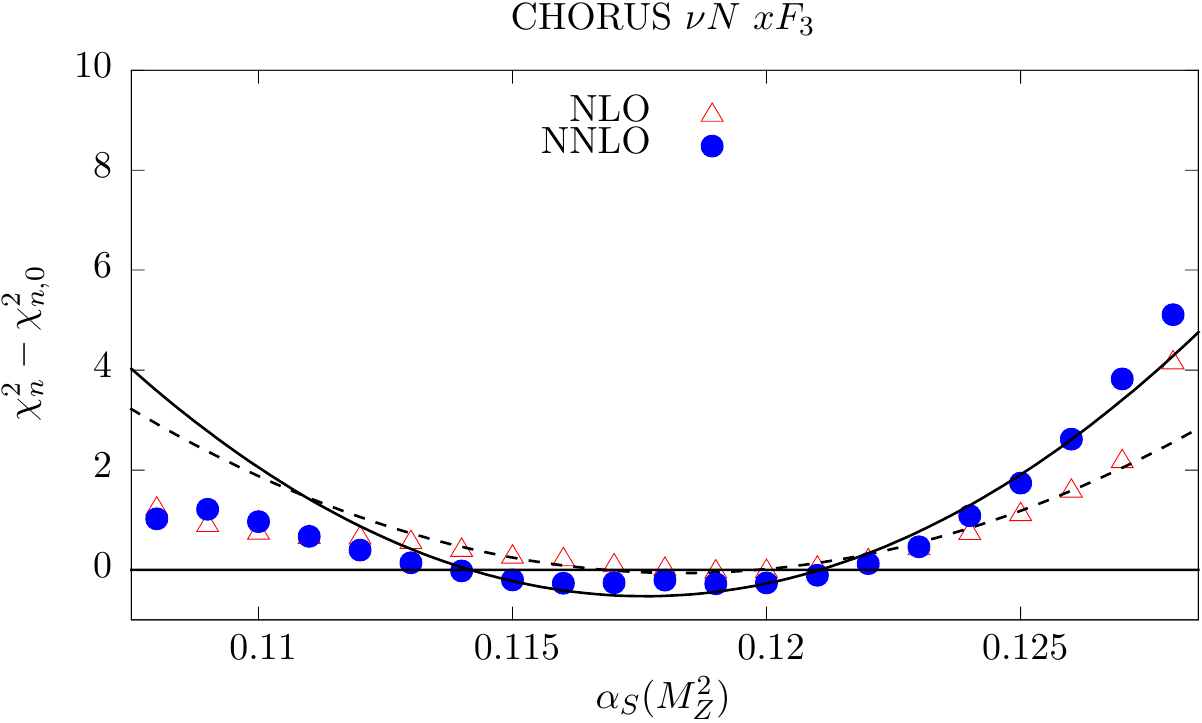}
    \includegraphics[width=0.45\textwidth]{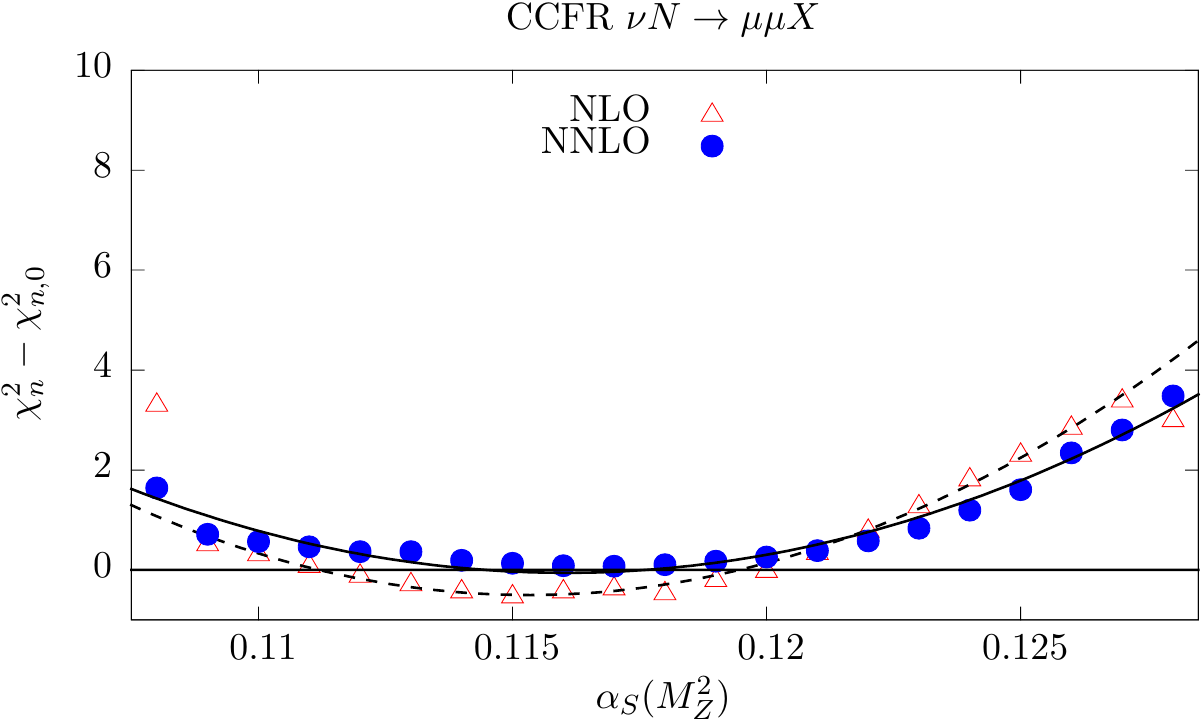}\qquad
      \includegraphics[width=0.45\textwidth]{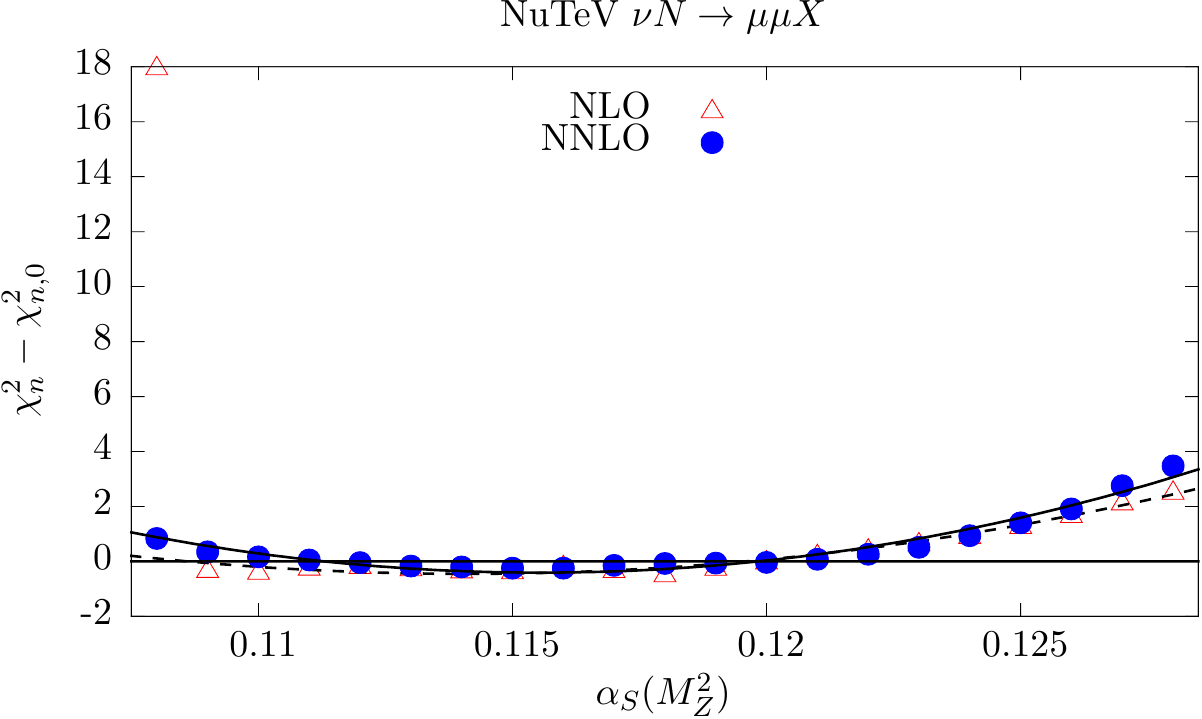}
  \caption{$\chi_n^2$ profiles obtained when varying $\alpha_S(M_Z^2)$, for the subset of data from deep--inelastic fixed--target experiments. The results from the NNLO global fits are shown by bullet points (and a continuous curve), while those from the NLO global fits are shown by triangles (and a dashed curve). (Continued from the previous figure.)}
  \label{fig:nnlofixed2}
\end{figure}


\begin{figure}
  \centering
\includegraphics[width=0.45\textwidth]{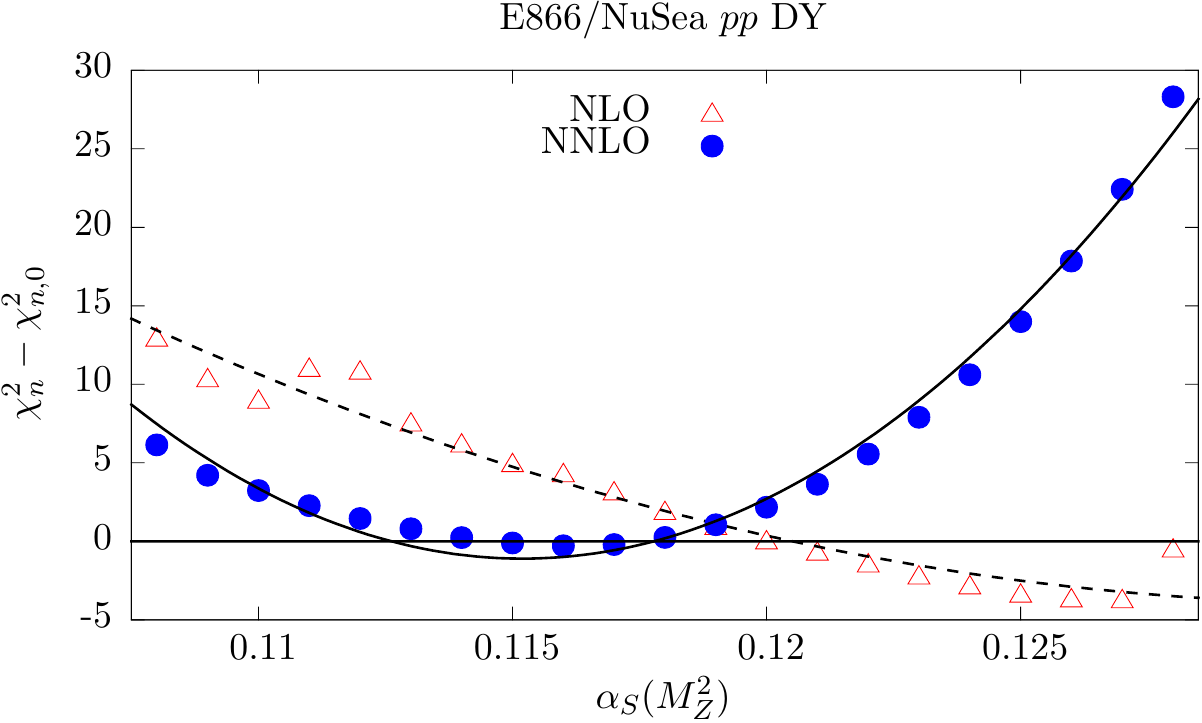}\qquad
    \includegraphics[width=0.45\textwidth]{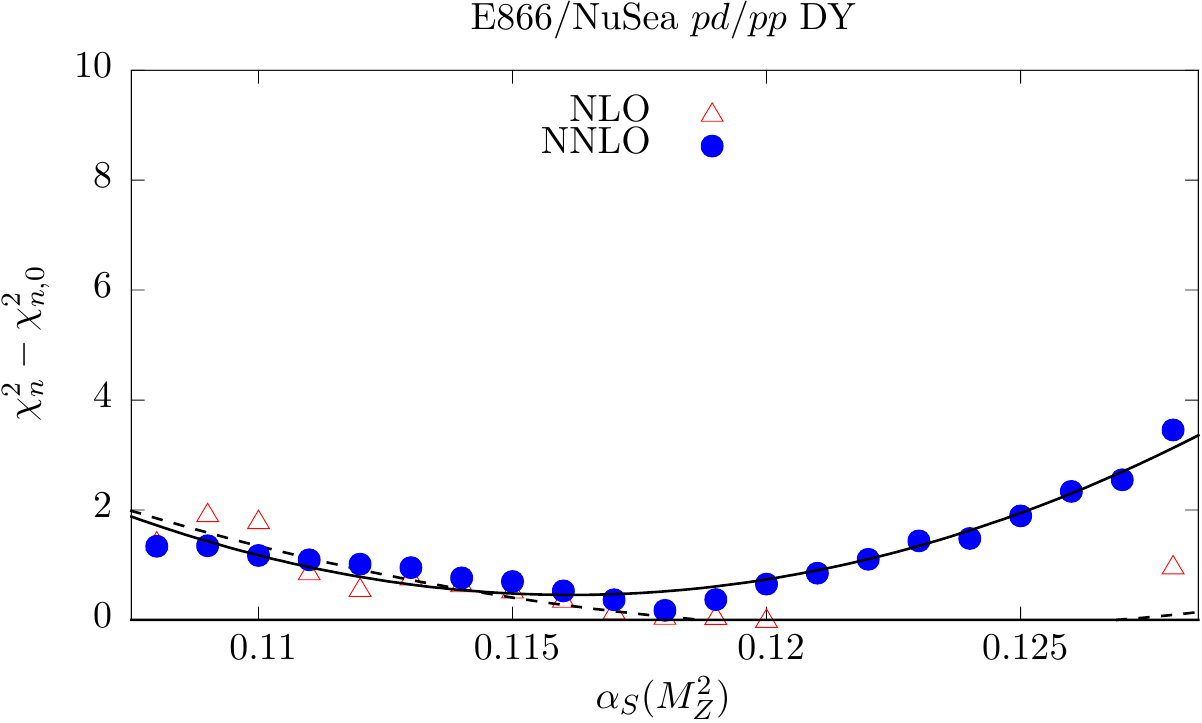}
    
  \includegraphics[width=0.45\textwidth]{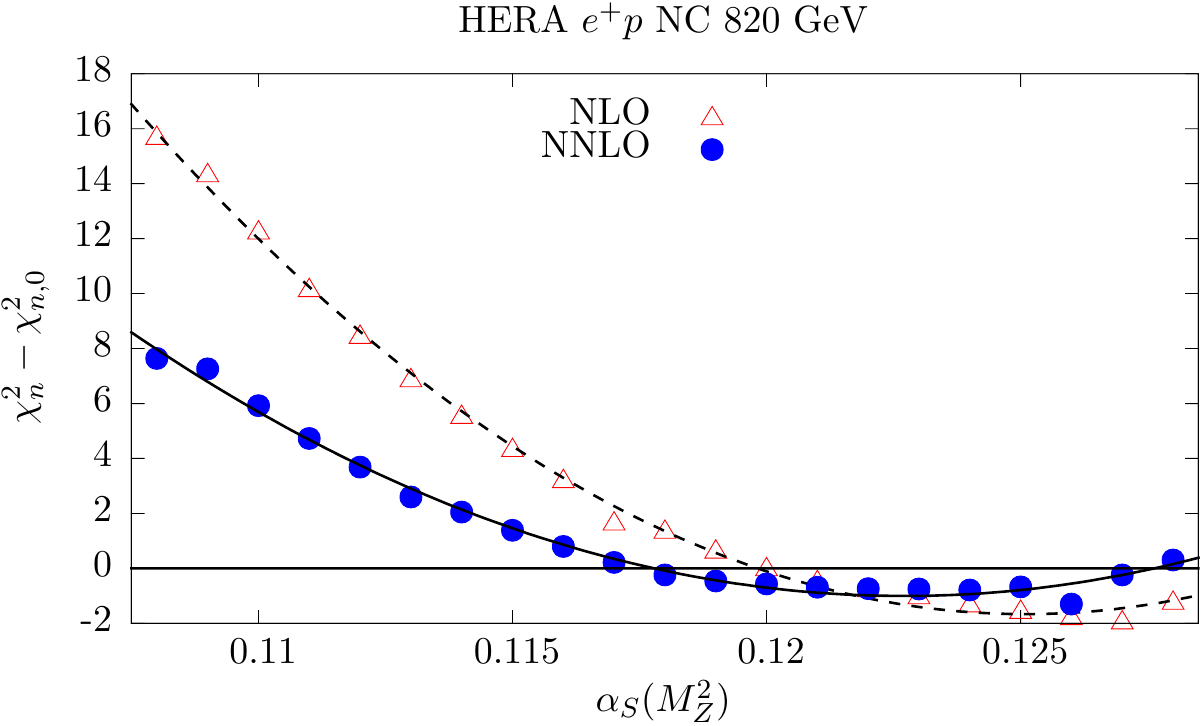}\qquad
    \includegraphics[width=0.45\textwidth]{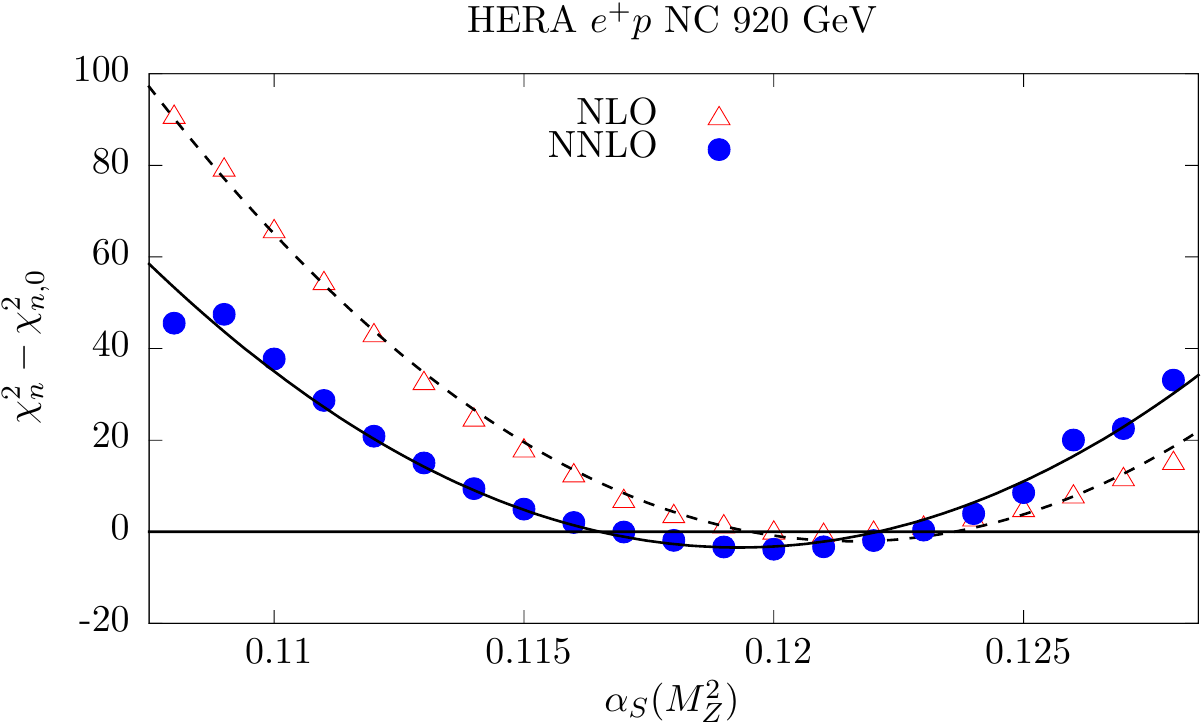}
      \includegraphics[width=0.45\textwidth]{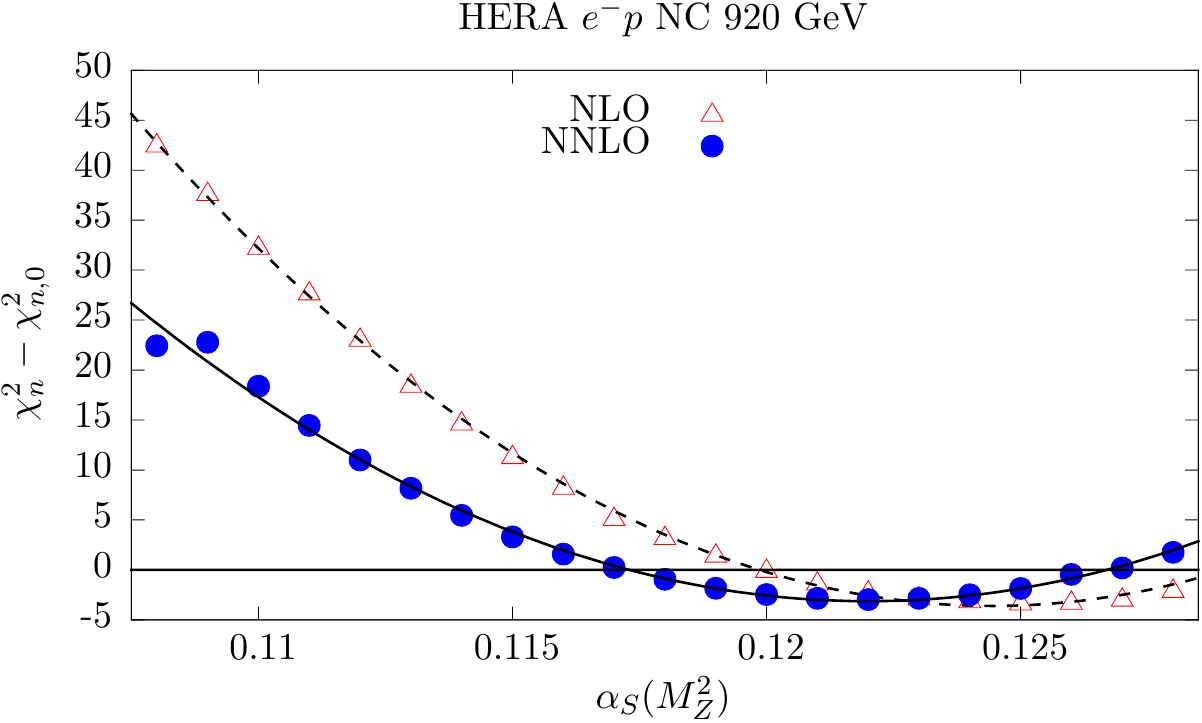}\qquad
    \includegraphics[width=0.45\textwidth]{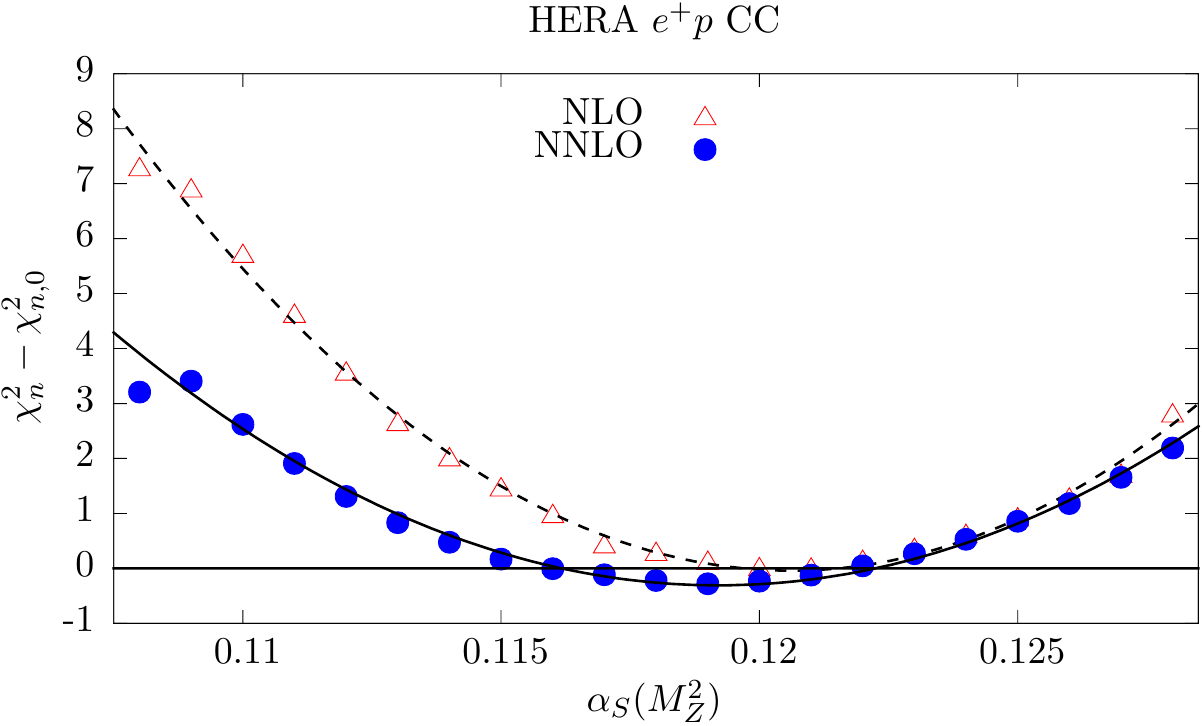}
      \includegraphics[width=0.45\textwidth]{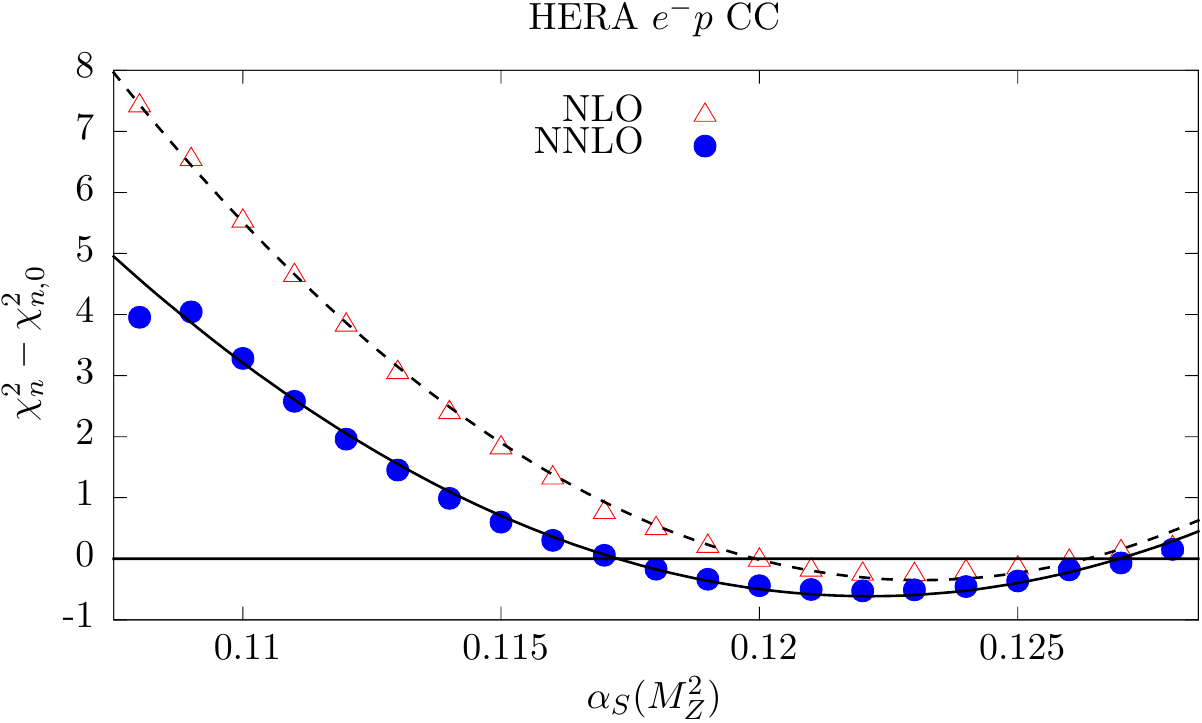}\qquad
    \includegraphics[width=0.45\textwidth]{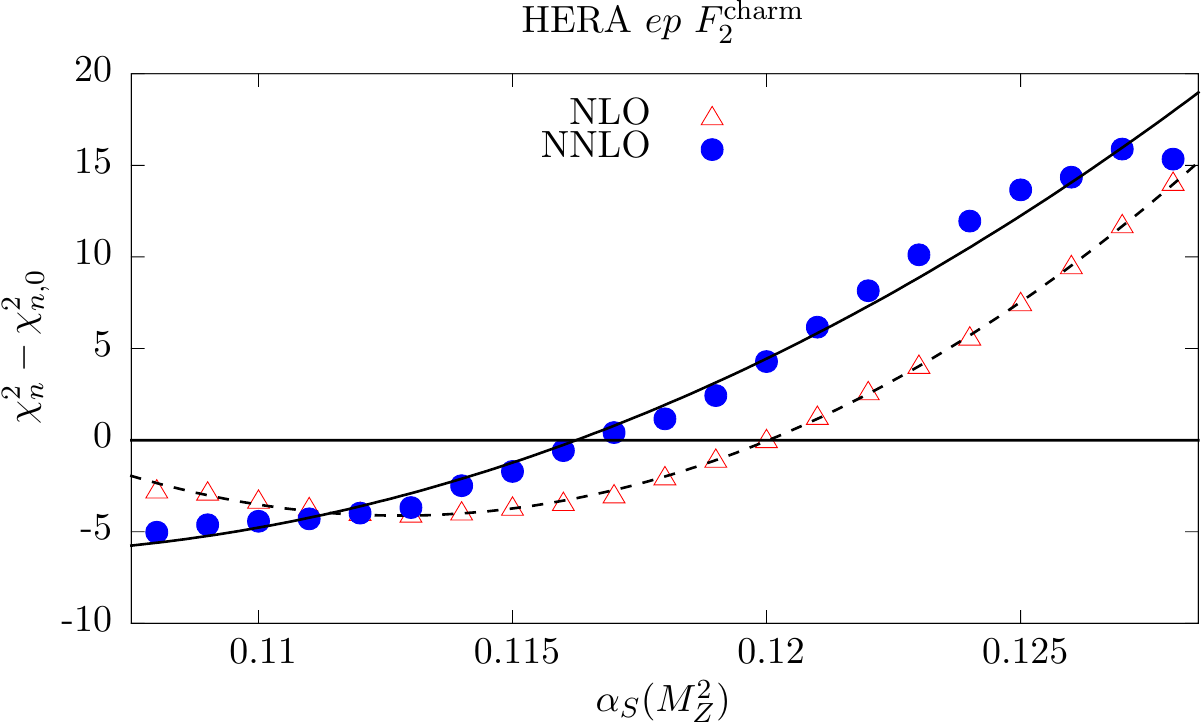}
  \caption{$\chi_n^2$ profiles obtained when varying $\alpha_S(M_Z^2)$ coming from the Drell--Yan fixed--target experiments and from the combined H1 and ZEUS measurements at HERA. The results from the NNLO global fits are shown by bullet points (and a continuous curve), while those from the NLO global fits are shown by triangles (and a dashed curve).}
  \label{fig:nnlohera}
\end{figure}

\begin{figure}
  \centering
   \includegraphics[width=0.45\textwidth]{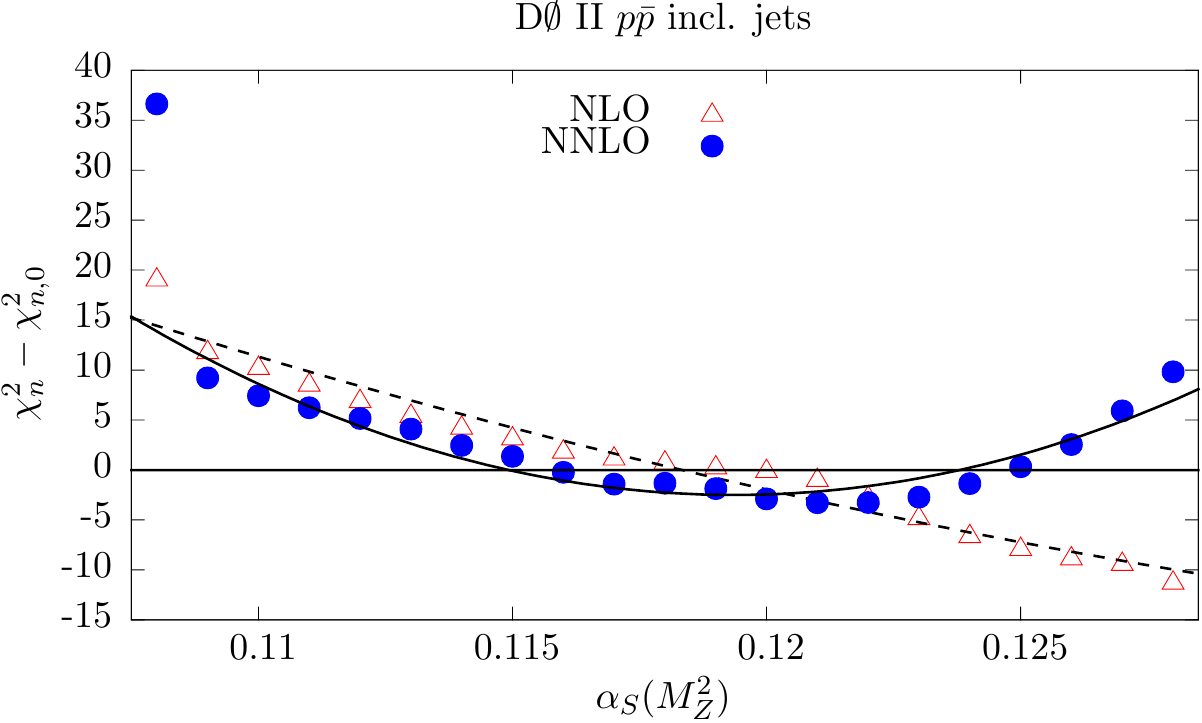}\qquad
      \includegraphics[width=0.45\textwidth]{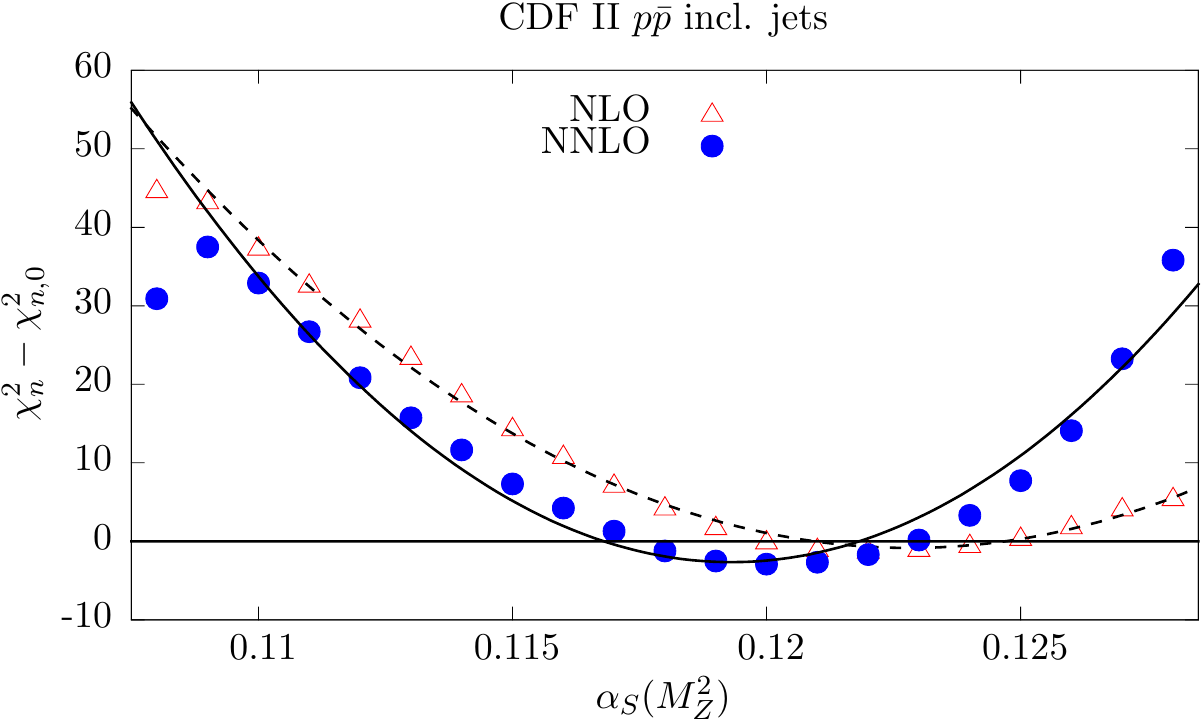}
     \includegraphics[width=0.45\textwidth]{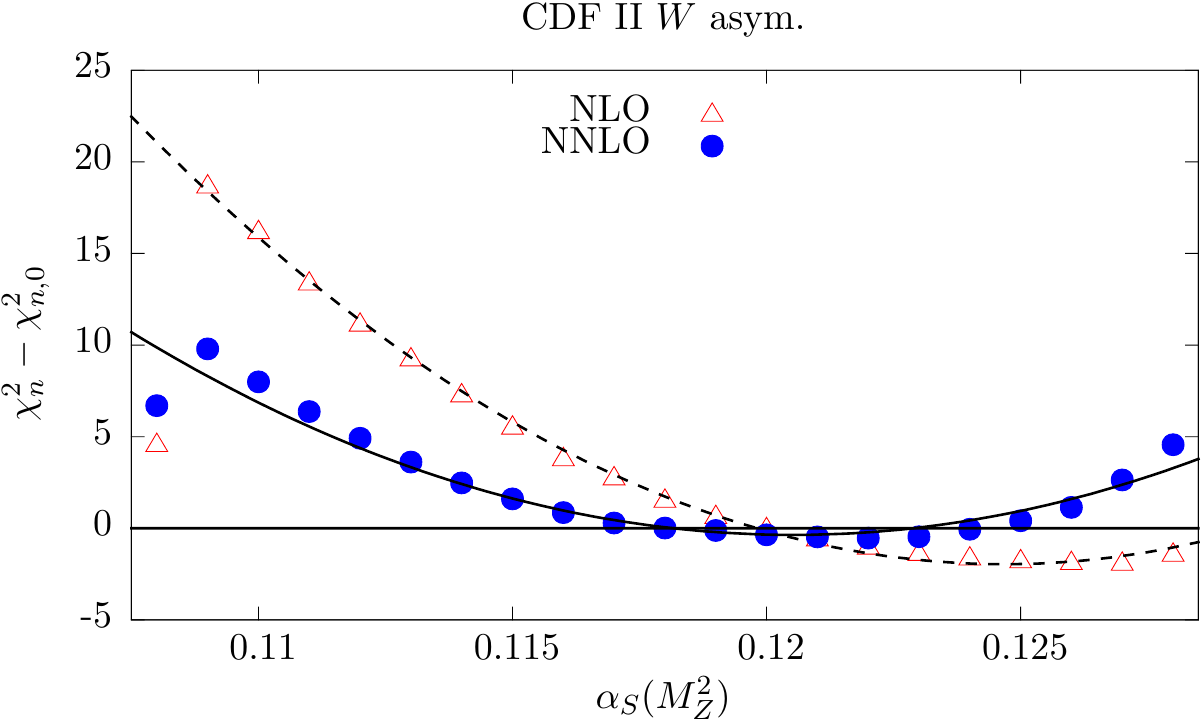}\qquad
    \includegraphics[width=0.45\textwidth]{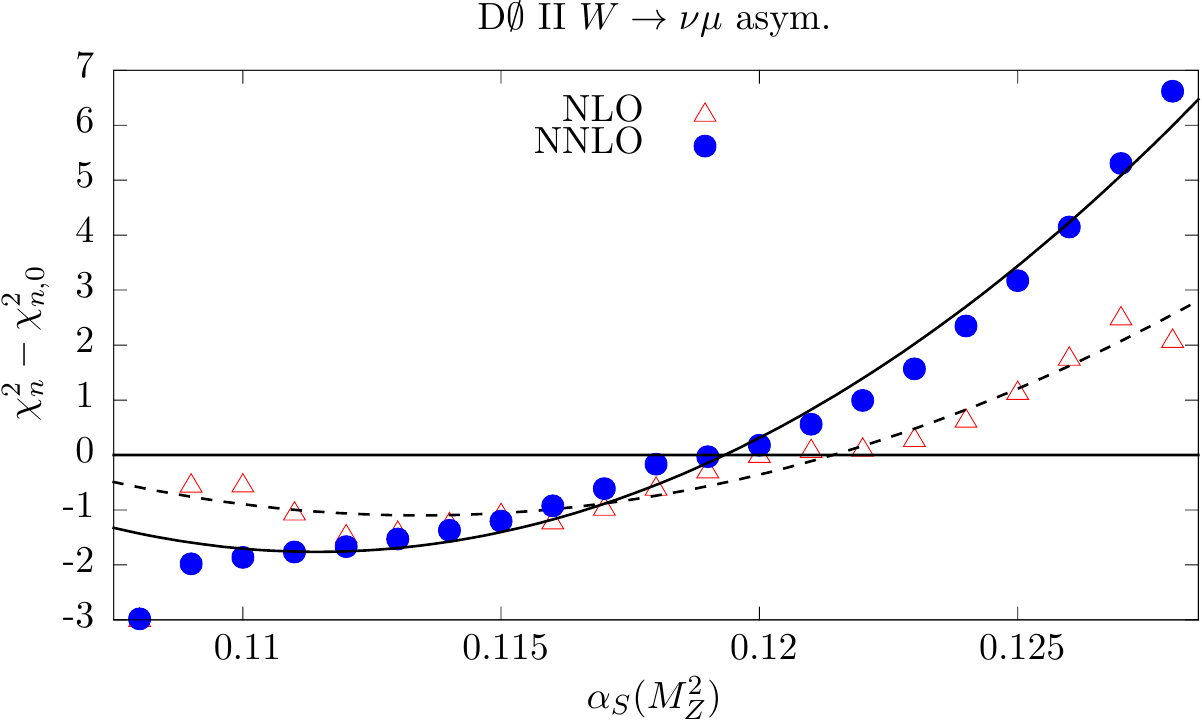}
     \includegraphics[width=0.45\textwidth]{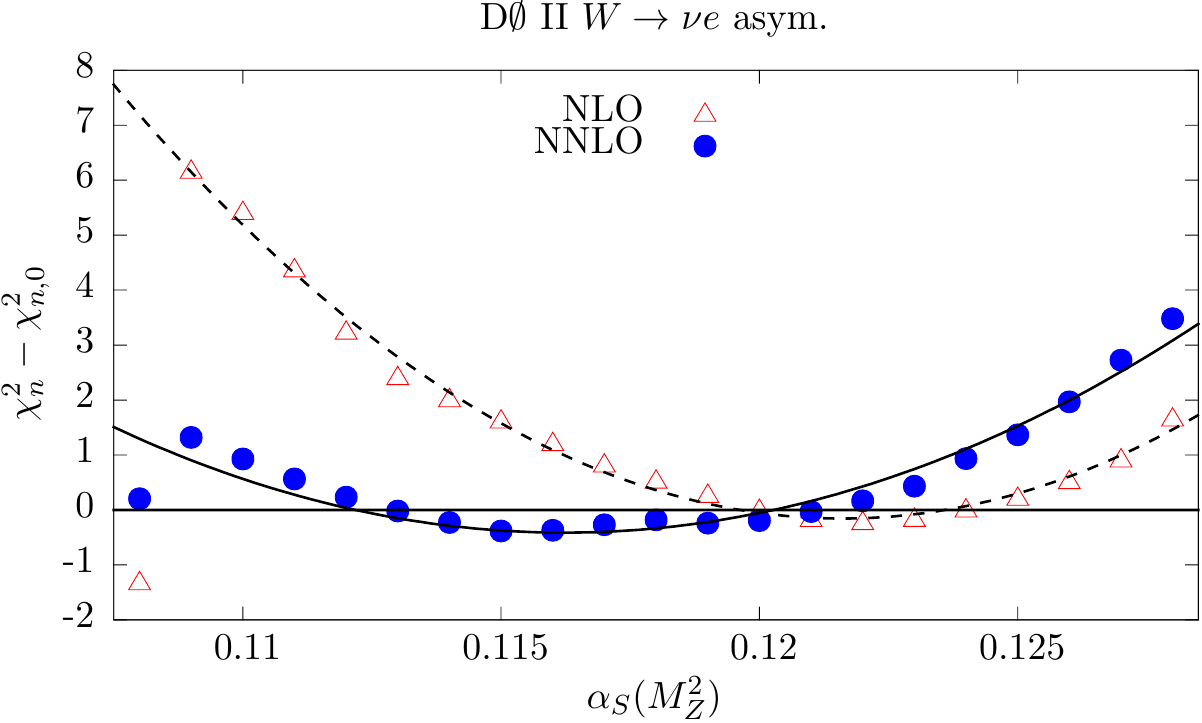}\qquad
    \includegraphics[width=0.45\textwidth]{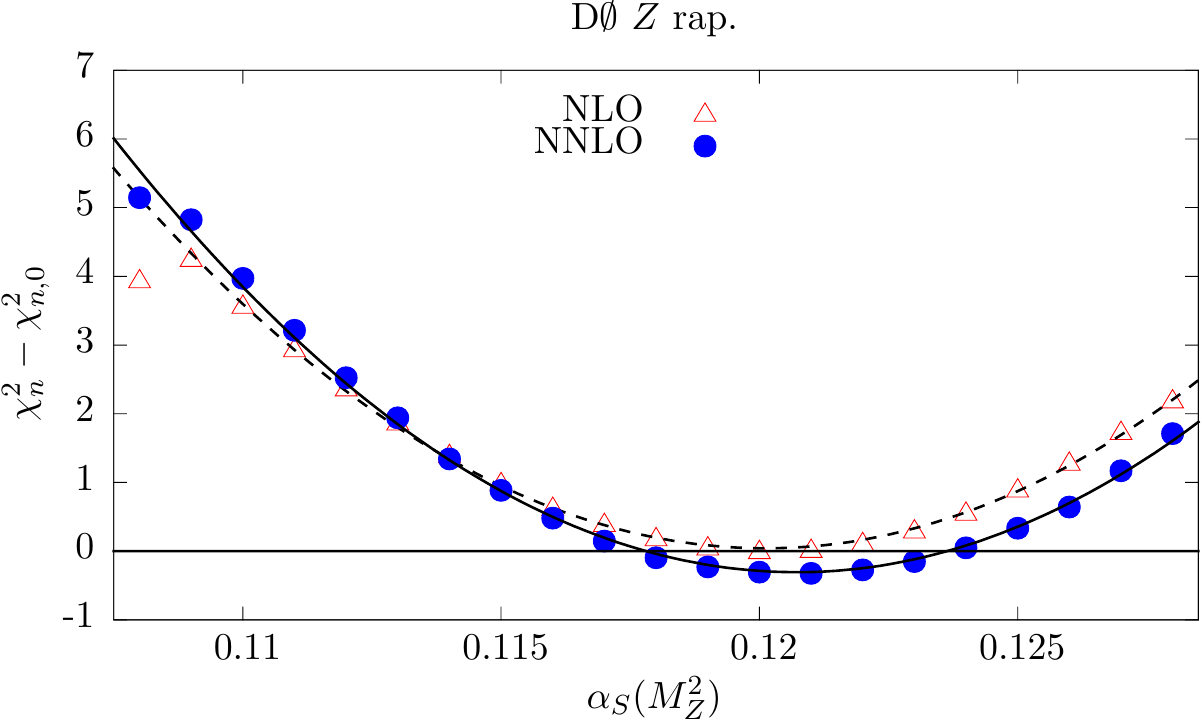}
     \includegraphics[width=0.45\textwidth]{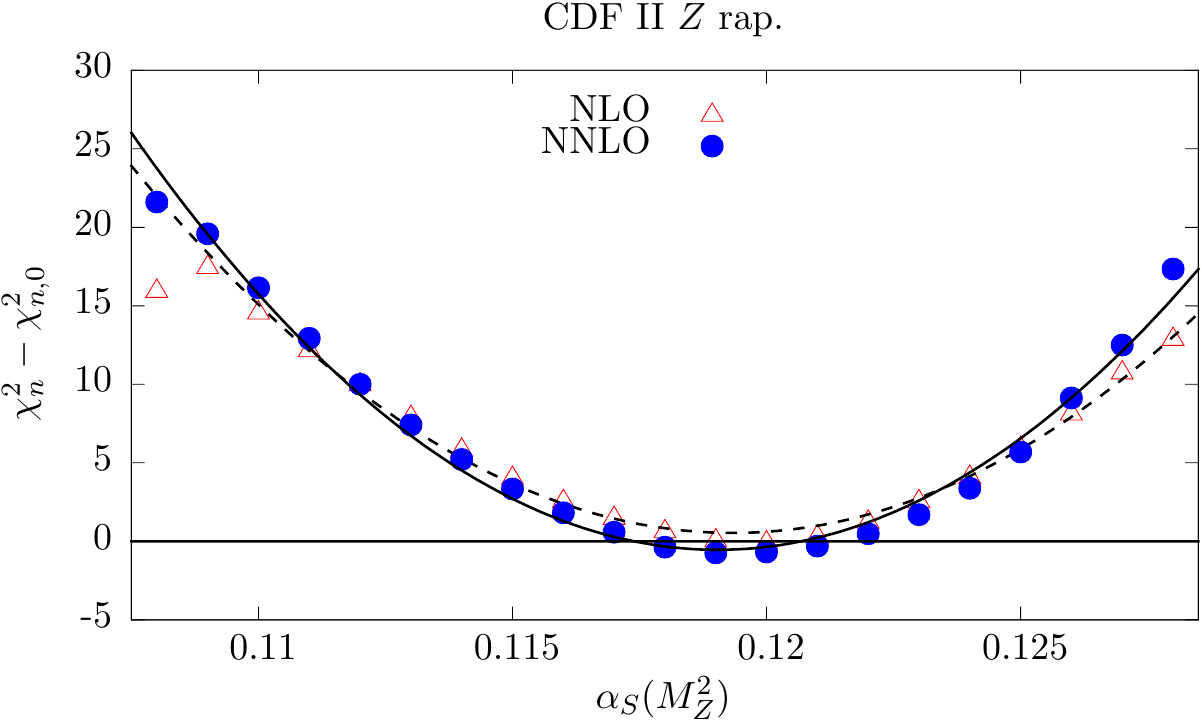}\qquad
       \includegraphics[width=0.45\textwidth]{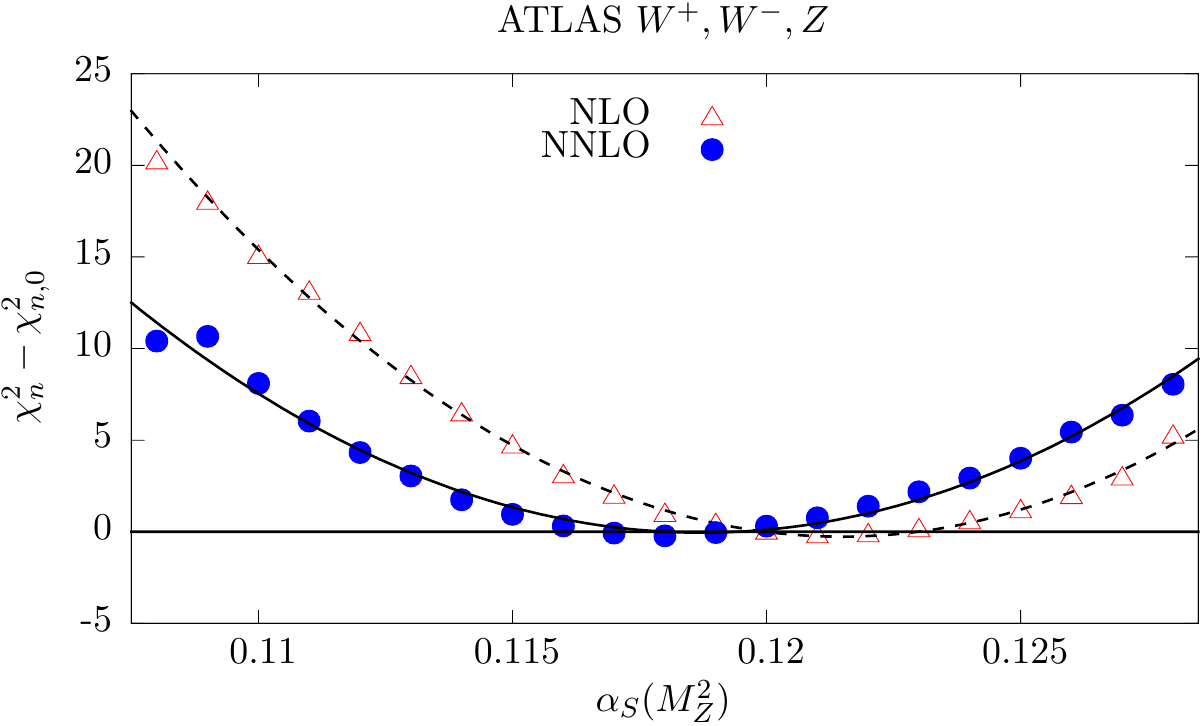}
  \caption{$\chi_n^2$ profiles obtained when varying $\alpha_S(M_Z^2)$, coming from the subset of data of the CDF and D0 Tevatron experiments, together with the plot for the ATLAS $W$ and $Z$ production data. The results from the NNLO global fits are shown by bullet points (and a continuous curve), while those from the NLO global fits are shown by triangles (and a dashed curve).}
  \label{fig:nnlotev}
\end{figure}

\begin{figure}
  \centering
             \includegraphics[width=0.45\textwidth]{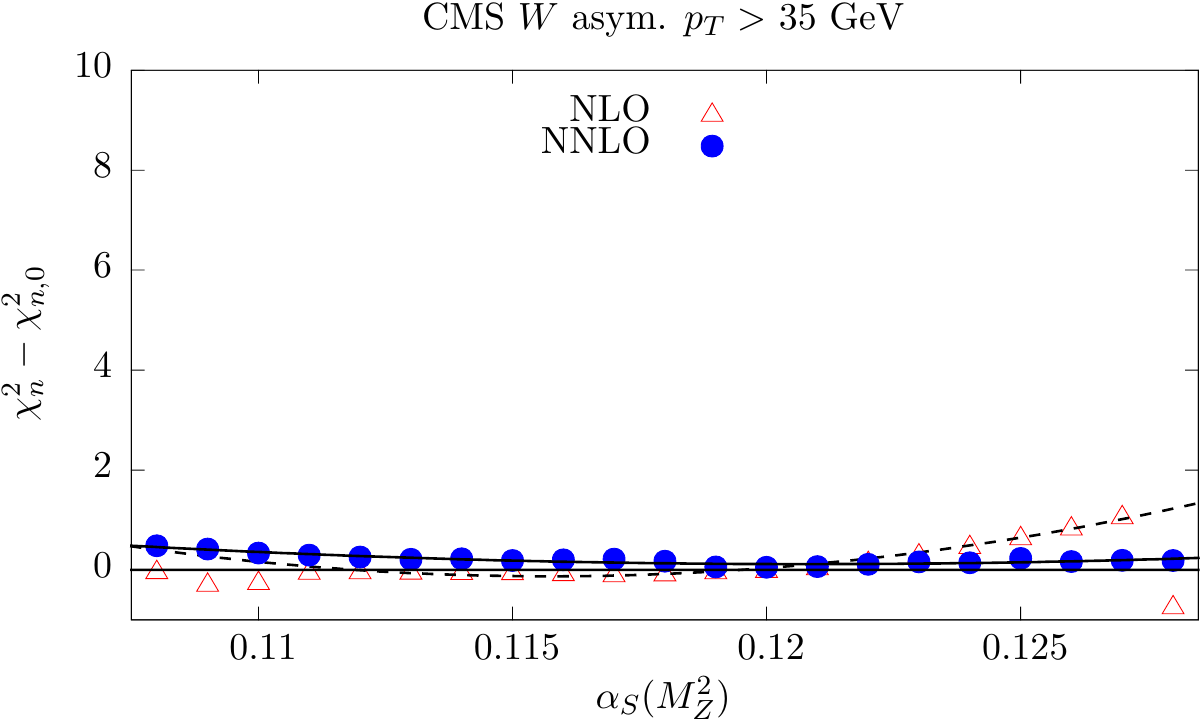}\qquad
       \includegraphics[width=0.45\textwidth]{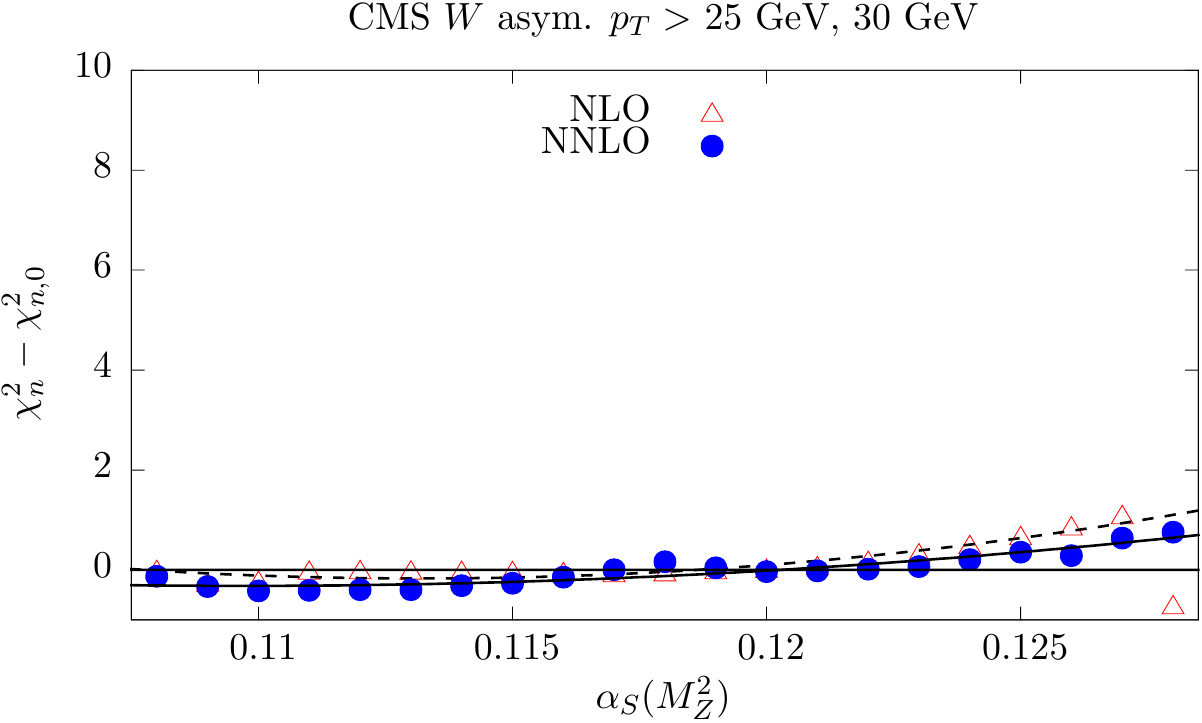}
    \includegraphics[width=0.45\textwidth]{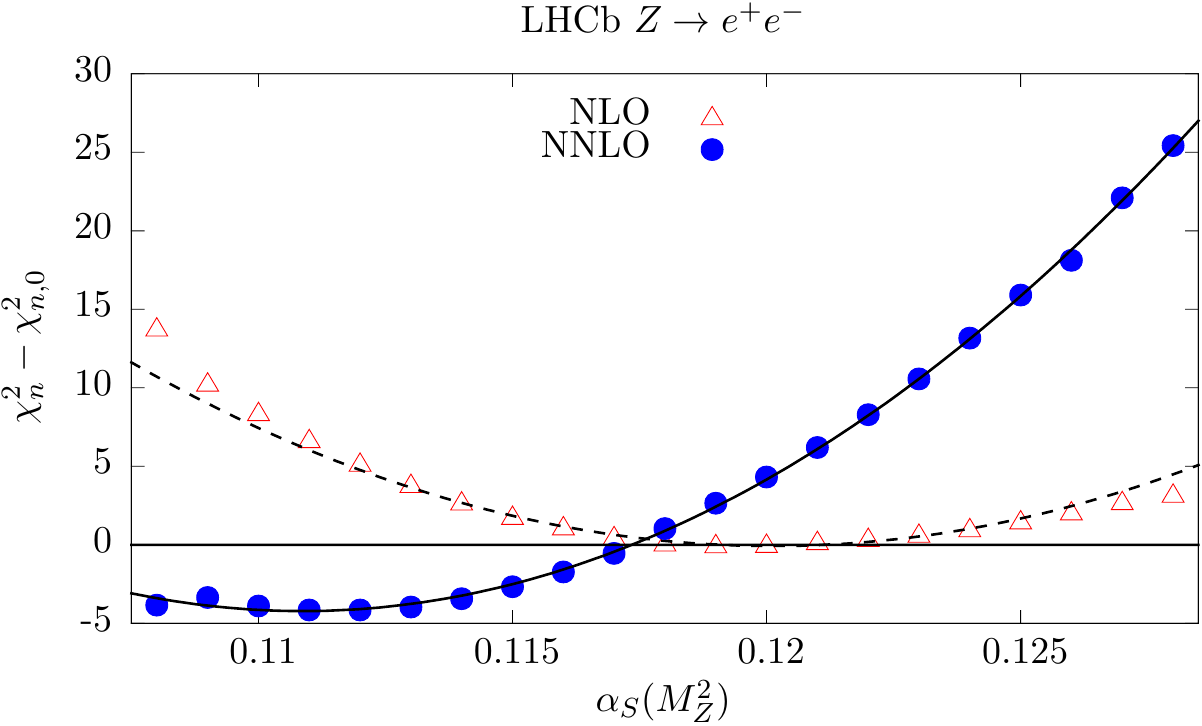}\qquad
     \includegraphics[width=0.45\textwidth]{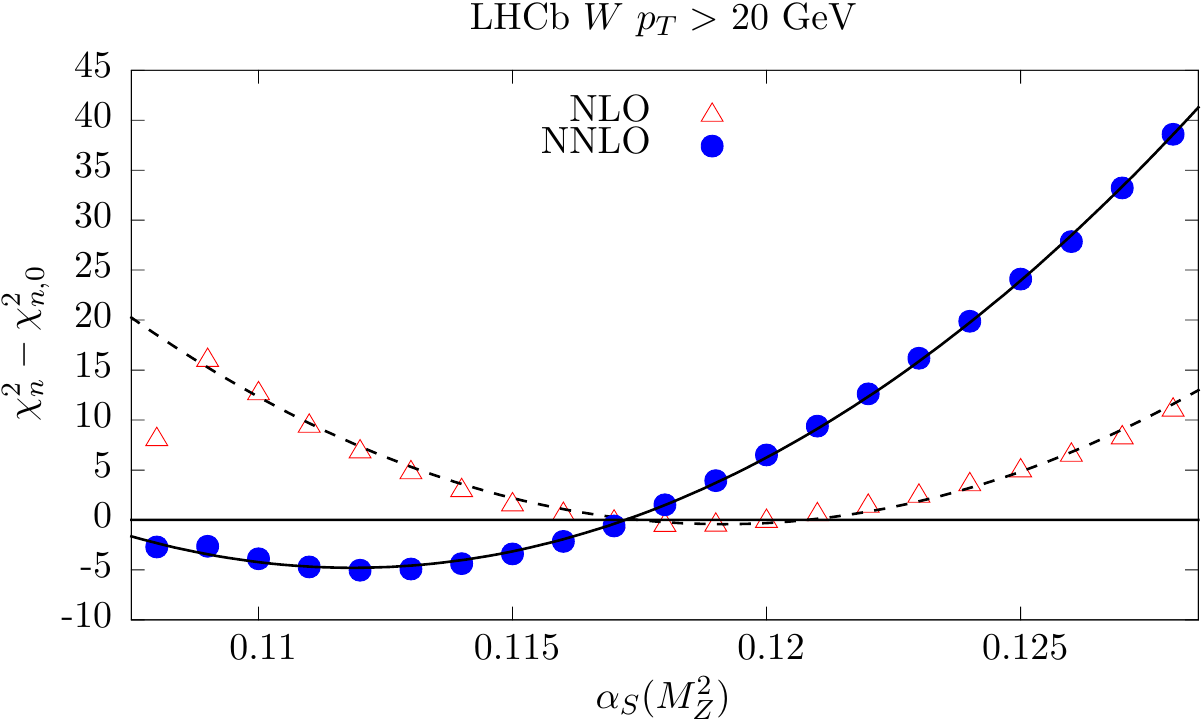}
    \includegraphics[width=0.45\textwidth]{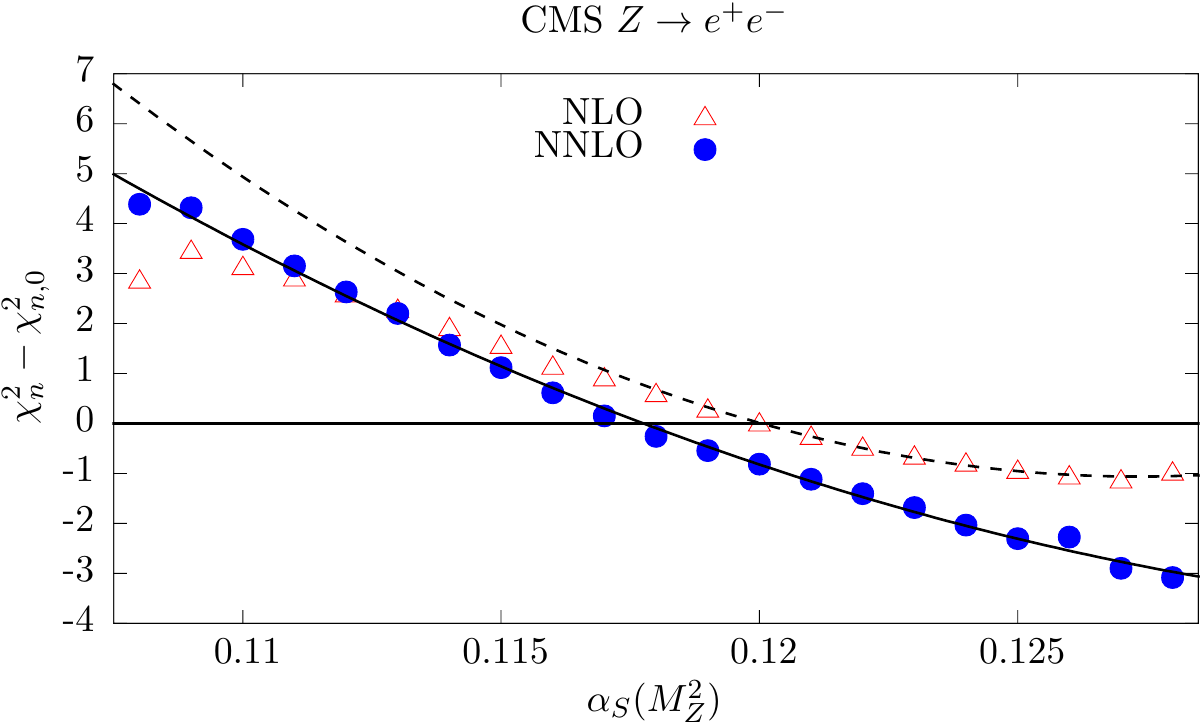}\qquad
     \includegraphics[width=0.45\textwidth]{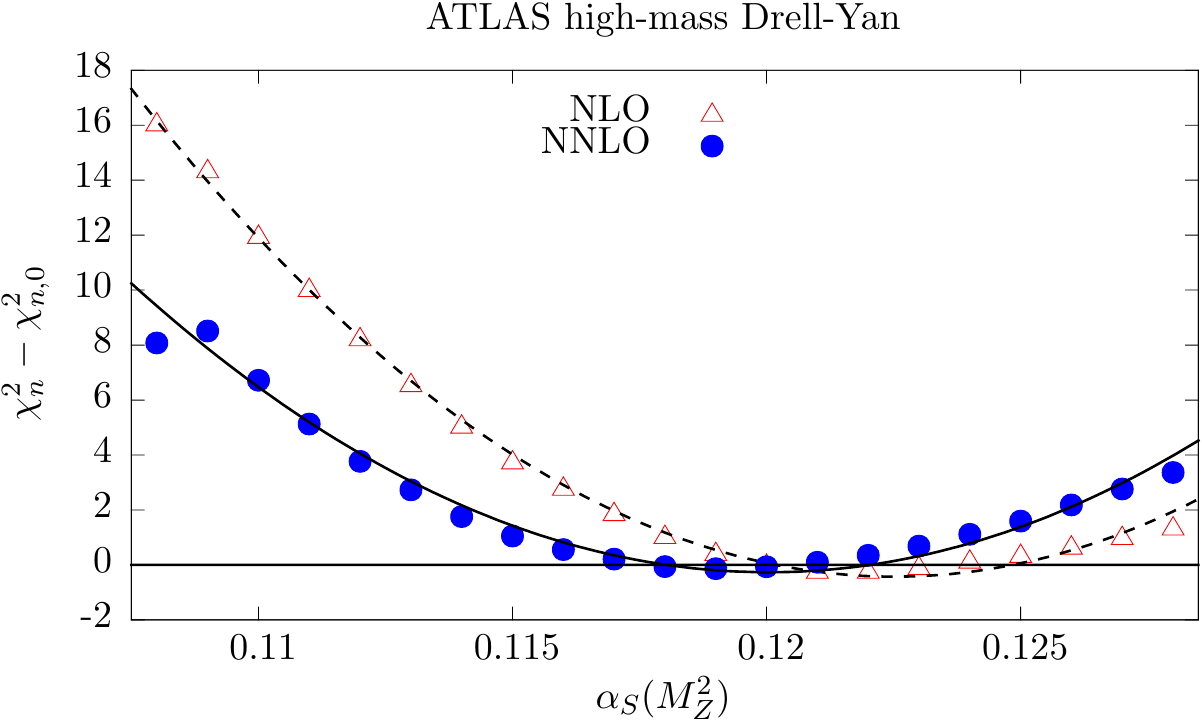}
    \includegraphics[width=0.45\textwidth]{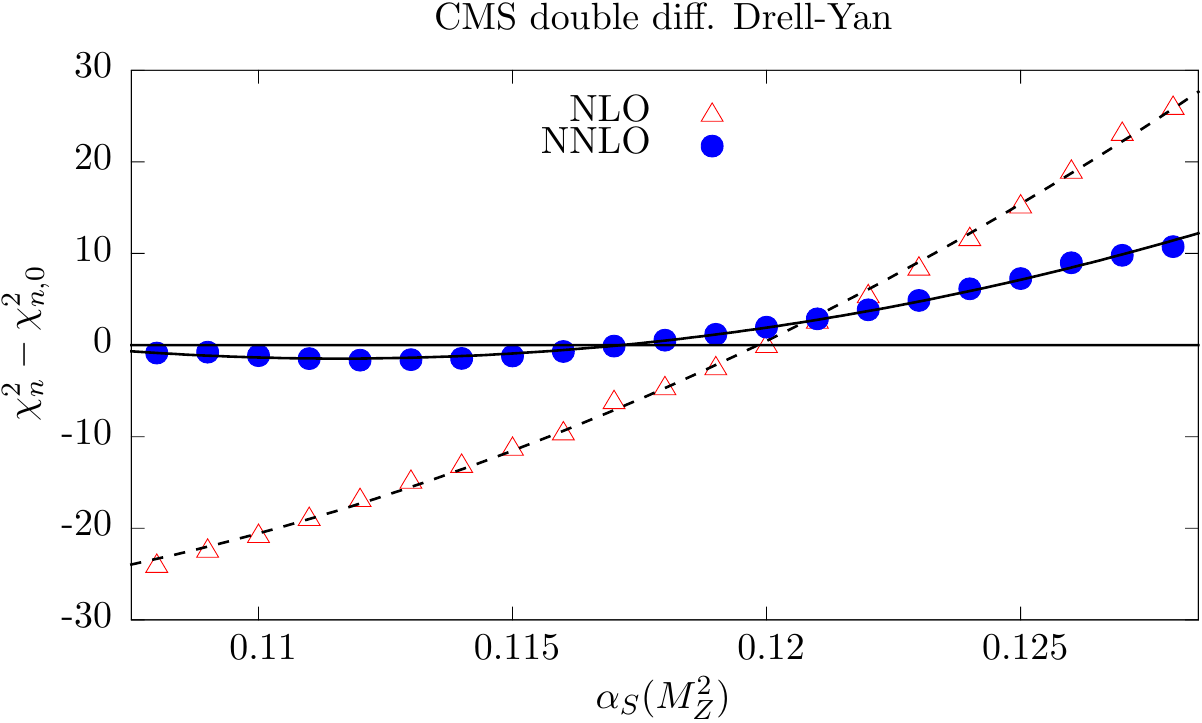}
  \caption{$\chi_n^2$ profiles obtained when varying $\alpha_S(M_Z^2)$, from the subset of data collected by the LHC experiments. The results from the NNLO global fits are shown by bullet points (and a continuous curve), while those from the NLO global fits are shown by triangles (and a dashed curve).  The $\chi_n^2$ profiles for $t\overline{t}$ data  are shown in Fig.~\ref{fig:nnlotop}  and discussed in Section \ref{sec:tt}. }
  \label{fig:nnloLHC}
\end{figure}

\begin{figure}
  \centering
             \includegraphics[width=0.45\textwidth]{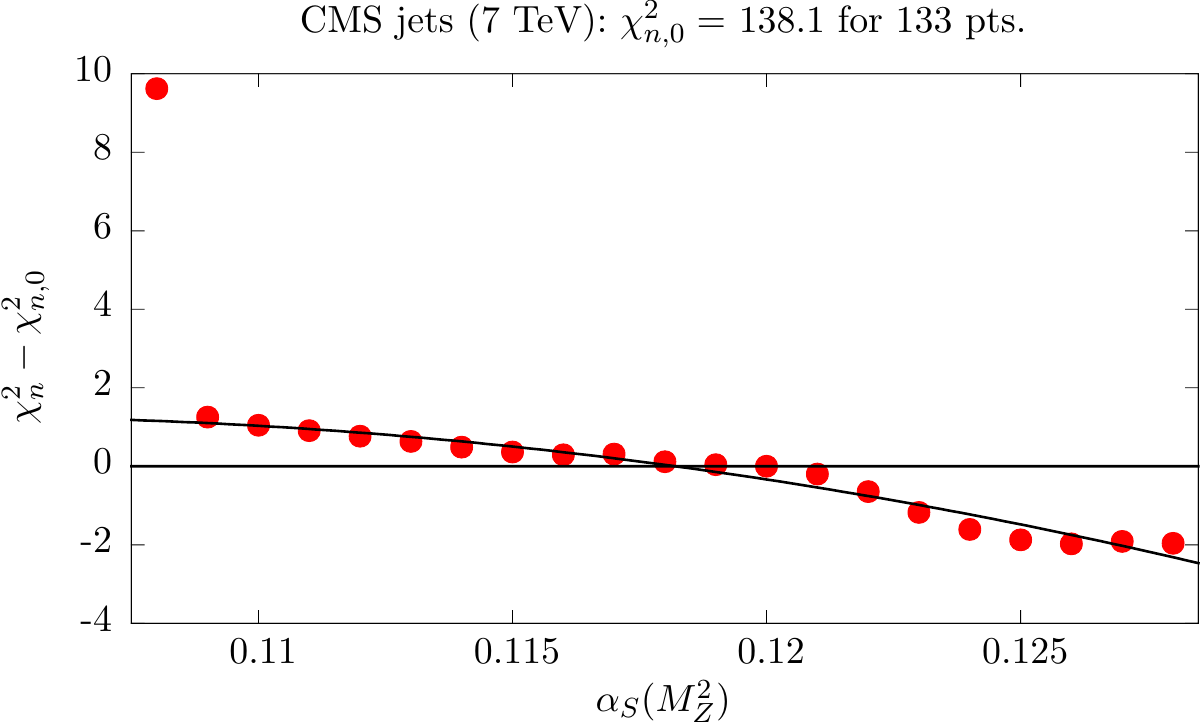}\qquad
       \includegraphics[width=0.45\textwidth]{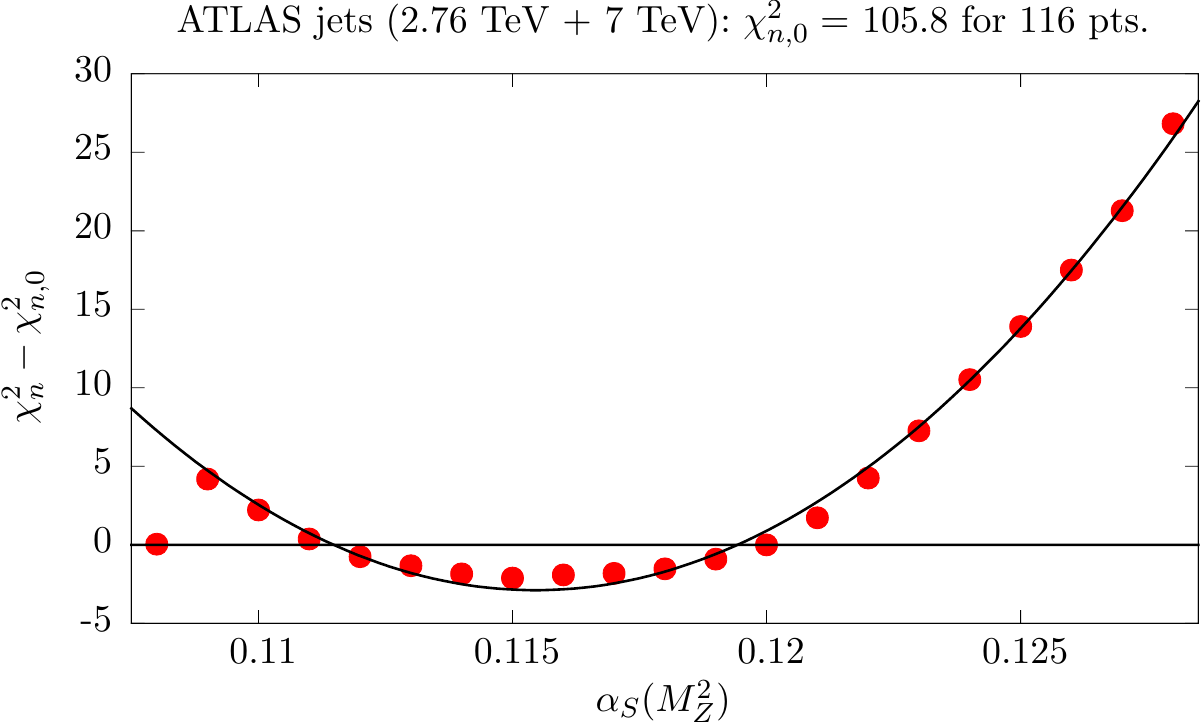}
    \includegraphics[width=0.45\textwidth]{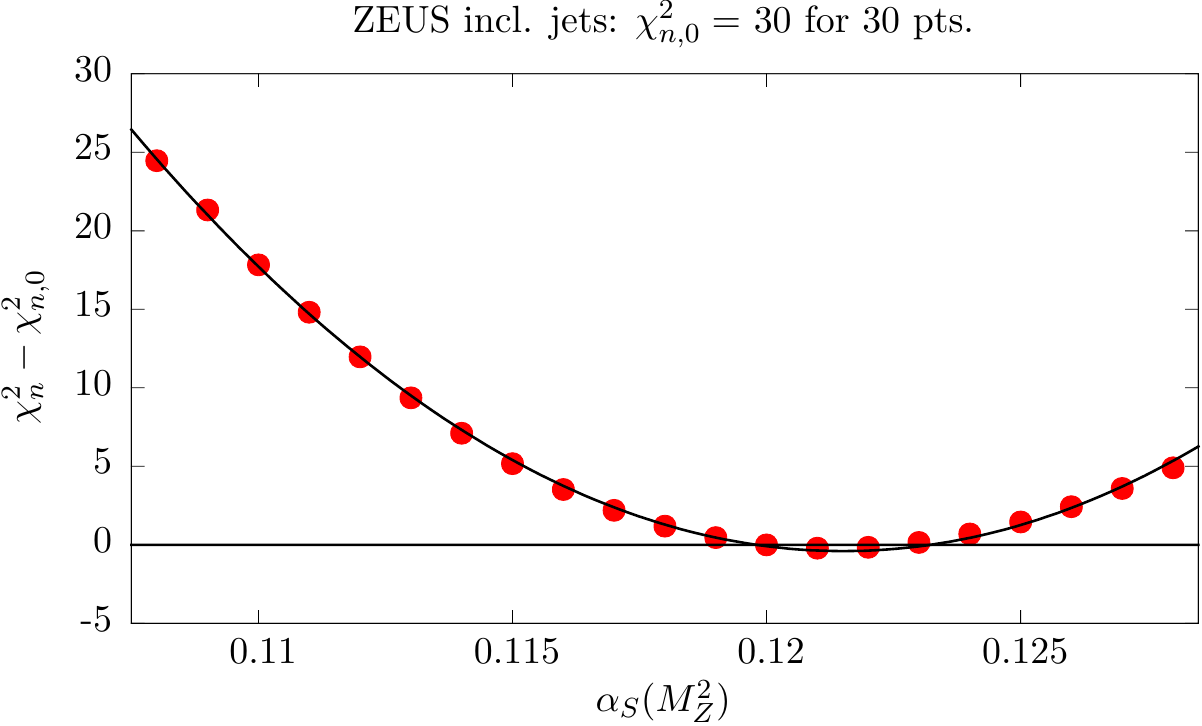}\qquad
     \includegraphics[width=0.45\textwidth]{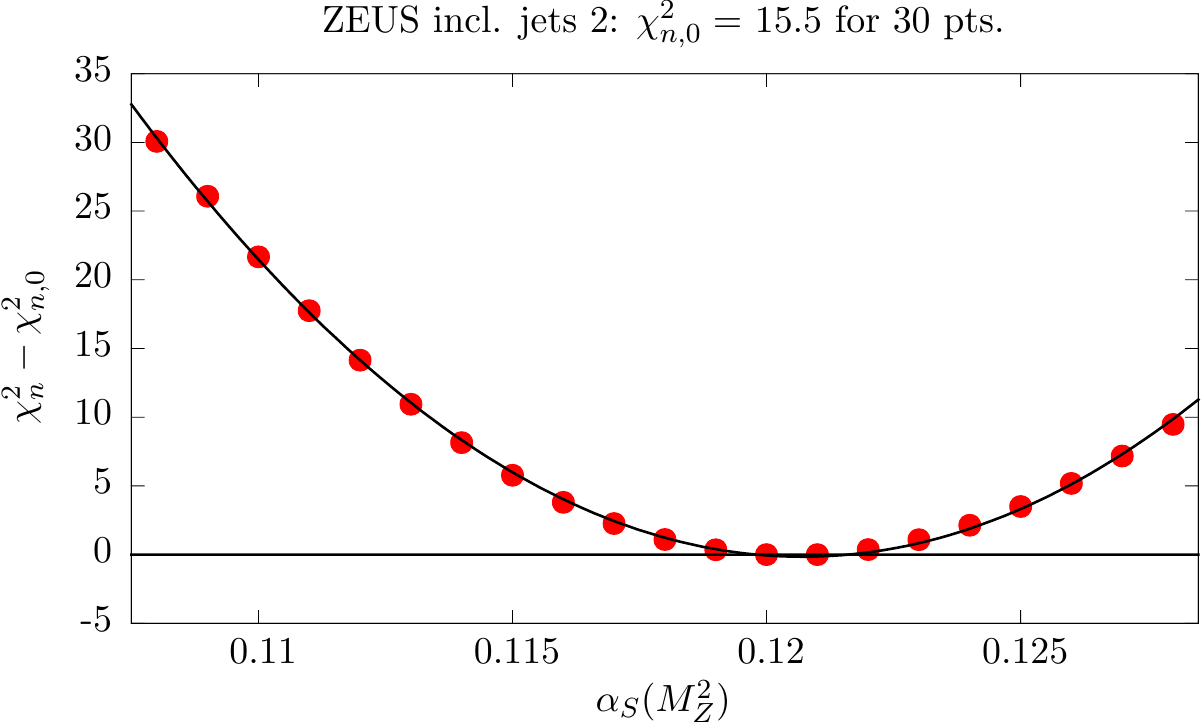}
    \includegraphics[width=0.45\textwidth]{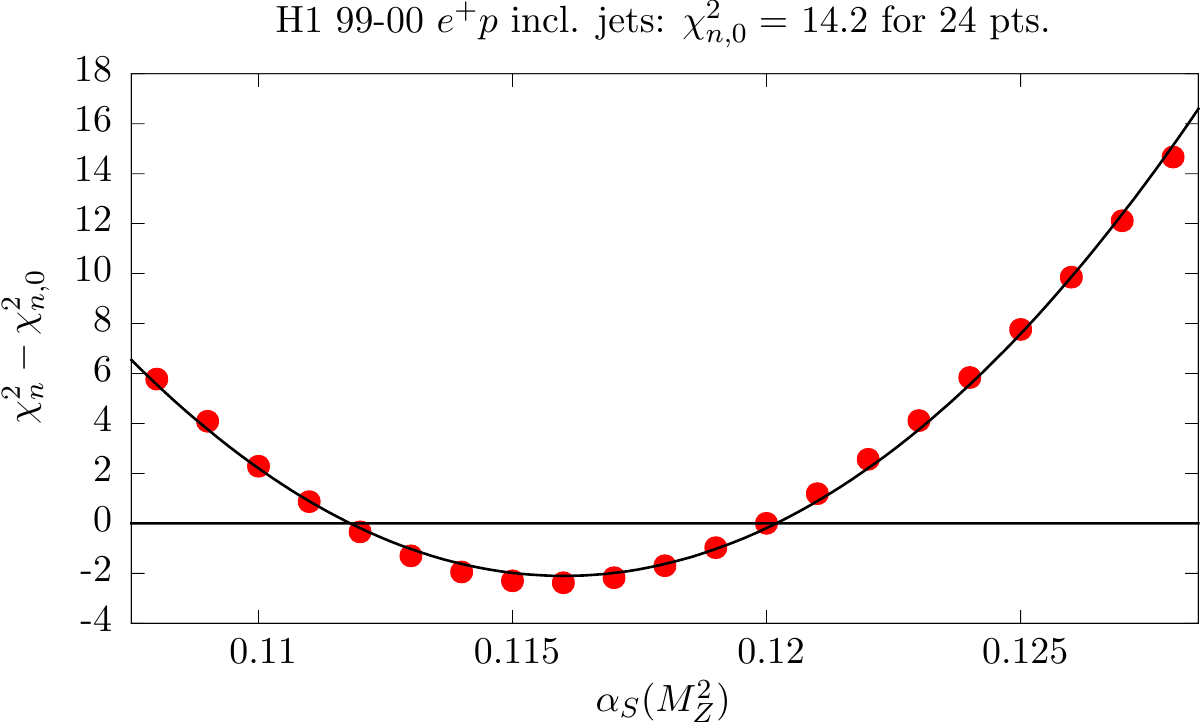}
  \caption{$\chi_n^2$ profiles for jet data sets, included in the NLO fit, but not in the NNLO fit, when varying $\alpha_S(M_Z^2)$.}
  \label{fig:nloonly}
\end{figure}

The longitudinal structure function $F_L$ leads off with an $\alpha_S$ term, 
and so the value of $(\chi^2_n-\chi^2_{n,0})$ depends more sensitively on $\a$. The NNLO plot shows an excellent quadratic 
dependence on  $\alpha_S(M_Z^2)$, centred at 0.118.
The NNLO coefficient functions for 
$F_L(x,Q^2)$~\cite{Moch:2004xu,Vermaseren:2005qc} are positive and significant,
and the NLO fit tries to mimic these with a higher value of $\alpha_S(M_Z^2)$.
Indeed, the data for $F_L$, and also the E866/NuSea $pp$ Drell--Yan cross 
sections data, are clearly more quadratic at NNLO than at NLO, with minima 
closer to the best--fit values.  This indicates a strong preference for the 
NNLO description, which is not so apparent if only the global best--fit values 
$\chi_{n,0}^2$ are known. As the E866/NuSea data for $pd/pp$ Drell--Yan 
production are a ratio of cross sections, the sensitivity to the value of 
$\alpha_S(M_Z^2)$ is small.

The Tevatron data, as well as the ATLAS $W^{\pm},Z$ production data and the 
ATLAS high--mass Drell--Yan data, show, at NNLO, $\alpha_S(M_Z^2)$ profiles with 
quadratic behaviour with minima close to the best fit values. Again, the 
profiles are improved to those at NLO. The counter example are the LHCb 
data, which have profiles which are more reasonable at NLO than at NNLO.
In general, the charge--lepton asymmetry measurements arising from $W^{\pm}$ 
production at the Tevatron and the LHC, which are a ratio of cross sections, 
have much less constraint on the value of $\alpha_S(M_Z^2)$.

Judging from the values of $(\chi^2_n-\chi^2_{n,0})$
away from the different minima of the various data sets or, rather, the 
steepness of the quadratic forms in $\alpha_S(M_Z^2)$, we see that there is a 
tendency for data at lower energies or lower $Q^2$ to have more constraint on 
the optimum global value of $\alpha_S(M_Z^2)$. This is to be anticipated, as we 
will see in Section \ref{sec:PDF}.

\section{$t\overline{t}$ data: $m_t$ -- $\alpha_S$ correlation  \label{sec:tt}}

\begin{figure}
  \centering
       \includegraphics[width=0.45\textwidth]{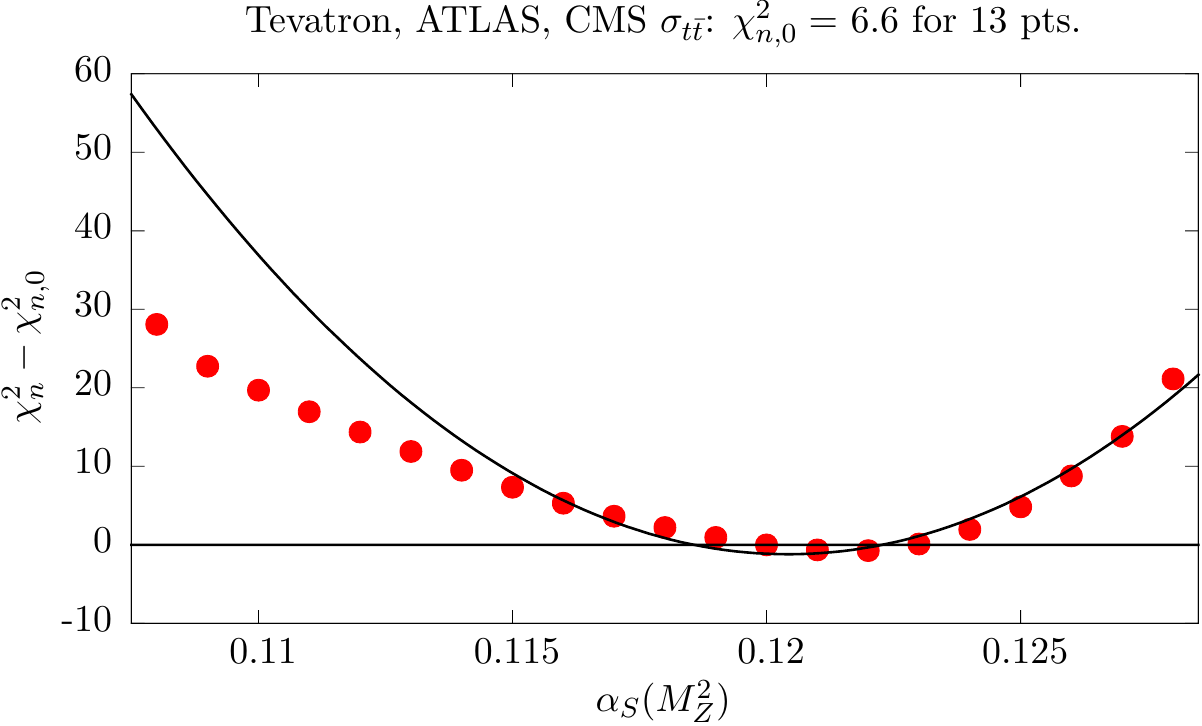}\qquad
              \includegraphics[width=0.45\textwidth]{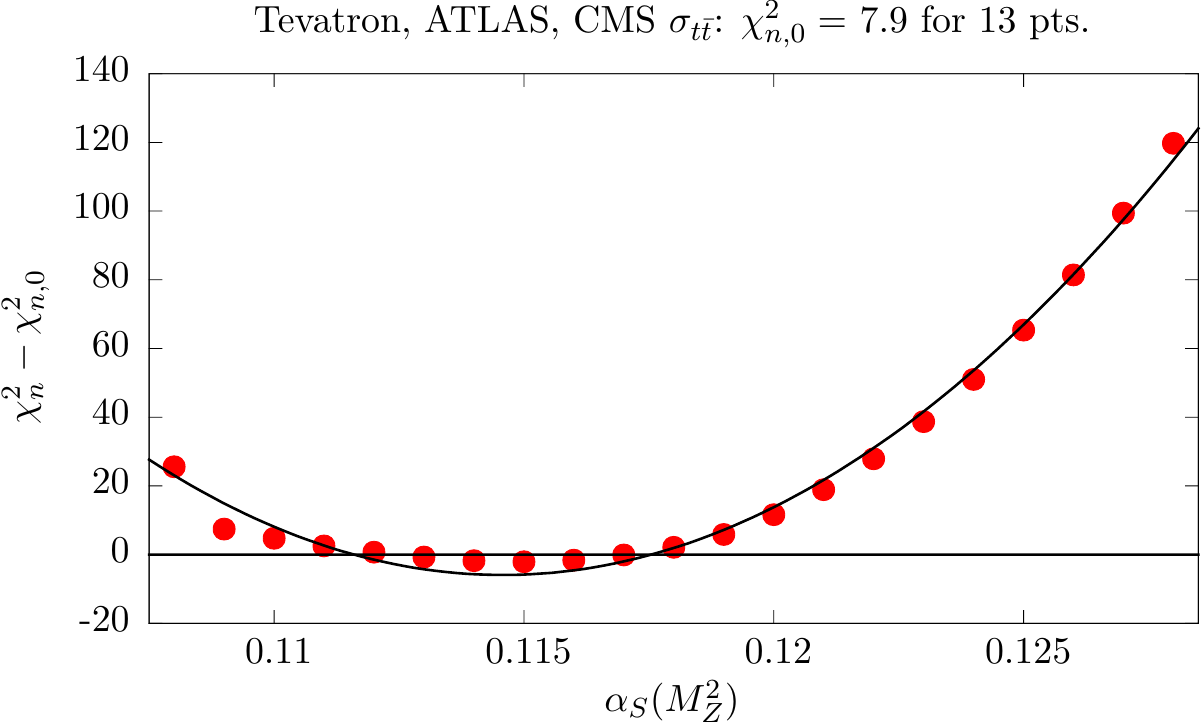}
  \caption{$\chi_n^2$ profiles for $t\overline{t}$ data in the NLO (left) and NNLO (right) fits, when varying $\alpha_S(M_Z^2)$.}
  \label{fig:nnlotop}
\end{figure}

There is a particularly strong, but also complicated, relationship
between the value of $\alpha_S(M_Z^2)$ and the fit to data on the inclusive cross section for  ${t \bar t}$ production,
$\sigma_{t \bar t}$. We show the $\chi^2$ profiles at NLO and NNLO 
in Fig.~\ref{fig:nnlotop}. Clearly there is a preference for a lower value
of $\alpha_S(M_Z^2)$ at NNLO than at NLO, and a strong constraint in both cases, with $\chi^2$ increasing by a large number of units, certainly compared to the number of data points, for
small changes in $\a$. 
Indeed, nominally $\sigma_{t\bar t}$ provides one of the strongest constraints
of any data set for the lower limit of $\a$ at NLO and the upper limit of 
$\a$ at NNLO. However, the picture is more complicated than for other data 
sets due to the very strong correlation with the value of the mass $m_t$ of the top quark.

In the global fits the theory calculation of $\sigma_{t \bar t}$ is performed 
with a preferred value of the top quark pole mass of $172.5~\GeV$, since this 
is the default in PYTHIA, used to extract 
the cross section in many of the measurements. Moreover, the majority of the cross
sections are quoted for this value of $m_t$. This value is also consistent 
with the world average of the measured value of $173.34$ GeV~\cite{PDG2014}. 
However, we allow a $1~\GeV$
uncertainty on the value of $m_t$ which can be thought of as accounting 
for the uncertainty in the value of $m_t$ itself and also for the small
variation in the extracted cross sections with $m_t$ used; in general this is 
about a third the size of the variation of the calculation of 
$\sigma_{t \bar t}$ with $m_t$, and the net effect is an effective uncertainty
a little lower than $1~\GeV$.  To be specific, $m_t$ is left as a free 
parameter in the fit, but there is a $\chi^2$ penalty of $\chi^2_{m_t} = 
(m_t-172.5~\GeV)^2$ applied to keep the value close to the preferred value.
This penalty is included in the values in Fig.~\ref{fig:nnlotop}.  
The allowed variation in $m_t$ away from the preferred central value of 
$172.5~\GeV$ results in the NLO fit preferring a low value of 
$m_t=171.7~\GeV$ and the NNLO fit preferring a high value of $m_t=174.2~\GeV$. 
The low value of $m_t$ in the global fit and the high value of $\a$ preferred by 
$\sigma_{t \bar t}$ when $\a$ is varied, both occur for the same reason. That is, the 
NLO cross section tends to undershoot the data, and raising $\a$ and lowering 
$m_t$ both raise the cross section, leading to better agreement.

The NNLO correction to the cross section
in the pole mass scheme is moderate, but large compared to the most precise data,
and hence the NNLO cross section tends to be too high. This leads to the 
opposite pulls to those at NLO, i.e NNLO prefers $\a$ low and $m_t$ high. 
Within the global fit we find that the allowed variation with accompanying 
penalty for deviations from $m_t=172.5~\GeV$ results in $m_t$ values at the 
best fit values of $\a$
which are of order $1-2 \sigma$ away from either our preferred 
value or the world
average, so have no particular inconsistency, but it is useful to examine the 
interplay between $\a$ and $m_t$ in rather more detail.

\subsection{Effect on $\chi^2_{\rm global}$ to changes of $m_t$ and $\a$}
\begin{figure}
  \centering
       \includegraphics[width=0.49\textwidth]{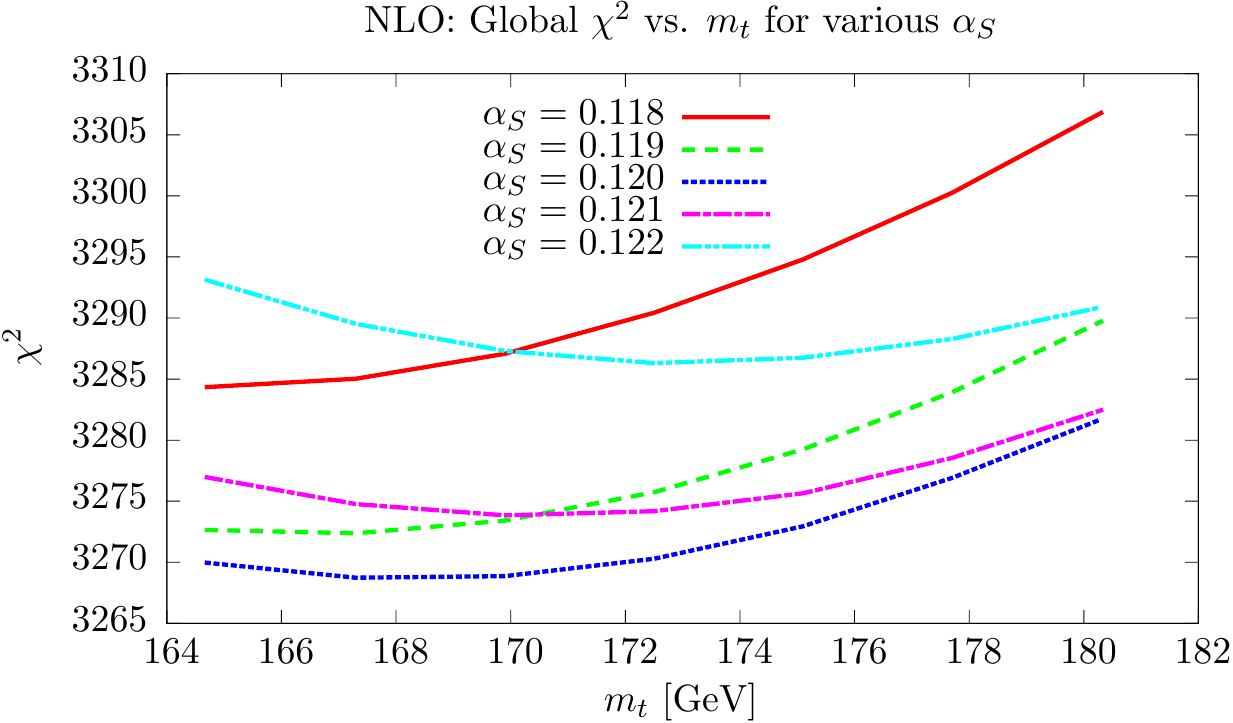}
             \includegraphics[width=0.49\textwidth]{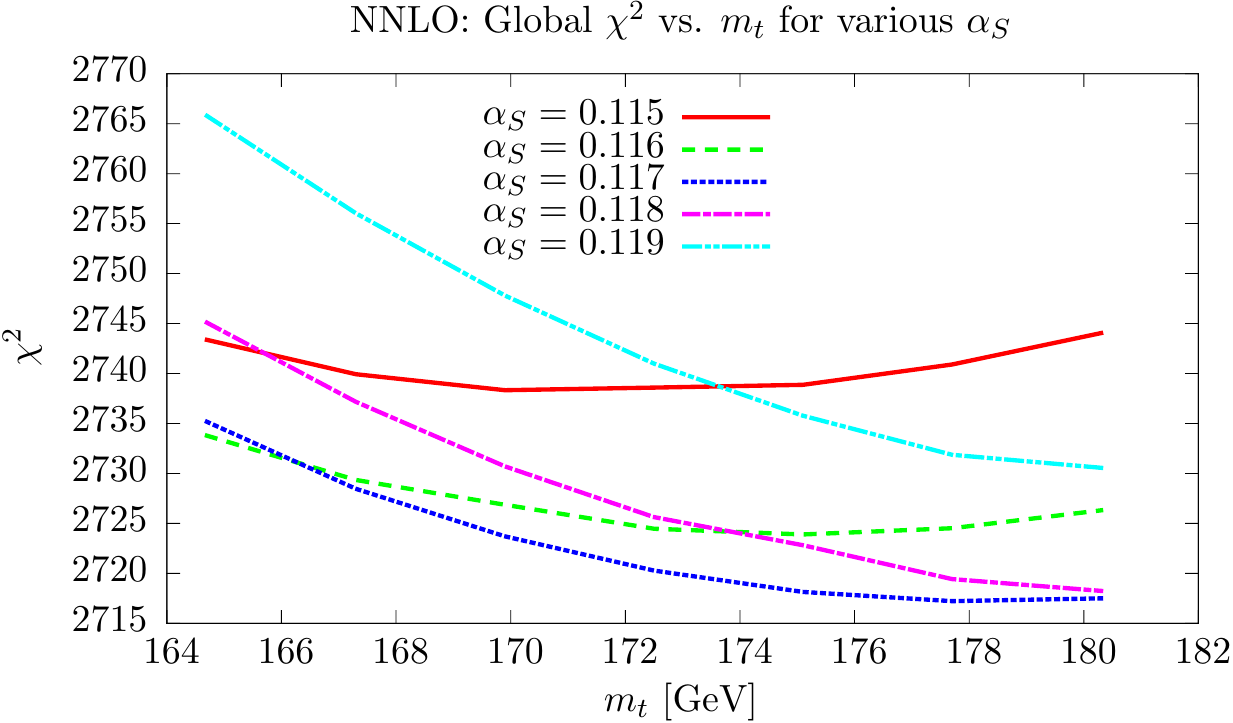}
           \caption{Global $\chi_n^2$ minima as a function of the top mass $m_t$, for different fixed values of $\alpha_S(M_Z^2)$. There is no $\chi^2$ penalty for varying $m_t$.}
  \label{fig:nnlotopglobal}
\end{figure}

First we investigate the quality of the global fit as a function of both
$\a$ and $m_t$. This is shown in Fig.~\ref{fig:nnlotopglobal}, where we plot $\chi^2_{\rm global}$ versus 
$m_t$ at several different values of $\a$ (In these plots $m_t$ is varied with no $\chi^2$ penalty for
deviations away from the ``preferred'' value). At NLO one can see
that regardless of $m_t$ the best global fit is always obtained quite clearly for $\a$ near to $0.120$,
with the fit quality for $\a=0.119$ or $\a=0.121$ each being a few units worse at all values of $m_t$.
It is only for $m_t > 180~\GeV$ that the quality for $\a=0.121$ approaches that of $0.120$ and the best 
fit would be for $\a\approx 0.1205$. At this mass the global $\chi^2$ is about 10 units above the minimum though. 
Similarly at NNLO $\a=0.117$ gives a lower $\chi^2_{\rm global}$ for all masses between about $166~\GeV$ and 
$181~\GeV$, when $\a=0.116$ and  $\a=0.118$ respectively give the same $\chi^2_{\rm global}$ values. Hence, 
even completely unreasonable variations of $\sim 7-10~\GeV$ result in changes of the best fit values of $\a$ of 
only $\sim 0.0005$. We do note, however, that without a penalty for $m_t$ variation the best global fits
are at $m_t=168~\GeV$ and $m_t=180~\GeV$ at NLO and NNLO respectively, so some penalty is clearly necessary. Ultimately, the value of $\a$ determined by the global fit is 
very insensitive to the value of $m_t$ used, and indeed, to the $\sigma_{t \bar t}$ data,
because these correspond to relatively few data points. Indeed, if these are left out of the global 
fit the change in the optimum value of $\a$ is only of order $0.0001-2$ at NLO and NNLO. 
However, the interplay between $\a$ and $m_t$ is more dramatic for the $\sigma_{t \bar t}$
data alone, as we will now show.
  
\subsection{Effect on $\chi^2_{t \bar t}$ to changes of $m_t$ and $\a$}
\begin{figure}
  \centering
       \includegraphics[width=0.49\textwidth]{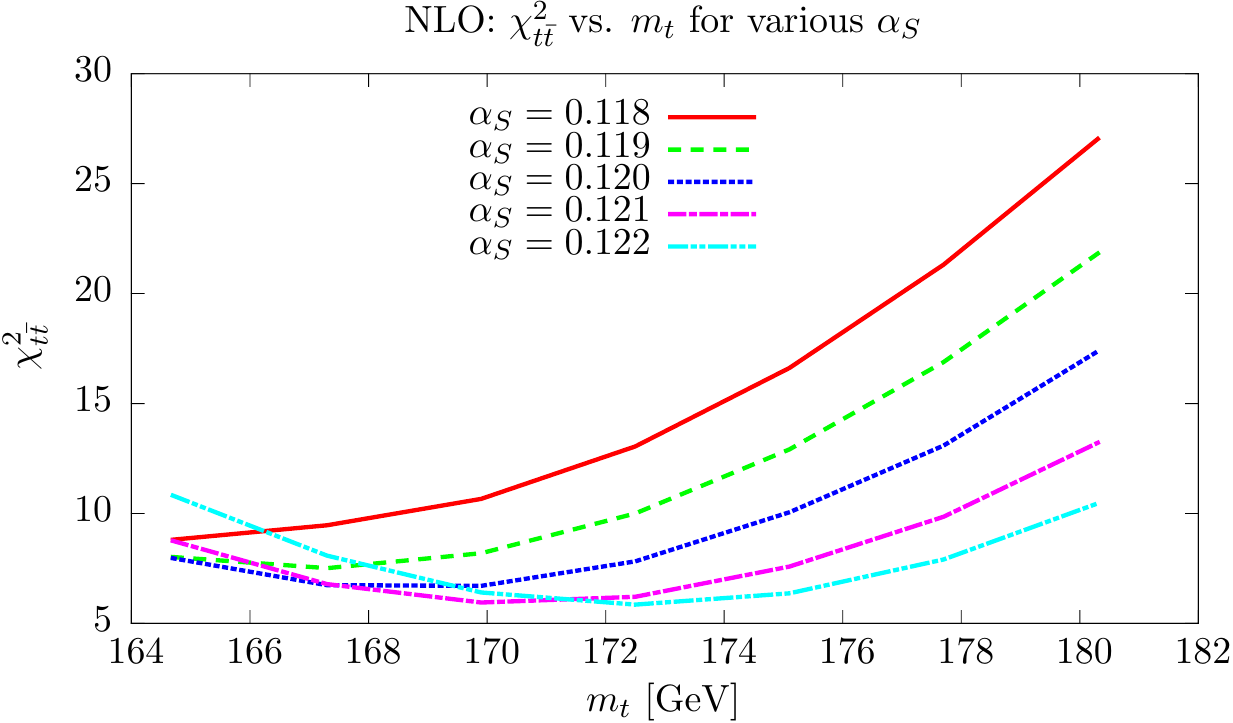}
             \includegraphics[width=0.49\textwidth]{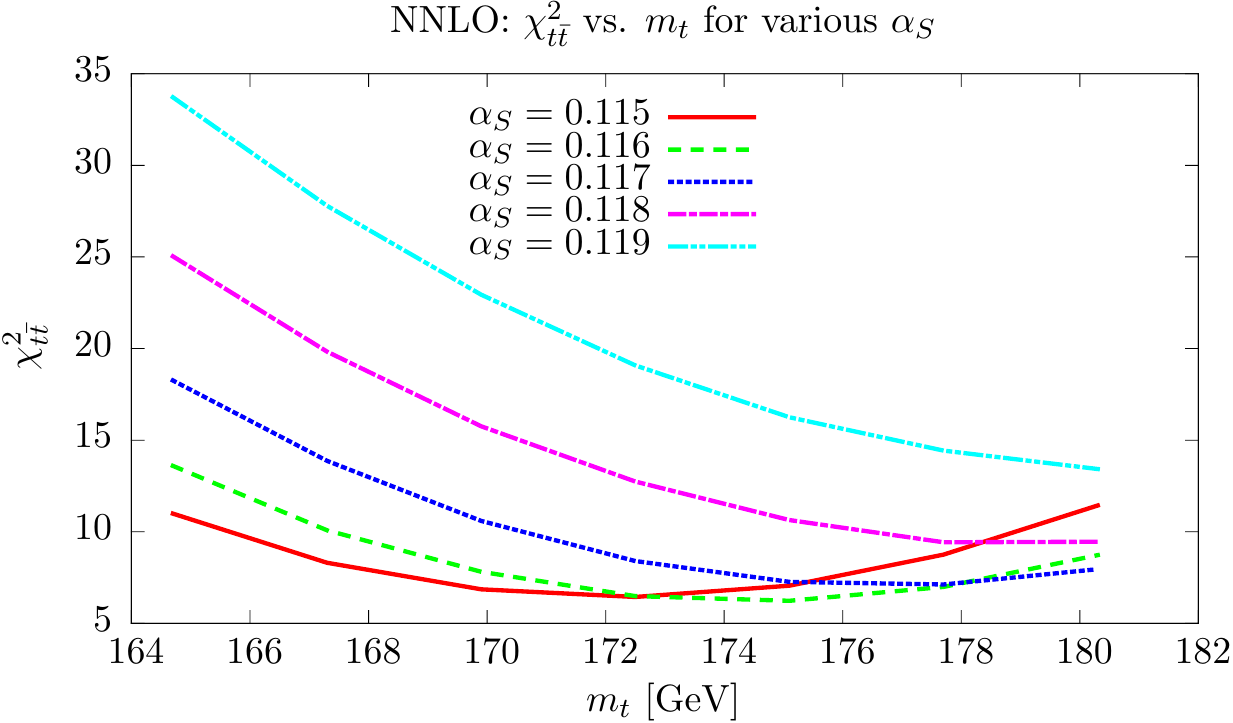}
           \caption{$\chi_n^2$ values for inclusive $t\overline{t}$ cross section data at the global minimum, as a function of the top mass $m_t$, for different fixed values of $\alpha_S(M_Z^2)$. There is no $\chi^2$ penalty for varying $m_t$.}
  \label{fig:nnlotoponly}
\end{figure}

The equivalent plots to Fig.~\ref{fig:nnlotopglobal} are shown in Fig.~\ref{fig:nnlotoponly} for the fit quality 
to the inclusive $\sigma_{t \bar t}$ cross section data. Again, there is no 
penalty applied for $m_t$ variation. At NLO it is clear 
that, except for very low values of $m_t$, the best fit is achieved for higher values of 
$\a$, i.e. $\a=0.121$ or for $m_t>172~\GeV$, $\a=0.122$. Indeed, the best possible fit
to the top cross section data is for $m_t\approx 172~\GeV$ and $\a=0.122$. However, 
the improvement in $\chi^2_{t\bar t}$ compared to $\a=0.120$ for this mass is only $\sim 2$
units -- far less than the deterioration in the $\chi^2$ for the rest of the data when going from 
$\a=0.120$ to $0.122$. Overall the minimum $\chi^2$ achieved for any $\a$ is quite flat with 
 $m_t$, changing by at most 2 units for $168~\GeV < m_t < 178~\GeV$. 
However, it is clear that the variation of $\chi^2_{t \bar t}$ is different for different
values of $\a$. As $\a$ decreases there is a preference for a smaller mass, hence if the 
central value of $m_t$ had been chosen higher than $172.5~\GeV$ for example, the best fit 
to $\sigma_{t \bar t}$ would be for a higher value of $\a$.
The constraint on $\a$ in the upper direction would be weakened slightly, however this data set does provide a significant constraint in this direction. If the penalty had been 
less severe, e.g. a increase in $\chi^2_{t \bar t}$ for $\Delta m_t =2~\GeV$ rather than 
$\Delta m_t=1~\GeV$, the best value of $m_t$ and $\a$ would not change significantly, as the 
fit quality does not improve for masses of $m_t<171.7$ for any $\a$, even discounting the penalty. However, the $\chi^2_{t \bar t}$ curves for lower values of $\a$, i.e. 0.119 and 0.118
are falling quite steeply as $m_t$ decreases in the vicinity of 
$m_t=172~\GeV$, so the increase in $\chi^2_{t\bar t}$ with decreasing $\a$ seen in Fig.~\ref{fig:nnlotop} (left) would be less severe if $m_t$ was allowed to choose smaller values, and the constraint on the lower
values of $\a$ would be reduced somewhat.    
Hence, at NLO, alternative treatments of $m_t$ would allow a slightly higher best fit $\a$ than the default treatment, and a  little scope for a relaxation of the lower limit on $\a$. 

At NNLO it is again clear that higher values of $\a$ prefer higher values of $m_t$. However, for $\a=0.118$ or $0.119$ 
the value of $m_t$ corresponding to the best fit is $m_t=180~\GeV$ or more. Again, there is little variation in the best value of $\chi^2_{t\bar t}$ for $168~\GeV$$ < m_t < 178~\GeV$, but the best fit is achieved for $\a=0.115$ or $0.116$ \footnote{Our constraint on $\a$ is very consistent with that in \cite{Chatrchyan:2013haa}.}, only 
becoming $\a=0.117$ at $m_t=178~\GeV$. In this case if the penalty for variations in $m_t$ away from the default central value were relaxed it would make little difference, as even for 
$\a=0.115$ the best fit is for $m_t\approx 172 ~\GeV$. It might 
allow slightly better fits for $\a\sim 0.110$, but this would have no influence on the overall constraint on $\a$, which is constrained by many data not to be much lower than $0.115$. 
A potential increase in $m_t$, either by change of 
default central value, or a relaxation of the penalty, would allow for a potentially a slightly higher value of $m_t$ for the best fit, as the minimum possible $\chi^2_{t\bar t}$ is almost completely flat between $172~\GeV$ $<m_t < 176~\GeV$. This would be accompanied by a slight increase in $\a$.  
It would also allow a 
little relaxation in the constraint on higher values of $\a$. The $\chi^2_{t \bar t}$ curves for $\a=0.118$ and 0.119 are decreasing with increasing $m_t$ in the vicinity of $m_t=174~\GeV$, and a higher allowed value of $m_t$ would enable the increase in $\chi^2$ with $\a$ in Fig.~\ref{fig:nnlotoponly} (right) to be less steep.  Hence, at NNLO alternative treatments of $m_t$ would allow a slightly higher best fit value of $\a$ than the default treatment, and a  little scope for a relaxation of the upper limit on $\a$. 
 
Hence, the overall conclusion is that some added freedom in $m_t$ would lead to potentially rather small changes in the minima of the $\chi^2$ curves in Fig.~\ref{fig:nnlotoponly}, but a reduced rate of increase of $\chi^2$ away from the minima. The implications of this will be discussed in the next section.

\section{Uncertainty on $\alpha_S(M_Z^2)$ and calculation of PDF+$\alpha_S(M_Z^2)$ uncertainty  \label{sec:uncsq}}
First, recall that in the MMHT2014 analysis \cite{MMHT} we determined the uncertainties of the PDFs using the Hessian approach with a dynamical tolerance procedure. We obtained PDF `error' eigenvector sets, each corresponding to 68$\%$ confidence level uncertainty, where the vectors are orthogonal to each other and span the PDF parameter space.

\begin{figure} 
\begin{center}
\vspace*{-1.0cm}
\includegraphics[height=9cm]{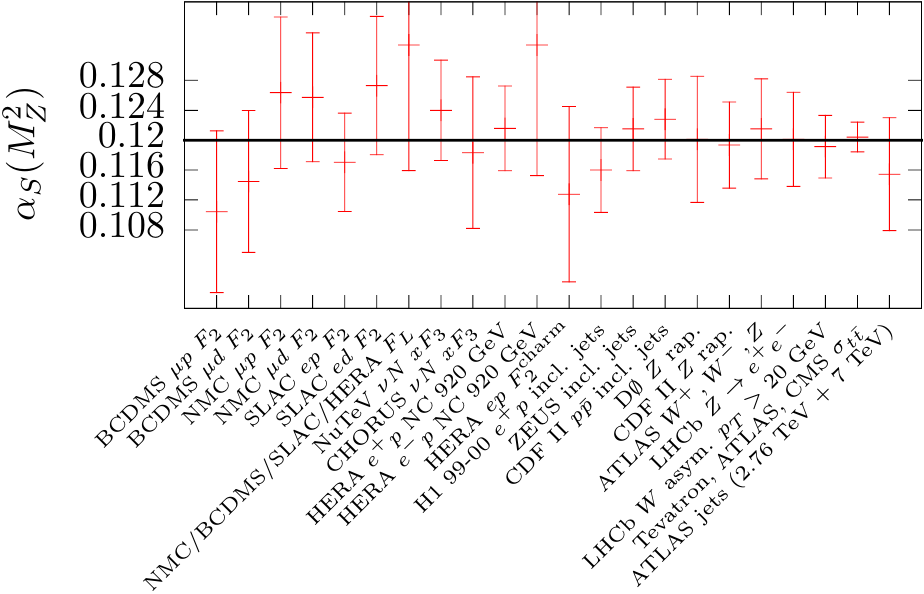}
\includegraphics[height=9cm]{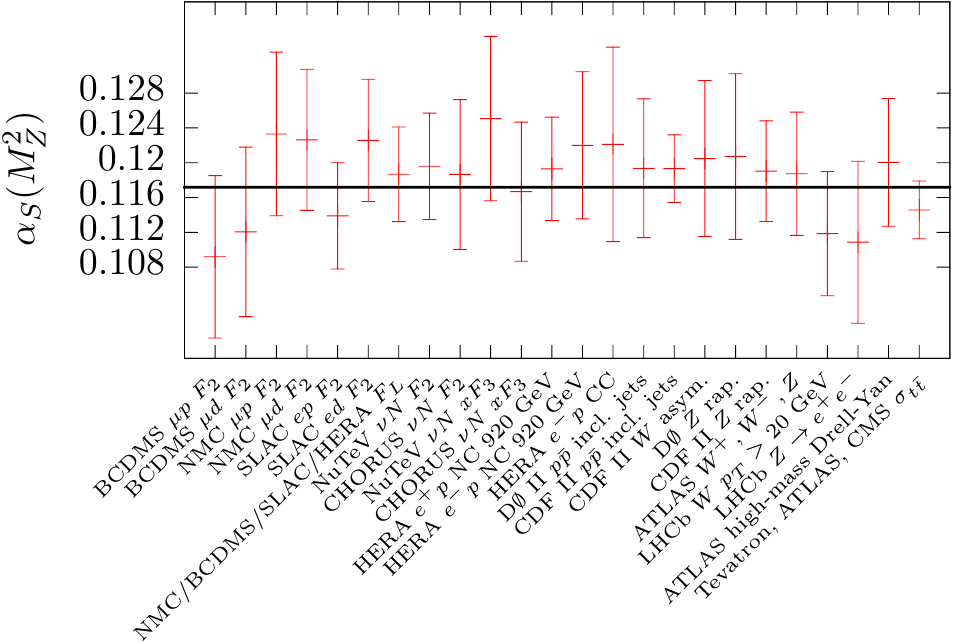}
\caption{\sf The upper and lower plots show the value of $\a$ corresponding to the best fit, together with
 the upper and lower  1$\sigma$ constraints on $\a$ from the more 
constraining data sets at NLO and NNLO respectively.}
\label{fig:limits}
\end{center}
\end{figure}

In order to determine the uncertainty on $\alpha_S(M_Z^2)$ at NLO and NNLO 
we begin by using the same technique as in the MSTW study of  Ref. \cite{MSTWalpha}; that is, for the `error'
eigenvectors 
we apply the 
tolerance procedure to determine the uncertainty in each direction away from
the value at the best fit when one data set goes beyond its $68\%$ confidence
level uncertainty. The values at which each data set does reach its 
$68\%$ confidence level uncertainty, plus the value of $\alpha_S(M_Z^2)$ 
for which each data set has its best fit (within the context of a global fit)
are shown at NLO and NNLO in Fig.~\ref{fig:limits}. However, unlike Fig.~7 of
\cite{MSTWalpha} we do not show all data, as with the increased number of sets there are now too many to show
clearly on a single figure. Moreover, as seen earlier, many data sets have very little
dependence, and hence produce very little constraint. Hence, we show those 
where both limits are within the range of $\alpha_S(M_Z^2)$ explicitly studied,
i.e. $0.108-0.128$ or where one limit is within 0.005 of the best fit
value of $\alpha_S(M_Z^2)$. None of the data sets omitted using these 
criteria have a significant pull on $\alpha_S(M_Z^2)$.  

The dominant constraint on $\a$ in the downwards direction at NLO is from the 
top pair cross section data and, using the dynamical tolerance procedure, gives 
an uncertainty of $\Delta \a =-0.0014$. In the upwards direction it is the BCDMSp data with 
an uncertainty of $\Delta \a = +0.0012$. At NNLO the dominant downward constraint 
comes from NuTeV $F_3(x,Q^2)$ data which gives $\Delta \a= - 0.0012$ and in the upwards 
direction it is the top pair cross section data, where the uncertainty is 
$\Delta \a=+0.0008$. 

There are a number of other data sets which give almost as strong 
constraints. For instance, at NLO in the downwards direction we find that SLAC deuterium 
data give $\Delta \a= -0.0018$ and in the upwards direction H1 jets give 
$\Delta \a= +0.0019$. At NNLO in the downwards direction SLAC deuterium 
data and CDF jet data give $\Delta \a \approx -0.0014$,   
and in the upwards direction, at NNLO, the BCDMSp data give $\Delta \a =+0.0014$. In all cases 
there are other data sets that are not much less constraining than those mentioned 
explicitly. Hence, in no case is it a single data set which is overwhelmingly 
providing the dominant constraint on the upper or lower limit of $\a$. 
Similarly, no single data sets would change the central value by more than 
$0.001$ if it were to be omitted. 

Two of the four dominant constraints nominally come from $\sigma_{t \bar t}$, and at 
NLO we have $\a =0.1201_{-0.0014}^{+0.0012}$ and at NNLO $\a = 0.1172_{-0.0012}^{+0.0008}$. 
However, in the previous section we highlighted the interplay between $\a$ and 
$m_t$ when examining the fit quality of the $\sigma_{t \bar t}$ data. We demonstrated 
that if some extra flexibility is allowed on the choice of central value of $m_t$ 
and/or on the 1-$\sigma$ uncertainty that is used, then the constraints are relaxed to some 
degree. Hence, we are reluctant to treat the constraint from the data on $\sigma_{t \bar t}$
completely rigorously. In order to see quite how we should deal with the constraints 
nominally due to these data we first check which data sets provide the next tightest 
constraint. 
If we were simply to ignore the constraints from $\sigma_{t \bar t}$ we would find 
a change in uncertainty at NLO of $\Delta \a = -0.0012 \to -0.0017$  and at NNLO 
$\Delta \a = +0.0008 \to +0.0014$. These are significant, but hardly dramatic changes, and
it would be no surprise if some alternative treatment of the default top mass 
resulted in changes of a similar type. Hence, it might be suitable to take these values 
as a simple alternative, arguing that the constraints from $\sigma_{t \bar t}$ are not 
sufficiently greater than those from other data sets {\it either} to ignore the possible effects 
of alternative treatments of the mass $m_t$ {\it or} to warrant a completely thorough investigation at 
this stage.\footnote{The constraint from $\sigma_{t \bar t}$ data does provide the 
dominant constraint in one direction for eigenvector 15 at NNLO. However, very nearly 
as strong a constraint is provided by other data sets and the eigenvector only provides
at the very most $40\%$ of the uncertainty on one distribution, the gluon, at 
any $x$ value, in practice at high $x$. Hence, a slightly increased tolerance for this 
eigenvector would have minimal impact on any PDF uncertainties.} 
However, there is the additional feature to note --  whichever criterion we use we 
have some, albeit not too dramatic, asymmetry in the $\a$ uncertainty. There is no strong 
reason to apply this slight asymmetry, as the $\chi^2$ profile for the global fit
follows the quadratic curve very well at both NLO and NNLO, and the degree of asymmetry 
obtained using the dynamical tolerance procedure is arguably within the 
``uncertainty of the uncertainty''. Hence at NLO and NNLO we average the two uncertainties 
(obtained without the $\sigma_{t \bar t}$ constraint) obtaining 
\bea
\alpha_{S,{\rm NLO}}(M_Z^2) & = & 0.1201 \pm 0.0015 \label{eq:optNLOunc}\\
\alpha_{S,{\rm NNLO}}(M_Z^2) & = & 0.1172 \pm 0.0013. \label{eq:optNNLOunc}
\eea
This corresponds to $\Delta^{\rm NLO} \chi^2_{\rm global}=10.3$ and 
$\Delta^{\rm NNLO} \chi^2_{\rm global}=7.2$. These are the sort of tolerance values typical of 
the majority of PDF eigenvectors. 
 
Each of these values of $\a$ is within 1$\sigma$ of the world average (without DIS data) 
of $0.1187 \pm 0.0007$, though in opposite directions. As noted earlier, the inclusion of 
$\a$ as a data point leads to values of $0.1178$ and $0.1195$ at NNLO and NLO 
respectively. These are somewhat closer to the world average, and very near to $0.118$ at NNLO, 
but still quite close to 0.120 at NLO.\footnote{Taking a weighted average of the 
values in eqs. (\ref{eq:optNLOunc}) and (\ref{eq:optNNLOunc}) would result in values
slightly nearer to the world average, reflecting the fact that the dynamical tolerance 
procedure used to determine the uncertainty results in a $\Delta \chi^2_{\rm global} > 1$.} 
Hence, we interpret the values in 
eqs. (\ref{eq:optNLOunc}) and (\ref{eq:optNNLOunc}) as independent measurements of 
$\a$, but acknowledge that at NNLO taking both this determination and the world average 
into account a round value of $\a=0.118$ is an appropriate one at which to present the 
PDFs. At NLO we would recommend the use of $\a=0.120$ as the preferred value for the 
PDFs, but have made eigenvector sets available at $\a=0.118$. If a value of 
$\a=0.119$ were desired the average of the results at $\a=0.118$ and $0.120$ would 
provide an excellent approximation.

When considering the uncertainty on the prediction for a physical quantity we should include the uncertainty on $\a$, as well as that on the PDFs. This is particularly important for cross sections that at leading order are proportional to a power of the coupling, such as $\sigma_{t{\bar t}}$ or $\sigma_{\rm Higgs}$, which are proportional to $\alpha_S^2$. A naive procedure would be to compute the error as
\be
\Delta \sigma~=~\sqrt{(\Delta \sigma_{\rm PDF})^2+(\Delta\sigma_{\alpha_S})^2}
\label{eq:PDFaunc}
\ee
where $\Delta\sigma_{\alpha_S}$ is variation of the cross section when $\a$ is allowed to vary over a given range. However, it is inconsistent to use different values of $\alpha_S$ in the partonic hard subprocess cross section and in the PDF evolution. Moreover, in a global PDF analysis, there are non--negligible correlations between the PDFs and the value of $\alpha_S$.

In the MSTW study \cite{MSTWalpha} of the PDF$+\a$ uncertainties arising from the 
MSTW2008 analysis we advocated using our best fit value of $\a$ as the central 
value for PDF predictions, and then provided additional eigenvector sets  
at $\pm 0.5 \sigma$ and $\pm 1 \sigma$ values of $\a$. The uncertainty was then 
calculated by taking the envelope of the predictions using all these eigenvector 
sets. This still seems like an appropriate algorithm for use with the dynamical
tolerance procedure of obtaining uncertainties. However, it can only really be applied
if the central prediction is obtained using the PDFs defined at the best fit 
value of $\a$, which is no longer the case, and moreover was a rather complicated and 
time--consuming procedure.

Since the MSTW study \cite{MSTWalpha} was undertaken  it has been shown
that, within the Hessian approach to PDF uncertainties,  
the correct PDF+$\a$ uncertainty on any quantity can be obtained by simply taking the PDFs
defined at $\a \pm \Delta \a$ and treating these two PDF sets (and their accompanying
value of $\a$) as an extra pair of eigenvectors \cite{CTEQalphas}. In short, the full uncertainty is obtained
by adding the uncertainty from this extra eigenvector pair in quadrature with the PDF
uncertainty. So we are back to the naive form (\ref{eq:PDFaunc}), but now, importantly, with the correlations between the PDFs and $\alpha_S$ included.  This has the advantages of both being very simple, but also separating
out the $\a$ uncertainty on a quantity explicitly from the purely PDF
uncertainty. Strictly speaking, the method only holds if the central PDFs are those 
obtained from the best fit when $\a$ is left free, and if the uncertainty $\Delta\a$ on 
$\a$, that is used, is the uncertainty obtained from the fit. If we use PDFs defined 
at $\a=0.118$ at NNLO we are still very near the best fit, and the error induced will
be very small. At NLO a larger error will be induced by using the PDFs defined at 
$\a=0.118$ than those at $\a=0.120$. Any choice of $\Delta \a$ of $0.001-0.002$ should
only induce a small error. Hence, overall we now advocate using this approach
with NLO PDFs defined at $\a=0.120$ and NNLO PDFs defined at $\a=0.118$. The value of 
$\Delta \a$ is open to the choice of the user to some extent, but it is recommended 
to stay within the range $\Delta \a$ that we have found. 

In Section \ref{sec:cxunc} we apply the above procedure to determine the PDF$+\a$ uncertainties on the predictions for the cross sections for benchmark processes at the Tevatron and the LHC, but first we examine the change in the PDF sets themselves with $\a$.

\section{Comparison of PDF sets  \label{sec:PDF}}

\begin{figure} 
\begin{center}
\includegraphics[height=5cm]{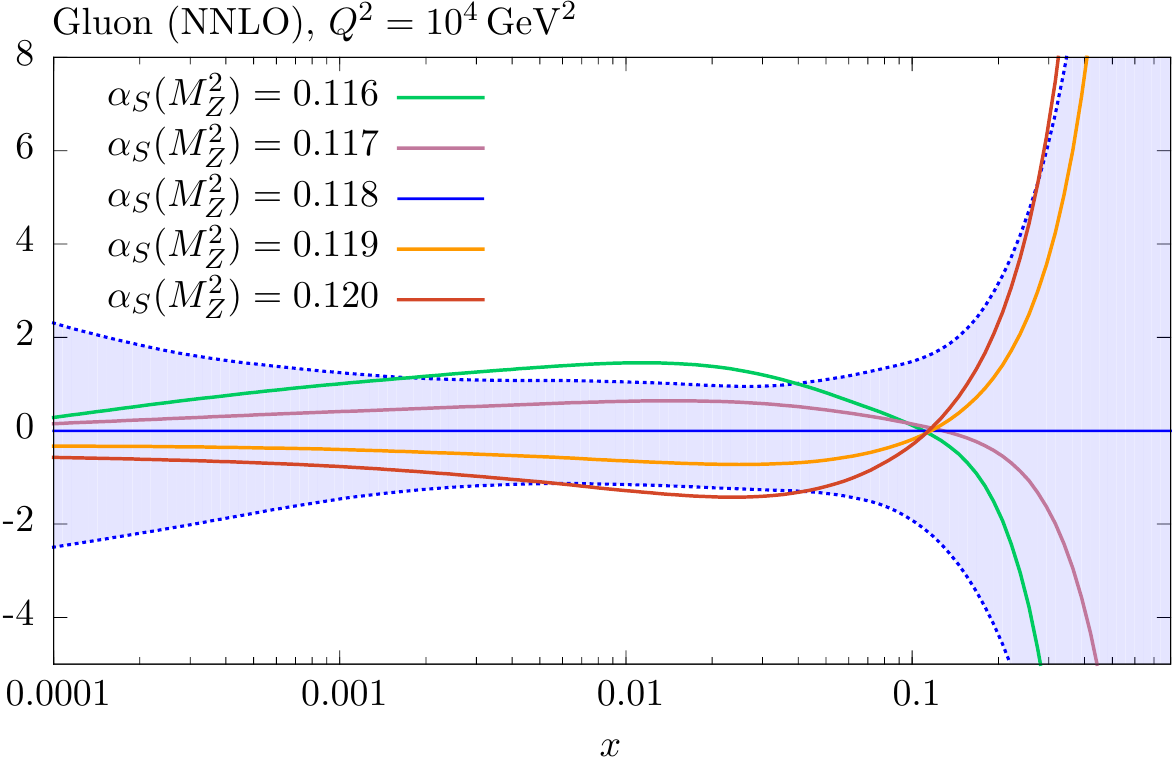}
\includegraphics[height=5cm]{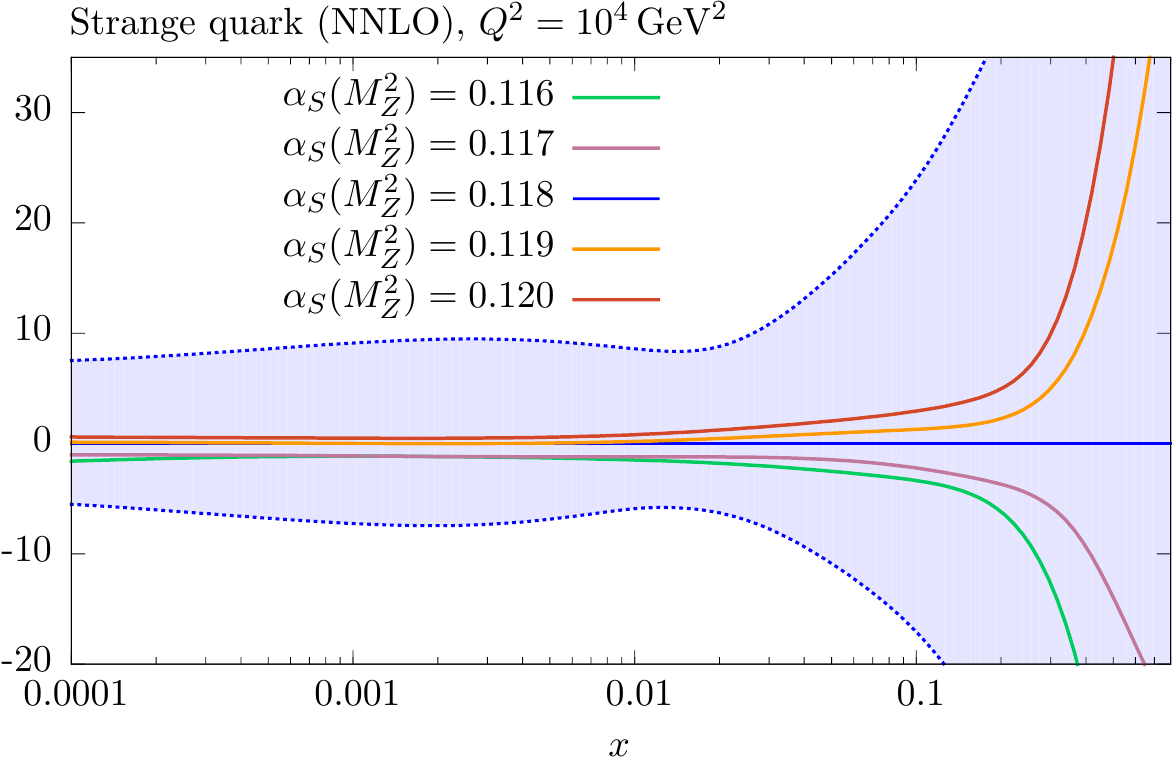}
\caption{Percentage difference in the NNLO gluon and strange quark PDFs at $Q^2=10^4$ ${\rm GeV}^2$ relative to central ($\alpha_S(M_Z^2)=0.118$) set for fits with different values of $\alpha_S$, with the percentage error bands for the central set also shown.}
\label{fig:pdfcomp1}
\end{center}
\end{figure}

\begin{figure} 
\begin{center}
\includegraphics[height=5cm]{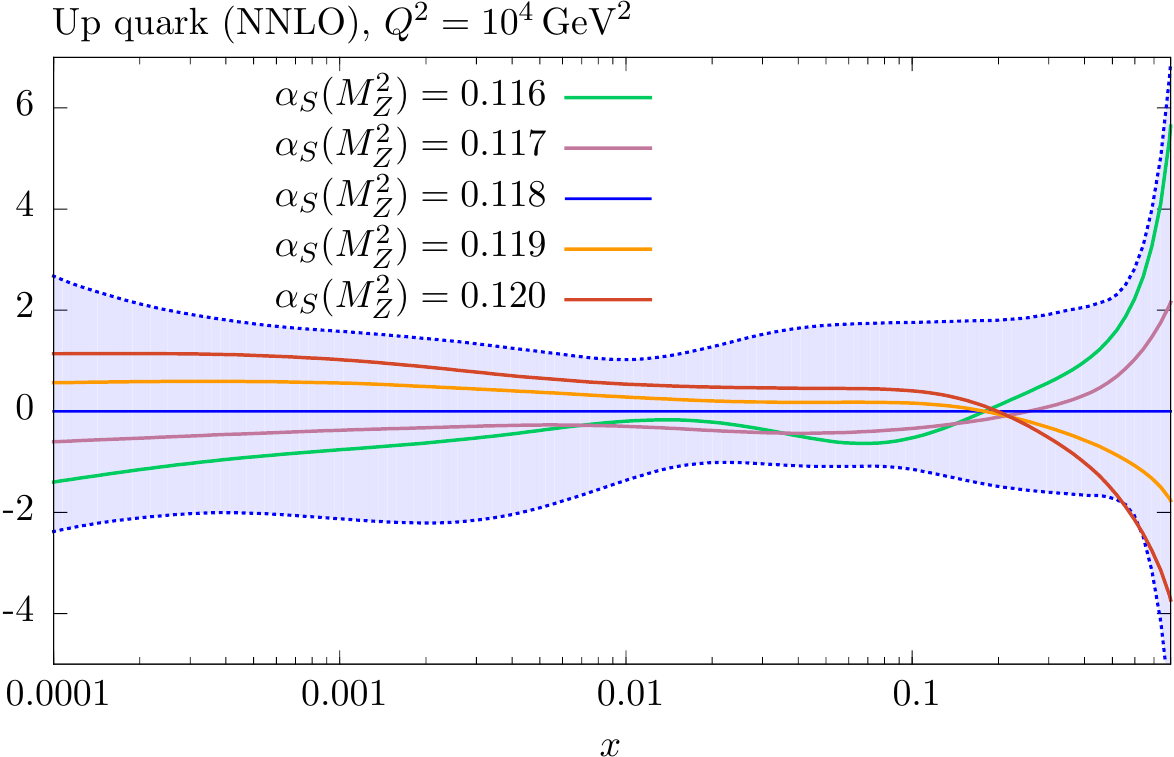}
\includegraphics[height=5cm]{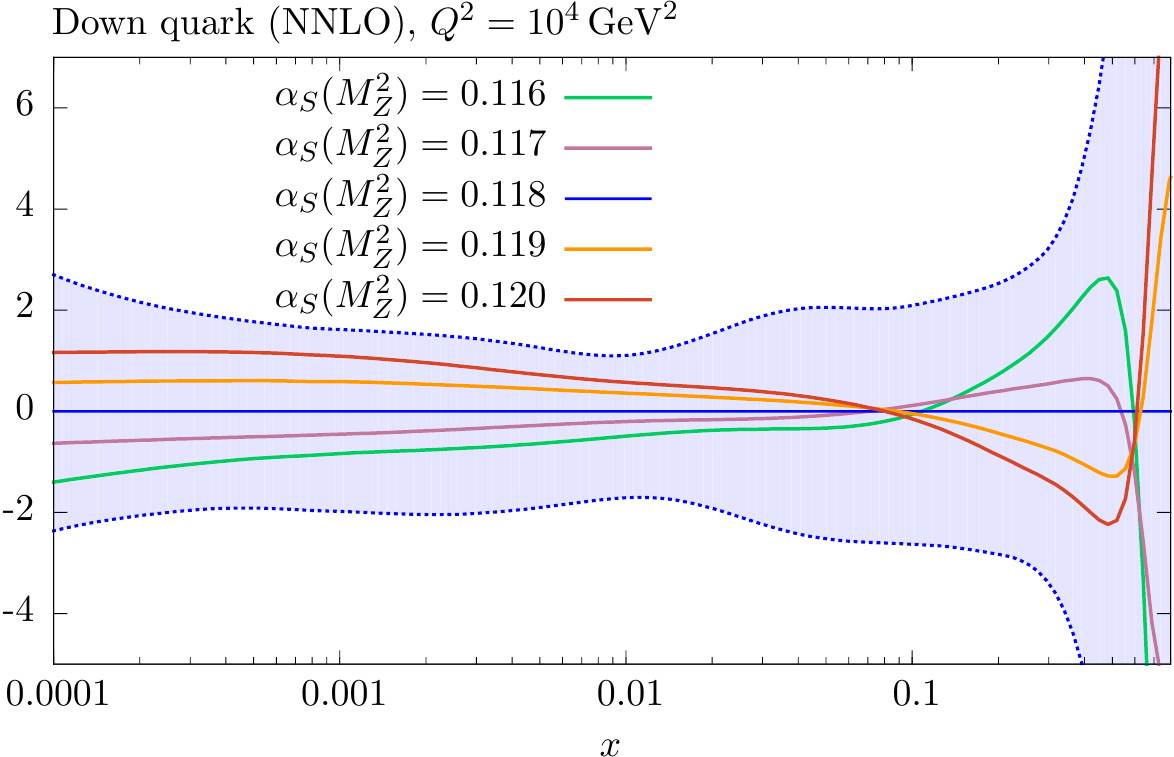}
\caption{Percentage difference in the NNLO up and down quark PDFs at $Q^2=10^4$ ${\rm GeV}^2$ relative to central ($\alpha_S(M_Z^2)=0.118$) set for fits with different values of $\alpha_S$, with the percentage error bands for the central set also shown.}
\label{fig:pdfcomp2}
\end{center}
\end{figure}

\begin{figure} 
\begin{center}
\includegraphics[height=5cm]{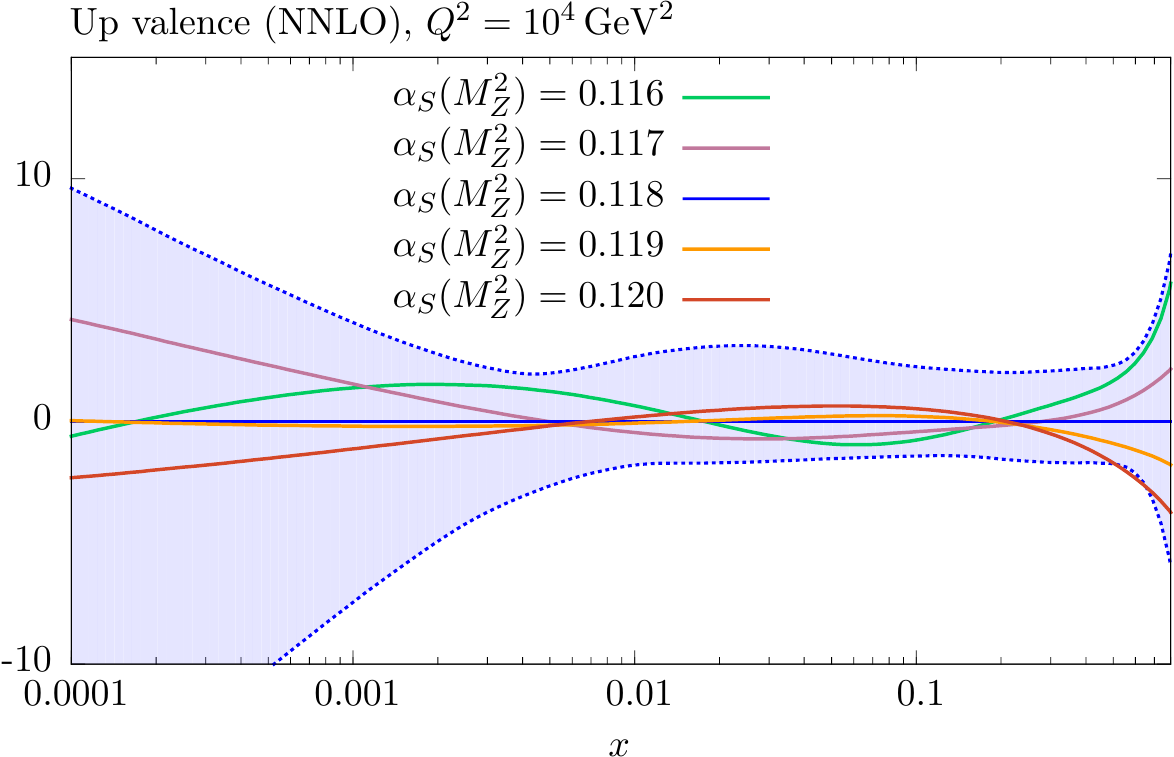}
\includegraphics[height=5cm]{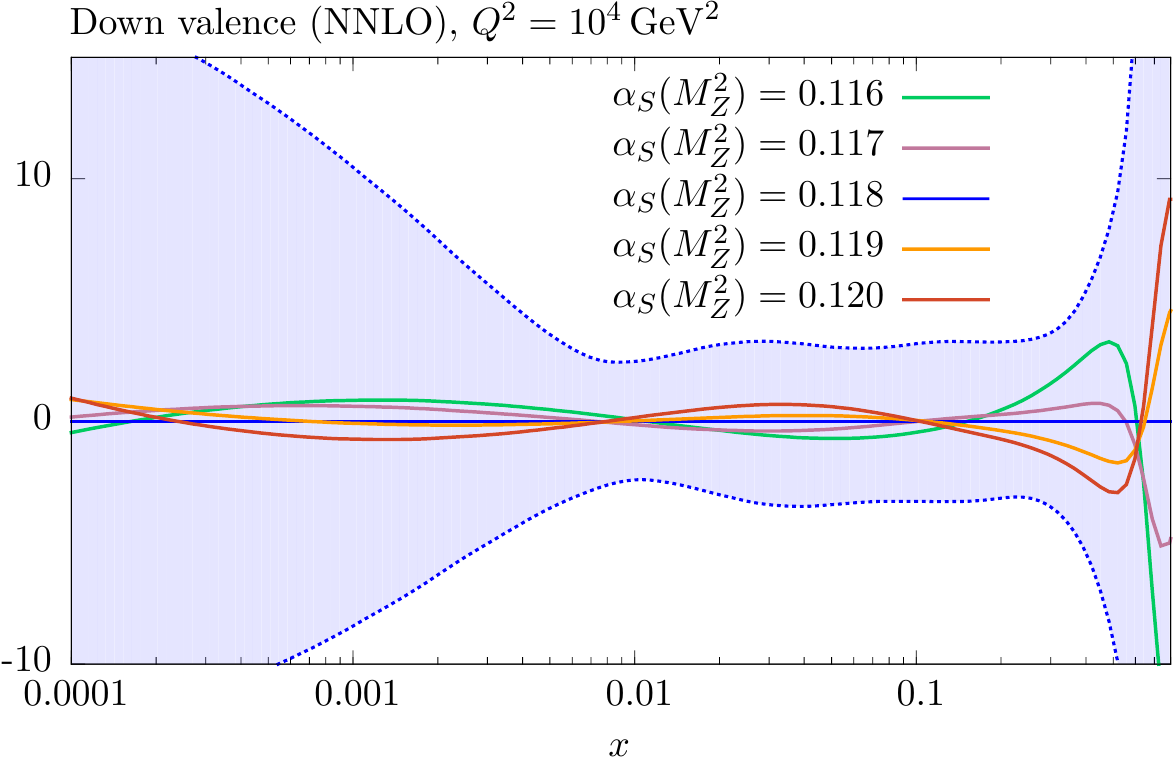}
\caption{Percentage difference in the NNLO up and down valence quark PDFs at $Q^2=10^4$ ${\rm GeV}^2$  relative to central ($\alpha_S(M_Z^2)=0.118$) set for fits with different values of $\alpha_S$, with the percentage error bands for the central set also shown.}
\label{fig:pdfcomp3}
\end{center}
\end{figure}

It is informative to see the changes in the PDFs obtained in global fits 
for fixed values of $\alpha_S(M_Z^2)$ relative to those obtained for the 
central value; we only consider the NNLO case here, but note that the NLO PDFs behave in a similar way.
These are shown in Figs.~\ref{fig:pdfcomp1}--\ref{fig:pdfcomp3} for the various PDFs as a function of $x$ for 
$Q^2=10^4$ GeV$^2$ -- a value of $Q^2$ relevant to data from the LHC. In 
almost every case the changes in the PDFs for the coupling varied in the 
range $0.116<\alpha_S(M_Z^2)<0.120$ are well within the PDF uncertainty bounds.

As expected, the gluon distribution for $x<0.1$ is larger for 
$\alpha_S(M_Z^2)=0.116$ and smaller for $\alpha_S(M_Z^2)=0.120$: a change 
which preserves the product $\alpha_S g$, which approximately determines the
evolution of $F_2(x,Q^2)$ with $Q^2$ at low $x$. This is the dominant 
constraint on the gluon, and a smaller low $x$ gluon leads to a larger high--$x$ 
gluon (and {\it vice versa}) due to the momentum sum rule. 
The $u$ and $d$ PDFs have 
the opposite trend as $\alpha_S(M_Z^2)$ changes. At small $x$ values this is a
marginal effect, due to the interplay of a variety of competing elements. 
At high $x$ the decreasing quark distribution with increasing $\alpha_S$ is 
due to the quicker evolution of quarks to lower $x$.  The insensitivity of 
the strange quark PDF to variations of $\alpha_S(M_Z^2)$ at low $x$ is 
partly just due to the relative insensitivity of all low--$x$ quarks, but 
is also partially explained by the comments in the previous section about the 
MMHT analysis \cite{MMHT} of dimuon production in neutrino interactions -- 
where the changes in $\alpha_S(M_Z^2)$ are, to some extent, compensated by 
changes in the $B(D\to \mu)$ branching ratio parameter.

\begin{figure} 
\begin{center}
\includegraphics[height=5cm]{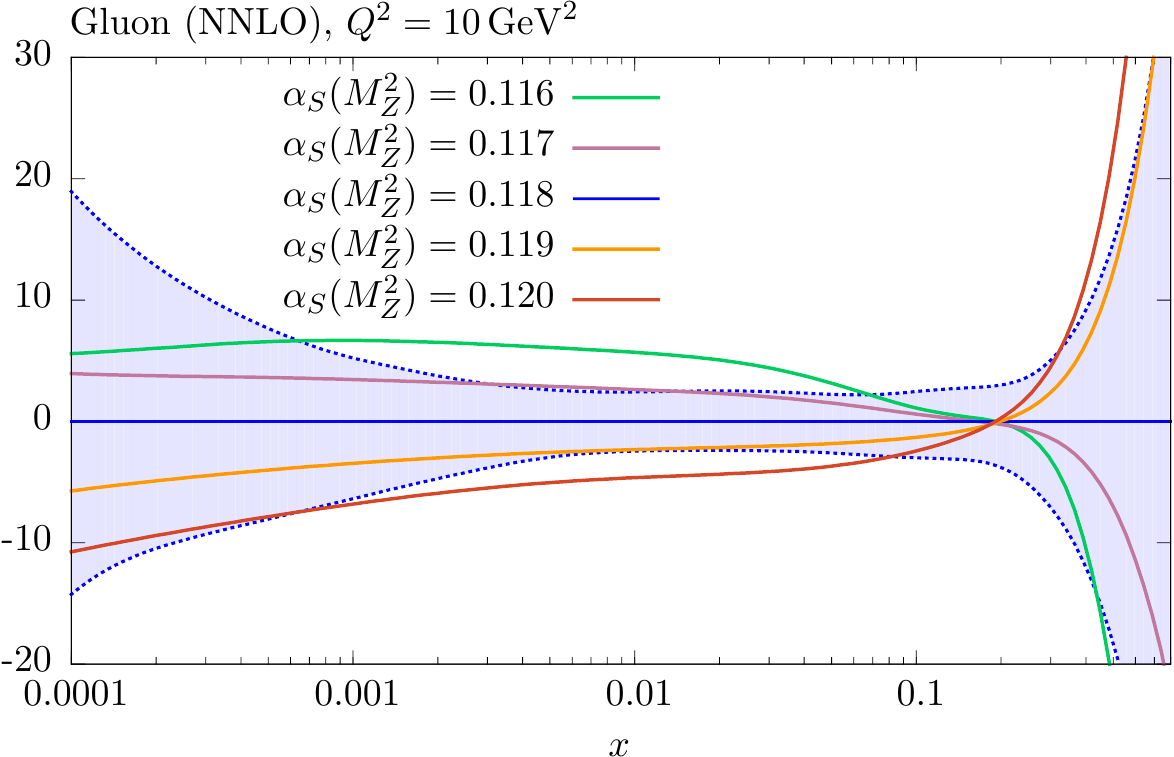}
\caption{Percentage difference in the NNLO gluon PDFs at $Q^2=10$ ${\rm GeV}^2$ relative to central ($\alpha_S(M_Z^2)=0.118$) set for fits with different values of $\alpha_S$, with the percentage error bands for the central set also shown.}
\label{fig:pdfcomp4}
\end{center}
\end{figure}

In Fig.~\ref{fig:pdfcomp4} we compare the changes in the gluon PDF for 
different fixed values of $\alpha_S(M_Z^2)$ at a much lower value of $Q^2$, 
namely $Q^2=10$ GeV$^2$.  Here the gluon PDF is much more sensitive to the 
value of $\alpha_S(M_Z^2)$, and the changes in the gluon PDF lie outside its 
uncertainty bounds. The message is clear. At the high value of $Q^2=10^4$ GeV$^2$ the long evolution length means that the gluon PDF in the relevant broad $x$ range about $x\sim 0.01$ is determined by PDFs at larger $x$, and is relatively insensitive to the parameters of the starting distributions.

\section{Benchmark cross sections \label{sec:cxunc}}

In this section  we show uncertainties for cross sections at the 
Tevatron, and for $7~\TeV$ and $14~\TeV$ at the LHC. Uncertainties for  
$8~\TeV$ and $13~\TeV$ will be very similar to those at $7~\TeV$ and $14~\TeV$
respectively. We calculate the cross sections for $W$ and $Z$ boson, Higgs boson via gluon--gluon fusion and top--quark pair production.

We calculate the PDF and $\a$ uncertainties for the MMHT2014
PDFs \cite{MMHT} at the default values of $\a$. We use a value of $\Delta \a=0.001$ as an
example, simply because PDF sets are readily available with $\a$ changes in units 
of $0.001$. However, for values similar to $\Delta \a=0.001$ a linear scaling
of the uncertainty can be applied to a very good approximation. As explained in 
Section \ref{sec:uncsq}, the full PDF+$\a$ uncertainty may then be obtained by adding the 
two uncertainties in quadrature. 

 To calculate the cross section we use the same procedure as was used in
\cite{MMHT}. That is, for $W, Z$ and Higgs production we use the code provided by W.J. Stirling, based on the
calculation in \cite{WZNNLO}, \cite{HiggsNNLO1} and \cite{HiggsNNLO2}, and for top pair production
we use the procedure and code of \cite{topNNLO}. Here our primary aim is not to present definitive predictions or to compare in detail to other PDF sets, as both these results are frequently provided in the literature with very specific choices of codes, scales and parameters which may differ from those used here.  Rather, our main objective is to illustrate the procedure for estimating realistic PDF$+\a$ uncertainties.

\begin{table}
\begin{center}
\renewcommand\arraystretch{1.25}
\begin{tabular}{|l|c|c|c|}
\hline
& $\sigma$& PDF unc.& $\alpha_S$ unc.  \\
\hline
$\!\! W\,\, {\rm Tevatron}\,\,(1.96~\TeV)$   & 2.782    & ${}^{+0.056}_{-0.056}$ $\left({}^{+2.0\%}_{-2.0\%}\right)$ & ${}^{+0.018}_{-0.020}$  $\left({}^{+0.65\%}_{-0.72\%}\right)$    \\   
$\!\! Z \,\,{\rm Tevatron}\,\,(1.96~\TeV)$   & 0.2559& ${}^{+0.0052}_{-0.0046}$  $\left({}^{+2.0\%}_{-1.8\%}\right)$ &${}^{+0.0015}_{-0.0018}$ $\left({}^{+0.59\%}_{-0.70\%}\right)$ \\    
\hline 
$\!\! W^+ \,\,{\rm LHC}\,\, (7~\TeV)$        &6.197    & ${}^{+0.103}_{-0.092}$ $\left({}^{+1.7\%}_{-1.5\%}\right)$  &  ${}^{+0.058}_{-0.065}$ $\left({}^{+0.94\%}_{-1.0\%}\right)$  \\    
$\!\! W^- \,\,{\rm LHC}\,\, (7~\TeV)$        & 4.306  & ${}^{+0.067}_{-0.076}$ $\left({}^{+1.6\%}_{-1.8\%}\right)$  & ${}^{+0.043}_{-0.043}$ $\left({}^{+1.0\%}_{-1.0\%}\right)$ \\    
$\!\! Z \,\,{\rm LHC}\,\, (7~\TeV)$          & 0.9638   & ${}^{+0.014}_{-0.013}$ $\left({}^{+1.5\%}_{-1.3\%}\right)$ & ${}^{+0.0091}_{-0.010}$ $\left({}^{+0.94\%}_{-1.0\%}\right)$  \\ \hline    
$\!\! W^+ \,\,{\rm LHC}\,\, (14~\TeV)$       & 12.48      & ${}^{+0.22}_{-0.18}$ $\left({}^{+1.8\%}_{-1.4\%}\right)$  & ${}^{+0.12}_{-0.14}$ $\left({}^{+0.97\%}_{-1.1\%}\right)$  \\    
$\!\! W^- \,\,{\rm LHC}\,\, (14~\TeV)$       & 9.32     &${}^{+0.15}_{-0.14}$ $\left({}^{+1.6\%}_{-1.5\%}\right)$   & ${}^{+0.098}_{-0.11}$ $\left({}^{+1.1\%}_{-1.2\%}\right)$    \\    
$\!\! Z \,\,{\rm LHC}\,\, (14~\TeV)$         & 2.065   & ${}^{+0.035}_{-0.030}$ $\left({}^{+1.7\%}_{-1.5\%}\right)$  & ${}^{+0.020}_{-0.025}$ $\left({}^{+0.97\%}_{-1.2\%}\right)$   \\ 
\hline
    \end{tabular}
\end{center}
\caption{\sf Predictions for $W^\pm$ and $Z$ cross sections (in nb), including leptonic branching, obtained with the NNLO MMHT2014 parton sets. The PDF and $\alpha_S$ uncertainties are also shown, where the $\alpha_S$ uncertainty corresponds to a variation of $\pm 0.001$ around its central value.  The full PDF$+\a$ uncertainty is obtained by adding these two uncertainties in quadrature, as explained in Section~\ref{sec:uncsq}.}
\label{tab:sigmaWZNNLO}   
\end{table}

\subsection{$W$ and $Z$ production}
We begin with the predictions for the $W$ and $Z$ production cross sections. 
The results at NNLO are shown in Table \ref{tab:sigmaWZNNLO}.
In this case the cross sections contain zeroth--order contributions in $\alpha_S$, with positive NLO corrections of about 
$20\%$, and  much smaller NNLO contributions. Hence a smaller than $1\%$ change in $\a$ will only directly 
increase the cross section by a small fraction of a percent. The PDF uncertainties on the cross 
sections are $2\%$ at the Tevatron and slightly smaller at the LHC -- the lower beam energy at the
Tevatron meaning the cross sections have more contribution from higher $x$ where PDF uncertainties
increase. The $\alpha_S$ uncertainty is small, about $0.6\%$ at the Tevatron and close to $1\%$ at 
the LHC, being slightly larger at 14~TeV than at 7~TeV. Hence, the $\alpha_S$ uncertainty is small, but
more than the small fraction of a percent expected from the direct change in the cross section with 
$\alpha_S$. In fact the main increase in cross sections with $\alpha_S$ is due to the change in 
the PDFs with the coupling, rather than its direct effect on the cross section. From Fig.~\ref{fig:pdfcomp2} 
 we see that the up and down quark PDFs increase with $\alpha_S$ below 
$x\sim 0.1-0.2$ due to increased speed of evolution. From Fig.~\ref{fig:pdfcomp1} we note that the strange quark PDF  increases a little
with $\alpha_S$ at all $x$ values. As already mentioned the Tevatron cross sections are more sensitive to the high--$x$ 
quarks, which decrease with increasing $\alpha_S$, so this introduces a certain amount of anti--correlation 
of the cross section with $\alpha_S$. However, the main contribution is from low enough $x$ that
the distributions increase with $\alpha_S$, so the net effect is an increase with $\alpha_S$ a little 
larger than that coming directly from the $\alpha_s$ dependence of the cross section. As the energy increases at the LHC the contributing 
quarks move on average to lower $x$ and the increase of the cross section with $\alpha_S$ increases --
very slightly more so at 14~TeV than at 7~TeV. However, even at 14~TeV the total PDF+$\alpha_S$ uncertainty 
obtained by adding the two contributions in quadrature, is only a maximum of about $25\%$ greater (for $W^-$) 
than the PDF uncertainty alone if $\Delta \a = 0.001$ is used.

\begin{table}
\begin{center}
\renewcommand\arraystretch{1.25}
\vspace{0.5cm}
\begin{tabular}{|l|c|c|c|}
\hline
& $\sigma$& PDF unc.& $\alpha_S$ unc.  \\
\hline
$t\overline{t}$ $ {\rm Tevatron}\,\,(1.96~\TeV)$   & 7.51    & ${}^{+0.21}_{-0.20}$ $\left({}^{+2.8\%}_{-2.7\%}\right)$ & ${}^{+0.17}_{-0.15}$  $\left({}^{+2.3\%}_{-2.1\%}\right)$    \\   
$t\overline{t}$  ${\rm LHC}\,\, (7~\TeV)$        &175.9    & ${}^{+3.9}_{-5.5}$ $\left({}^{+2.2\%}_{-3.1\%}\right)$  &  ${}^{+4.1}_{-3.3}$ $\left({}^{+2.3\%}_{-1.9\%}\right)$  \\    
$t\overline{t}$ ${\rm LHC}\,\, (14~\TeV)$       & 969.9   &${}^{+16}_{-20}$ $\left({}^{+1.6\%}_{-2.1\%}\right)$  & ${}^{+16}_{-14}$ $\left({}^{+1.6\%}_{-1.4\%}\right)$  \\    
\hline
    \end{tabular}
\end{center}
\caption{\sf Predictions for $t\overline{t}$ cross sections (in nb), obtained with the NNLO MMHT2014 parton sets. The PDF and $\alpha_S$ uncertainties are also shown, where the $\alpha_S$ uncertainty corresponds to a variation of $\pm 0.001$ around its central value. The full PDF$+\a$ uncertainty is obtained by adding these two uncertainties in quadrature, as explained in Section~\ref{sec:uncsq}.}
\label{tab:sigmatNNLO}   
\end{table}

\subsection{Top-quark pair production}
In Table \ref{tab:sigmatNNLO} we show the analogous results for the top--quark pair  production cross section. 
At the Tevatron the PDFs are probed in the region $x\approx 0.4/1.96\approx 0.2$, and the main production is from the $q{\bar q}$ channel.  As we saw, the quark distributions are reasonably insensitive to $\a$ in this region of 
$x$, as it is the approximate pivot point of the PDFs. Hence, there is only a small change in cross section due to changes 
in the PDFs with $\alpha_S$. However, the cross section for $t{\bar t}$ production begins at order 
$\alpha_S^2$, and there is a significant positive higher--order correction at NLO and 
still an appreciable one at NNLO. Therefore, a change in $\alpha_S$ a little lower than $1\%$ should give a direct 
change in the cross section of about $2\%$. This is roughly the change that is observed. This is compared 
to a PDF only uncertainty of nearly $3\%$ due to sensitivity to higher $x$ quarks that occurs for $W,Z$ production. 

At the LHC the dominant production at higher energies (and with a proton--proton rather than 
proton--antiproton collider) is gluon--gluon fusion, with the central $x$ value probed being
$x\approx 0.4/7 \approx 0.06$ at 7~TeV, and $x\approx 0.4/14\approx 0.03$ at 14~TeV. As seen from the left plot of Fig. \ref{fig:pdfcomp1} the 
gluon decreases with increasing $\a$ below $x=0.1$ and the maximum decrease is for $x\sim 0.02-0.03$. 
The $\a$ uncertainty on $\sigma_{t \bar t}$ for 7~TeV is about $2\%$, almost as large as at the 
Tevatron, with the gluon above the pivot point still contributing considerably to the cross section, so 
the indirect $\a$ uncertainty due to PDF variation largely cancels. For 14~TeV the lower $x$ probed 
means that most contribution is below the pivot point and there is some anti--correlation between    
the direct $\alpha_S$ variation and the indirect, with a reduced $\alpha_S$ uncertainty of $1.5\%$. 
At this highest energy the PDF only uncertainty has also reduced to about $2\%$ due to the decreased sensitivity
to the uncertainty in high--$x$ PDFs, the gluon in this case. At the Tevatron and for 7~TeV at the LHC the 
$\a$ uncertainty is a little smaller than the PDF uncertainty, and the total is about 1.3 times the
PDF uncertainty alone. At 14~TeV they are very similar in size, so the total uncertainty, for
 $\Delta \a = 0.001$  is about $\sqrt{2}$ that of the PDF uncertainty.      

\begin{table}
\begin{center}
\renewcommand\arraystretch{1.25}
\begin{tabular}{|l|c|c|c|}
\hline
& $\sigma$& PDF unc.& $\alpha_S$ unc.  \\
\hline
Higgs $ {\rm Tevatron}\,\,(1.96~\TeV)$   & 0.874    & ${}^{+0.024}_{-0.030}$ $\left({}^{+2.7\%}_{-3.4\%}\right)$ & ${}^{+0.022}_{-0.018}$  $\left({}^{+2.5\%}_{-2.1\%}\right)$    \\   
Higgs  ${\rm LHC}\,\, (7~\TeV)$        &14.56    & ${}^{+0.21}_{-0.29}$ $\left({}^{+1.4\%}_{-2.0\%}\right)$  &  ${}^{+0.23}_{-0.22}$ $\left({}^{+1.6\%}_{-1.5\%}\right)$  \\    
Higgs ${\rm LHC}\,\, (14~\TeV)$       & 47.69   &${}^{+0.63}_{-0.88}$ $\left({}^{+1.3\%}_{-1.8\%}\right)$  & ${}^{+0.71}_{-0.70}$ $\left({}^{+1.5\%}_{-1.5\%}\right)$  \\    
\hline
    \end{tabular}
\end{center}
\caption{\sf Predictions for the Higgs boson cross sections (in nb), obtained with the NNLO MMHT 2014 parton sets. The PDF and $\alpha_S$ uncertainties are also shown, where the $\alpha_S$ uncertainty corresponds to a variation of $\pm 0.001$ around its central value. The full PDF$+\a$ uncertainty is obtained by adding these two uncertainties in quadrature, as explained in Section~\ref{sec:uncsq}.}
\label{tab:sigmahNNLO}   
\end{table}

\subsection{Higgs boson production}
In Table \ref{tab:sigmahNNLO} we show the uncertainties in the rate of Higgs boson production from gluon--gluon fusion. Again, the cross section starts at order $\alpha^2_S$ and there are large positive NLO and NNLO contributions. Hence,
changes in $\alpha_S$ of about $1\%$ would be expected to lead to direct changes in the cross section
of about $3\%$. However, even at the Tevatron the dominant $x$ range probed, i.e. $x\approx 0.125/1.96 \approx 0.06$, corresponds to a region where
the gluon distribution falls with increasing $\a$ and at the LHC where $x \approx 0.01-0.02$ at central rapidity the 
anti--correlation between $\a$ and the gluon distribution is near its maximum. Hence, at the Tevatron the
total $\a$ uncertainty is a little less than the direct value at a little more than $2\%$, and at the LHC it is reduced to
$1.5\%$. In the former case this is a little less than the PDF uncertainty of $\sim 3\%$, with some sensitivity to 
the relatively poorly constrained high--$x$ gluon, while at the LHC the PDF uncertainty is much reduced due to the 
smaller $x$ probed, and is similar to the $\a$ uncertainty. Hence for $\Delta \a = 0.001$ the uncertainty on the 
Higgs boson cross 
section from gluon--gluon fusion  is about $\sqrt{2}$ that of the PDF uncertainty alone.

\begin{table}
\begin{center}
\renewcommand\arraystretch{1.25}
\vspace{0.5cm}
\begin{tabular}{|l|c|c|c|}
\hline
& $\sigma$& PDF unc.& $\alpha_S$ unc.  \\
\hline
Higgs $ {\rm Tevatron}\,\,(1.96~\TeV)$   & 0.644    & ${}^{+0.021}_{-0.022}$ $\left({}^{+3.3\%}_{-3.4\%}\right)$ & ${}^{+0.011}_{-0.0088}$  $\left({}^{+1.7\%}_{-1.4\%}\right)$    \\   
Higgs  ${\rm LHC}\,\, (7~\TeV)$        &11.28    & ${}^{+0.21}_{-0.20}$ $\left({}^{+1.9\%}_{-1.8\%}\right)$  &  ${}^{+0.15}_{-0.14}$ $\left({}^{+1.3\%}_{-1.2\%}\right)$  \\    
Higgs ${\rm LHC}\,\, (14~\TeV)$       & 37.63   &${}^{+0.67}_{-0.59}$ $\left({}^{+1.8\%}_{-1.6\%}\right)$  & ${}^{+0.51}_{-0.50}$ $\left({}^{+1.4\%}_{-1.3\%}\right)$  \\    
\hline
    \end{tabular}
\end{center}
\caption{\sf Predictions for Higgs Boson cross sections (in nb), obtained with the NLO MMHT 2014 parton sets.The PDF and $\alpha_s$ are shown, with the $\alpha_s$ uncertainty corresponding to a variation of $\pm 0.001$ around the central value ($\alpha_S(M_Z^2)=0.120$). The full PDF$+\a$ uncertainty is obtained by adding these two uncertainties in quadrature, as explained in Section~\ref{sec:uncsq}.}
\label{tab:sigmahNLO120}   
\end{table}

We also repeat the study at NLO for the Higgs cross section. The results are shown in 
Tables \ref{tab:sigmahNLO120} and \ref{tab:sigmahNLO118} for the central values of 
$\a=0.120$ and $\a=0.118$ respectively. The uncertainties are very different in the two cases, with the central values of the cross sections being
about $3\%$ lower for $\a=0.118$ than for $\a=0.120$. Both sets of predictions are 
about $30\%$ lower than at NNLO, highlighting the large NNLO correction for this process. 
The PDF uncertainties are very similar to those at NNLO, though a little larger in detail. 
However, the $\a$ uncertainties are noticeably reduced, as the large 
variation in the NNLO (${\cal O}(\alpha_S^4)$) cross section with $\alpha_S$ is now absent.

\begin{table}
\begin{center}
\renewcommand\arraystretch{1.25}
\vspace{0.5cm}
\begin{tabular}{|l|c|c|c|}
\hline
& $\sigma$& PDF unc.& $\alpha_S$ unc.  \\
\hline
Higgs $ {\rm Tevatron}\,\,(1.96~\TeV)$   & 0.625    & ${}^{+0.022}_{-0.019}$ $\left({}^{+3.5\%}_{-3.0\%}\right)$ & ${}^{+0.0096}_{-0.0086}$  $\left({}^{+1.5\%}_{-1.4\%}\right)$    \\   
Higgs  ${\rm LHC}\,\, (7~\TeV)$        &11.00    & ${}^{+0.21}_{-0.19}$ $\left({}^{+1.9\%}_{-1.7\%}\right)$  &  ${}^{+0.15}_{-0.14}$ $\left({}^{+1.4\%}_{-1.3\%}\right)$  \\    
Higgs ${\rm LHC}\,\, (14~\TeV)$       & 36.61   &${}^{+0.65}_{-0.56}$ $\left({}^{+1.8\%}_{-1.5\%}\right)$  & ${}^{+0.51}_{-0.50}$ $\left({}^{+1.4\%}_{-1.4\%}\right)$  \\    
\hline
    \end{tabular}
\end{center}
\caption{\sf Predictions for Higgs Boson cross sections (in nb), obtained with the NLO MMHT 2014 parton sets.The PDF and $\alpha_s$ are shown, with the $\alpha_s$ uncertainty corresponding to a variation of $\pm 0.001$ around the central value ($\alpha_S(M_Z^2)=0.118$). The full PDF$+\a$ uncertainty is obtained by adding these two uncertainties in quadrature, as explained in Section~\ref{sec:uncsq}.}
\label{tab:sigmahNLO118}   
\end{table}

\section{Conclusions}

The PDFs  determined from global fits to deep--inelastic and related hard--scattering data are highly correlated to the value of $\a$ used, and any changes in 
the values of $\a$ must be accompanied by changes in the PDFs such that 
the optimum fit to data is still obtained. In~\cite{MMHT} we produced PDF and uncertainty eigenvector sets for specific values of $\a$, guided by the values obtained when it was left as a free parameter in the fit. In this article we explicitly present PDF sets
and the global fit quality 
at NLO and NNLO for a wide variety of $\a$ values, i.e. $\a=0.108$ to 
$\a=0.128$ in steps of $\Delta \a =0.001$. Hence, we illustrate in more 
detail the origin of our best fit $\a$ values of  
\begin{align}
  \text{NLO:}\qquad\alpha_S(M_Z^2) &= 0.1201 \pm 0.0015 \text{ (68\% C.L.)},\\
  \text{NNLO:}\qquad\alpha_S(M_Z^2) &= 0.1172 \pm 0.0013 \text{ (68\% C.L.)},
\end{align}
already presented in \cite{MMHT}, but also present the uncertainties. 
We show the variation of the fit quality with $\a$ of each data set, 
within the context of the global fit, and see which are the more and less 
constraining sets, and which prefer higher and lower values. We see that 
most data sets show a systematic trend of preferring a slightly lower 
$\a$ value at NNLO than at NLO, but note that no particular type of 
data strongly prefers a high or low value of $\a$. HERA and Tevatron data 
tend to prefer higher values, but are not the most constraining data.
There are examples of 
fixed target DIS data which prefer either high or low values and 
similarly for the LHC data sets, which are new compared to our previous 
analysis~\cite{MSTWalpha}. Indeed our best values of 
$\a$ are almost unchanged from $\a=0.1202$ (NLO) and $\a=0.1171$ (NNLO). 
They are also very similar to the values obtained by NNPDF of 
$\a=0.1191$ (NLO)\cite{Lionetti:2011pw} and $\a=0.1173$ 
(NNLO)~\cite{Ball:2011us}. However, our extraction disagrees with the recent 
value $\a=0.1132$ (NNLO) in \cite{Alekhin:2013nda}. 
We find agreement at the level of one sigma or less
with the world average value of $\a=0.1187 \pm 0.0005$, and this improves when 
we include the world average (without the DIS determinations included) 
as a data point in our fit, when we obtain $\a=0.1195$ (NLO) and 
$\a=0.1178$ (NNLO). Hence, our NNLO value including $\a$ as an external
constraint is in excellent agreement with the preferred value, $\a=0.118$, for
which eigenvector sets are made available. The PDF sets obtained at the 21
different values of $\a$ at NLO and NNLO can be found at~\cite{UCLsite} 
and are available from the LHAPDF library~\cite{LHAPDF}. They should be useful in studies of 
$\a$ by other groups. 

In order to calculate the PDF$+\a$ uncertainty we now advocate the 
approach pioneered in \cite{CTEQalphas} of treating PDFs with 
$\a\pm \Delta \a$ as an extra eigenvector set. As shown in \cite{CTEQalphas},
provided certain conditions are met (at least approximately), the 
$\a$ uncertainty may be correctly added to the PDF uncertainty by simply adding 
in quadrature the variation of any quantity under a change in coupling 
$\Delta \a$ as long as the
change in $\a$ is accompanied by the appropriate change in PDFs required by the 
global fit. As examples, we have calculated the total cross sections for the production 
of $W$, $Z$, top quark pairs and Higgs bosons at the Tevatron and LHC.  
For $W$ and $Z$ production, where the LO subprocess is 
${\cal O}(\alpha_S^0)$ and is quark-initiated, the combined ``PDF+$\alpha_S$'' 
uncertainty is not much larger than the PDF-only uncertainty with a fixed 
$\alpha_S$.  However, the additional uncertainty due to $\alpha_S$ is more 
important for top quark pair production and Higgs boson production via 
gluon--gluon fusion, since the LO subprocess now is ${\cal O}(\alpha_S^2)$,
though the details depend on the correlation between $\a$ and the contributing 
PDFs. 

In addition, we note that for any particular process the details of the 
uncertainty can now be explicitly calculated in a straightforward way using 
the PDFs we have provided in this paper, together with the procedure 
for combining PDF and $\a$ uncertainty discussed in Section~\ref{sec:uncsq}.

 Moreover, it is 
 also straightforward to apply the procedure to determine the uncertainties  coming from combinations of PDF sets obtained by global analyses of different groups. 
Using techniques given in 
\cite{WattThorne,Gao:2013bia,Carrazza:2015hva,Carrazza:2015aoa} it is possible 
to combine different PDF sets at a preferred value of $\a$ such that the central 
value and the uncertainty of the combination is correctly obtained. The procedure to determine the uncertainty due to variations of $\a$ is as follows.
If each group used in the combination also makes available sets of 
PDFs obtained by repeating their global fits\footnote{For instance, if $\a=0.118$ is the preferred value then repeating global fits at $\a=0.117$ and $\a=0.119$ would be sufficient to quantify the uncertainty due variations of $\alpha_S$.} with $\a \pm \Delta \a$, then an additional pair 
of PDF sets representing the $\a$ variation of the combination can  be obtained 
just by taking 
the average of the PDFs from each group obtained at 
$\a+ \Delta \a$,  and by taking the average at $\a- \Delta \a$. As a result the PDF$+\a$
uncertainty for any quantity calculated using the combined set is just the PDF 
induced uncertainty at the 
preferred value of $\a$ added in quadrature to the $\a$ uncertainty 
determined from the two 
combined sets defined at $\a \pm \Delta \a$. Hence, a user may determine for any process
the optimum prediction, the PDF uncertainty, the $\a$ uncertainty and the complete PDF$+\a$ uncertainty arising from the combination of a whole collection of different PDFs.

\section*{Acknowledgements}

We particularly thank W. J. Stirling  and G. Watt for numerous discussions on PDFs
and for previous work without which this study would not be possible. We would like
to thank  Mandy Cooper--Sarkar, Albert de Roeck, Stefano Forte, Joey Huston, Pavel Nadolsky and Juan Rojo 
for various discussions on the relation between PDFs and $\alpha_S$.  
This work is supported partly by the London Centre for Terauniverse Studies (LCTS),
using funding from the European Research Council via the Advanced
Investigator Grant 267352. RST would also like to thank the IPPP, Durham, for
the award of a Research Associateship held while most of this
work was performed. We thank the
Science and Technology Facilities Council (STFC) for support via grant
awards ST/J000515/1 and ST/L000377/1.

\bibliography{references}{}
\bibliographystyle{h-physrev}

\end{document}